\documentclass[useAMS,usenatbib,usegraphicx]{mn2e}

\newcommand{\uchii}{{UC H{\scriptsize II} }}
\newcommand{\hii}{{H{\scriptsize II} }}

\title[Mopra mapping of NGC\,6334\,I and I(N)]{Mopra line survey mapping of NGC\,6334\,I and I(N) at 3\,mm}
\author[A.~J. Walsh]{A. J. Walsh$^{1,2}$\thanks{E-mail:
Andrew.Walsh@jcu.edu.au}, S. Thorwirth$^{3,4}$, H. Beuther$^{5}$ and M.~G. Burton$^{1}$\\
$^{1}$School of Physics, University of New South Wales, Sydney, 2052 NSW, Australia\\
$^{2}$Centre for Astronomy, School of Engineering and Physical Sciences, James Cook University, Townsville, QLD 4811, Australia\\
$^{3}$I. Physikalisches Institut, Universit\"at zu K\"oln, Z\"ulpicher Str. 77, 50937 K\"oln, Germany\\
$^{4}$Max-Planck-Institut f\"ur Radioastronomie, Auf dem H\"ugel 69, 53121 Bonn, Germany\\
$^{5}$Max-Planck-Institut f\"ur Astronomie, K\"onigstuhl 17, 69117 Heidelberg, Germany}

\begin{document}



\pagerange{\pageref{firstpage}--\pageref{lastpage}} \pubyear{2007}

\maketitle

\label{firstpage}

\begin{abstract}
A 5\arcmin$\times$5\arcmin\ region encompassing NGC\,6334\,I and I(N) has been mapped at a
wavelength of 3\,mm (from 83.5 to 115.5\,GHz) with the Mopra telescope at an angular
resolution between 33\arcsec\ and 36\arcsec. This investigation has made use of the
recently installed 3\,mm MMIC receiver and the Mopra Spectrometer (MOPS) with broadband
capabilities permitting total coverage of the entire frequency range with just five
different observations. In total, the spatial distribution of nineteen different molecules,
ions and radicals, along with additional selected isotoplogues have been studied.
Whilst most species trace the sites of star formation, CH$_3$CN appears to be most closely
associated with NGC\,6334\,I and I(N). Both CN and C$_2$H appear to be widespread, tracing
gas that is not associated with active star formation. Both N$_2$H$^+$ and HC$_3$N closely
resemble dust continuum emission, showing they are reliable tracers of dense material,
as well as the youngest stages of high mass star formation.
Hot (E$_u/k>$100\,K) thermal CH$_3$OH emission is preferentially found towards NGC\,6334\,I,
contrasting with I(N), where only cold (E$_u/k<$22\,K) thermal CH$_3$OH emission is found.
\end{abstract}


\section{Introduction}

NGC\,6334 is a giant molecular cloud located at a distance of 1.7\,kpc \citep{neckel78}
in the southern Galactic plane. Lying along a gas filament of 11\,pc in length, NGC\,6334 exhibits
several luminous sites of massive star formation -- as seen in the far-infrared
(\citealt{mcbreen79}, five sources denoted using roman numerals I to V) and radio continuum
(\citealt{rodriguez82}, six sources denoted with letters from A to F) with source
I dominating the millimetre to the far infrared \citep{sandell00}.
In close proximity to NGC\,6334\,I, about two arcminutes to the north, another bright source is found 
in the millimetre- and submillimetre continuum, denoted NGC\,6334\,I(N)
\citep{cheung78,gezari82,sandell00}.
While this source is believed to be a comparably young object -- as concluded from the lack
of H\,{\sc ii} regions and mid-IR emission -- there is compelling evidence however,
that star formation is going on there \citep[e.g.][]{megeath99}.
It is this twin core system, NGC\,6334\,I and I(N), that offers the rare opportunity
to study the evolution of high mass stars from the same parental cloud and in a relatively
small spatial region.

Single dish molecular line observations show both cores to be chemically rich
\citep{thorwirth03} with source I being
comparable in line density to prototypical hot cores such as Orion-KL and SgrB2(N)
\citep{bachiller90,mccutcheon00,thorwirth03,schilke06}. 
High spatial resolution observations towards both cores have been conducted:
ATCA investigations of NH$_3$ emission up to the (6,6) inversion transition reveals the presence
of warm gas in both cores \citep{beuther05,beuther07} and
SMA continuum observations at 1.3\,mm \citep{hunter06} resolve each core into a
sample of sub-cores of several tens of solar masses each, four for source I and
seven for I(N), demonstrating the formation of star clusters.

Several line mapping studies using standard molecular tracers were carried out in the past
to obtain information about the large scale spatial distribution of molecular gas
\citep{kraemer99,mccutcheon00} additionally leading to the
detection of several molecular outflows associated with both cores
\citep{bachiller90,megeath99,leurini06}.

In the present investigation we have used the Mopra telescope for spectral line mapping of a
5\arcmin$\times$5\arcmin\ region around NGC\,6334\,I and I(N) throughout the entire 3\,mm range.

\section{Observations}
\label{obs}

Observations were carried out in 2006 on June 25th, using the 22m Mopra telescope near Coonabarabran,
New South Wales, Australia\footnote{For detailed information on the facility visit the Mopra www page at
http://www.narrabri.atnf.csiro.au/mopra/}.
We used the recently commissioned 3\,mm MMIC receiver in combination with the new Mopra spectrometer, MOPS,
in broadband mode\footnote{Detailed information on the available receivers and backends can be found online
in the ``Technical Summary of the Mopra Radiotelescope'' available at
http://www.narrabri.atnf.csiro.au/mopra/mopragu.pdf} resulting in an instantaneous bandwidth of 8\,GHz split over
four overlapping IFs of 2.2\,GHz each.
At the time of the observations MOPS provided 1024 channels per IF, giving a total of 8192 channels. The
spectral resolution of the observations was 2\,MHz corresponding to a velocity resolution of about 6\,km\,s$^{-1}$
per channel. For NGC\,6334, this velocity resolution is about the same, or slightly larger than
expected line widths, which
means while we do not expect to spectrally resolve many lines, we will not suffer a significant loss in signal
due to spectral smearing. Thus, this work will necessarily concentrate on morphologies and integrated intensities
of emission, rather than any analysis of kinematic structure.

The pointing centre for the maps was chosen to be RA(J2000) = 17 20 54.24, Dec(J2000) = -35 46 11.5,
which is approximately half way between NGC\,6334\,I and I(N). We used the on-the-fly mapping routine to map
an area of 5\arcmin$\times$5\arcmin~around this position. Due to incomplete sampling of the edges of the map,
a slightly smaller area is considered in this work. A scanning rate of 3.5 arcsec\,s$^{-1}$ was used.
Data were averaged every 2 seconds which gives an optimum data collection rate with minimal smearing of
the output. A spacing of 10 arcseconds was used between each scanning row. Assuming a beam of between 33
and 36\arcsec, depending on the frequency \citep{ladd05}, this observing mode resulted
in a fully sampled map. A single map was obtained in
about 90 minutes, including about 15\% of this time for T$_{\rm SYS}$ calibration, reference observations
and pointing on a nearby SiO maser (AH Sco). Poor weather (clouds) affected some of the data, which usually appears
as faint horizontal stripes in the data (see the following sections for details).

\section{Results}
Figure \ref{spectra} shows spectra at I and I(N) across the 3\,mm band.
Each spectrum was made by integrating the emission over a box approximately
equal to the size of the beam. The spectrum for NGC\,6334\,I was centred on
the methanol maser site that overlaps with the \uchii region (17 20 53.35, -35 47 1.6)
and the spectrum for NGC\,6334\,I(N) was centred on the methanol maser site
(17 20 54.58 -35 45 8.6). Across the band, we
detect a total of 52 transitions from 19 species. Details of each detection
are given in Table \ref{tab1}. For each of these transitions, we provide a map
of the emission, shown in Figure \ref{images}.

\begin{figure*}
\includegraphics[width=0.5\textwidth]{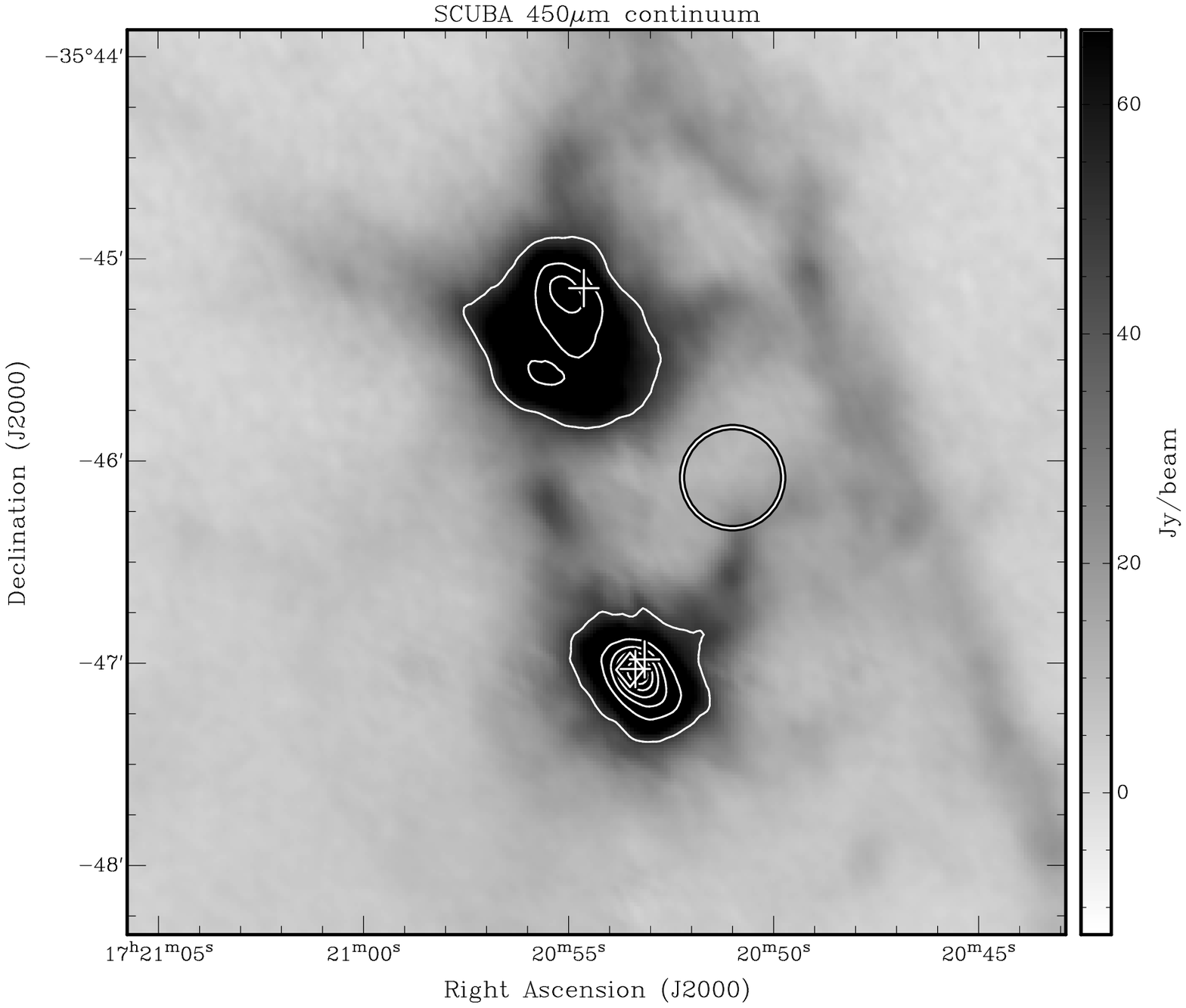}
\includegraphics[width=0.5\textwidth]{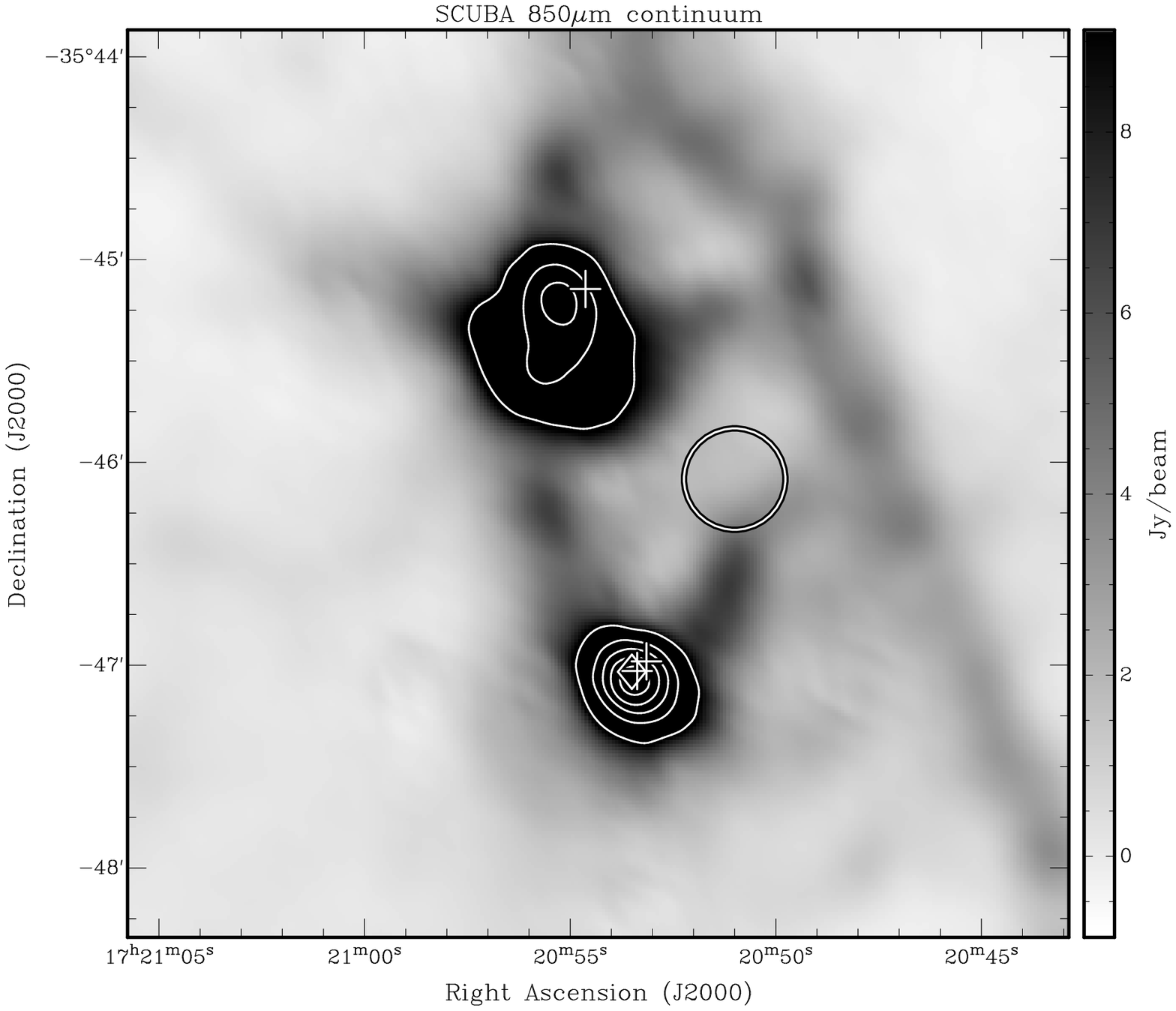}
\caption{Dust continuum images taken with SCUBA on the JCMT, from G\"{o}ren Sandell. The large circle
represents the approximate size and position of the \hii region NGC\,6334\,E \citep{carral02}.
The diamond represents the position of the \uchii region NGC\,6334\,F, also known as
NGC\,6334\,I. The two plus
symbols next to the diamond represent the position of two methanol maser sites
\citep{walsh98}. The single plus symbol in the northern half of each image represents
the position of NGC\,6334\,I(N), traced by a methanol maser site \citep{walsh98}.
{\bf (top)} 450$\mu$m
image. The lowest contour is at 50\,Jy\,beam$^{-1}$ and contours are spaced at
50\,Jy\,beam$^{-1}$. The peak flux density in the map is
369\,Jy\,beam$^{-1}$. {\bf (bottom)} 850$\mu$m image. The lowest contour is at
10\,Jy\,beam$^{-1}$ and contours are spaced at 10\,Jy\,beam$^{-1}$.
The peak flux density in the map is 59\,Jy\,beam$^{-1}$.}
\label{continuum}
\end{figure*}

Figure \ref{continuum} shows dust continuum emission in the region taken using SCUBA
on the JCMT (G\"{o}ran Sandell, {\em private communication}). These images were taken on June 26th and July 1st, 1998,
with measured half-power beam widths of $14.6''$ at 850$\mu$m and $8''$ at 450$\mu$m.
In this Figure, the plus symbols represent the positions of methanol maser emission \citep{walsh98}.
In the north, the methanol maser site marks the position of NGC\,6334\,I(N). In the south, two
maser sites are seen, spaced about $7''$ apart. Coincident with one of these maser sites, and
marked by the diamond is the \uchii region NGC\,6334\,I. The other, nearby maser site marks
the position of a hot core (eg. \citealt{beuther05}). The circle marks the position, and approximate
extent of the \hii region NGC\,6334\,E. Strong sources of submm continuum emission appear associated with
both NGC\,6334\,I and NGC\,6334\,I(N). At the position of NGC\,6334\,I, the dust continuum appears to peak at the \uchii
region position, rather than the hot core located to the north-west. The continuum emission also appears
to be essentially unresolved here. Higher resolution observations by \citet{hunter06} show the continuum
emission does fragment into multiple sources.

The continuum emission associated with NGC\,6334\,I appears
to be extended over almost an arcminute, corresponding to a projected length of nearly 0.5\,pc. The
continuum emission also appears to peak about $6''$ to the south-east of the maser site associated
with NGC\,6334\,I(N). The greyscale continuum images also show the low level dust emission, which exhibits
a prominent filamentary structure running from the south-western corner to the centre-top of
the field of view. Other, weaker filamentary structures appear as a bridge between NGC\,6334\,I and I(N),
as well as possibly connecting NGC\,6334\,I and NGC\,6334\,E, although very little continuum emission is
coincident with the \hii region NGC\,6334\,E. The very weak filamentary structure appears to extend
from NGC\,6334\,I(N) to the north-eastern corner of the field of view. Finally, there is some emission seen
directly to the north of NGC\,6334\,I(N), which may be an extension of the filament seen to bridge
the gap between NGC\,6334\,I and I(N).

\begin{figure*}
\includegraphics[width=\textwidth]{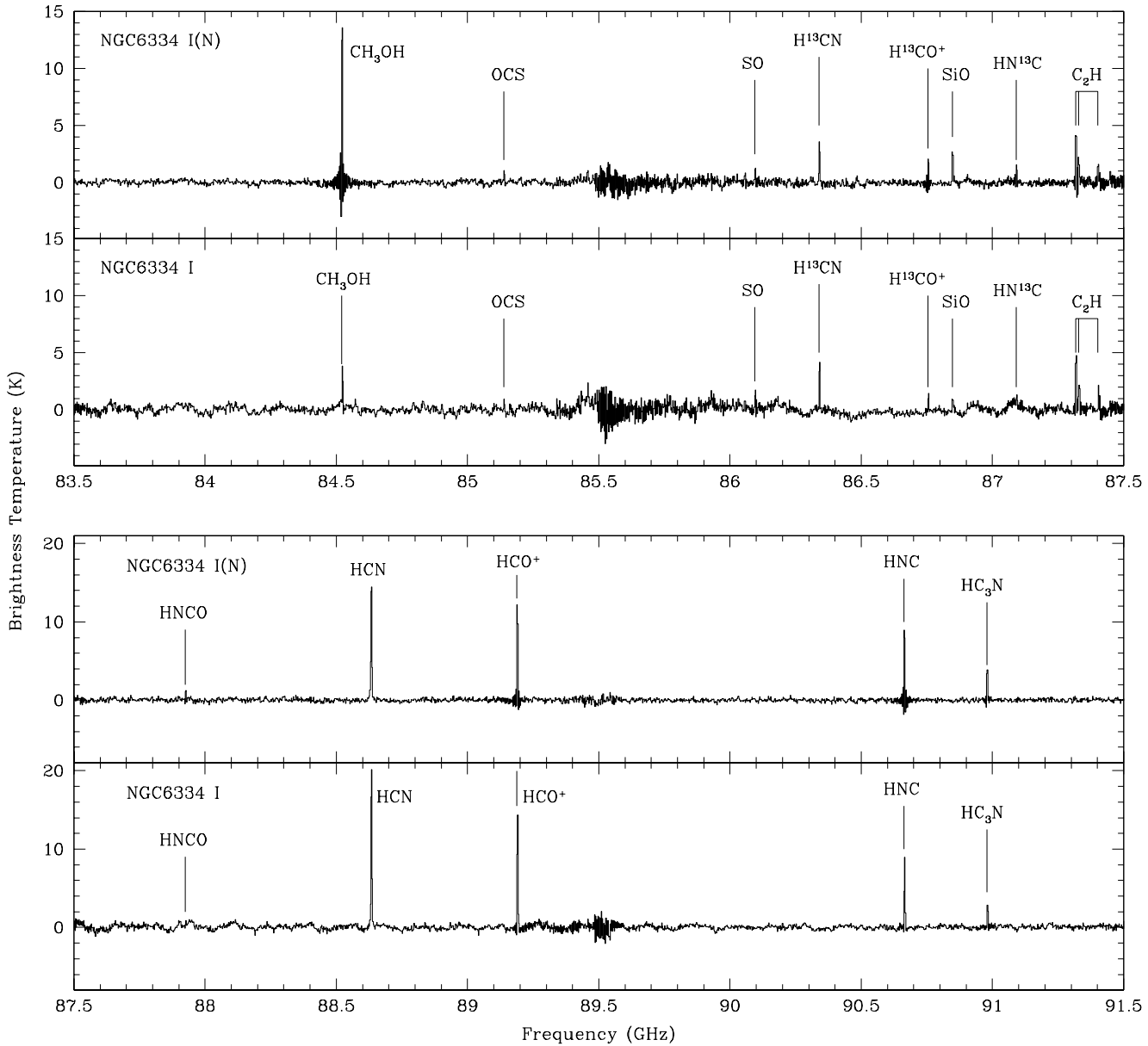}
\caption{Spectra of NGC\,6334\,I (lower) and NGC\,6334\,I(N) (upper)
shown between 83.5 and 91.5\,GHz. The frequencies of identified lines in each
spectrum are indicated with a solid vertical line. Each spectrum was produced by integrating over
a square box with each side approximately equal to the beam width ($36''$). Strong lines in the spectrum
occasionally produce a ringing artifact close to the line frequency, eg. CH$_3$OH in I(N). This is a result
of the Gibbs phenomenon and is most pronounced with narrow lines, such as masers.}
\label{spectra}
\end{figure*}

\begin{figure*}
\includegraphics[width=\textwidth]{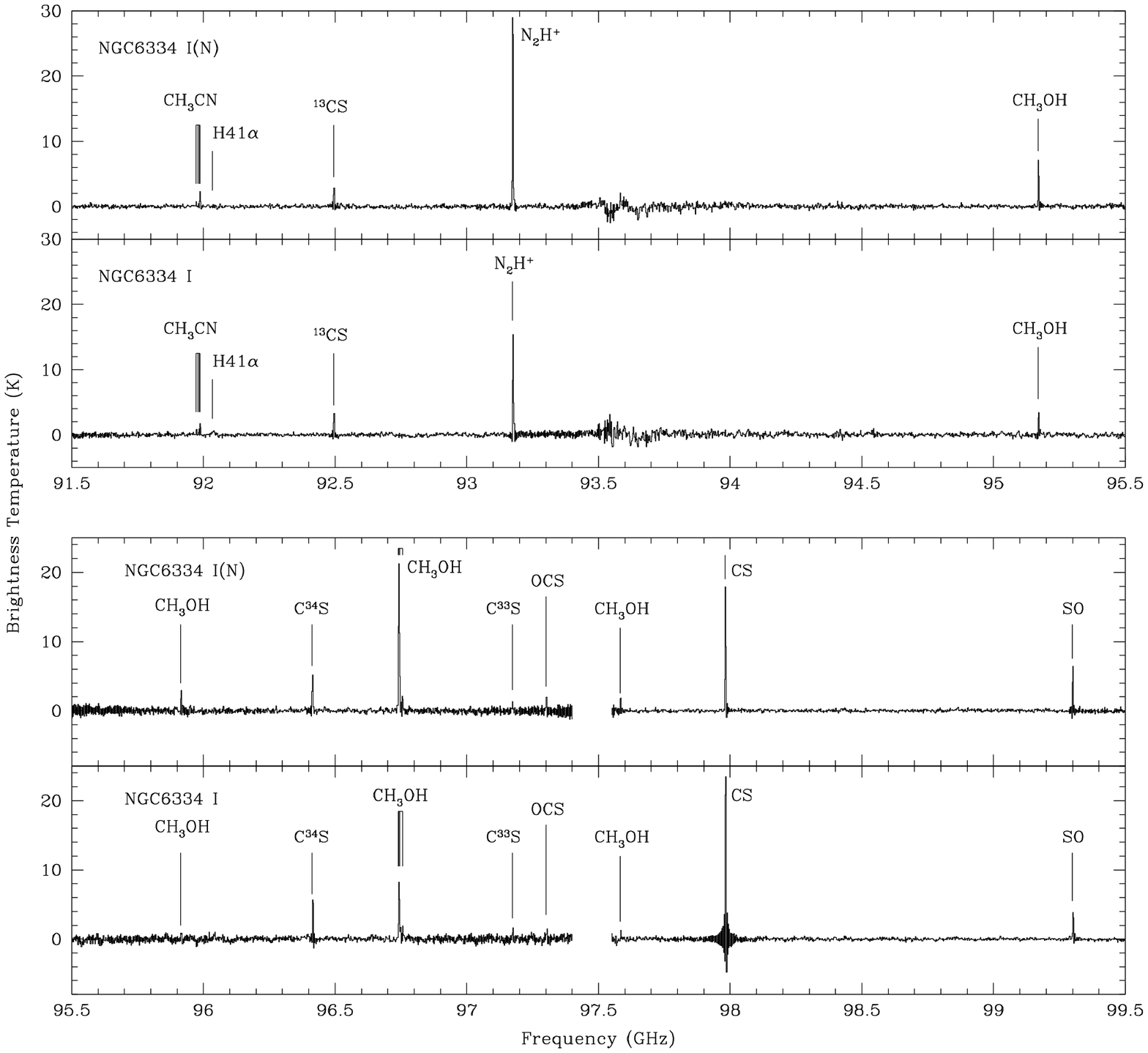}
\contcaption{... Frequency range is between 91.5 and 99.5\,GHz.}
\end{figure*}

\begin{figure*}
\includegraphics[width=\textwidth]{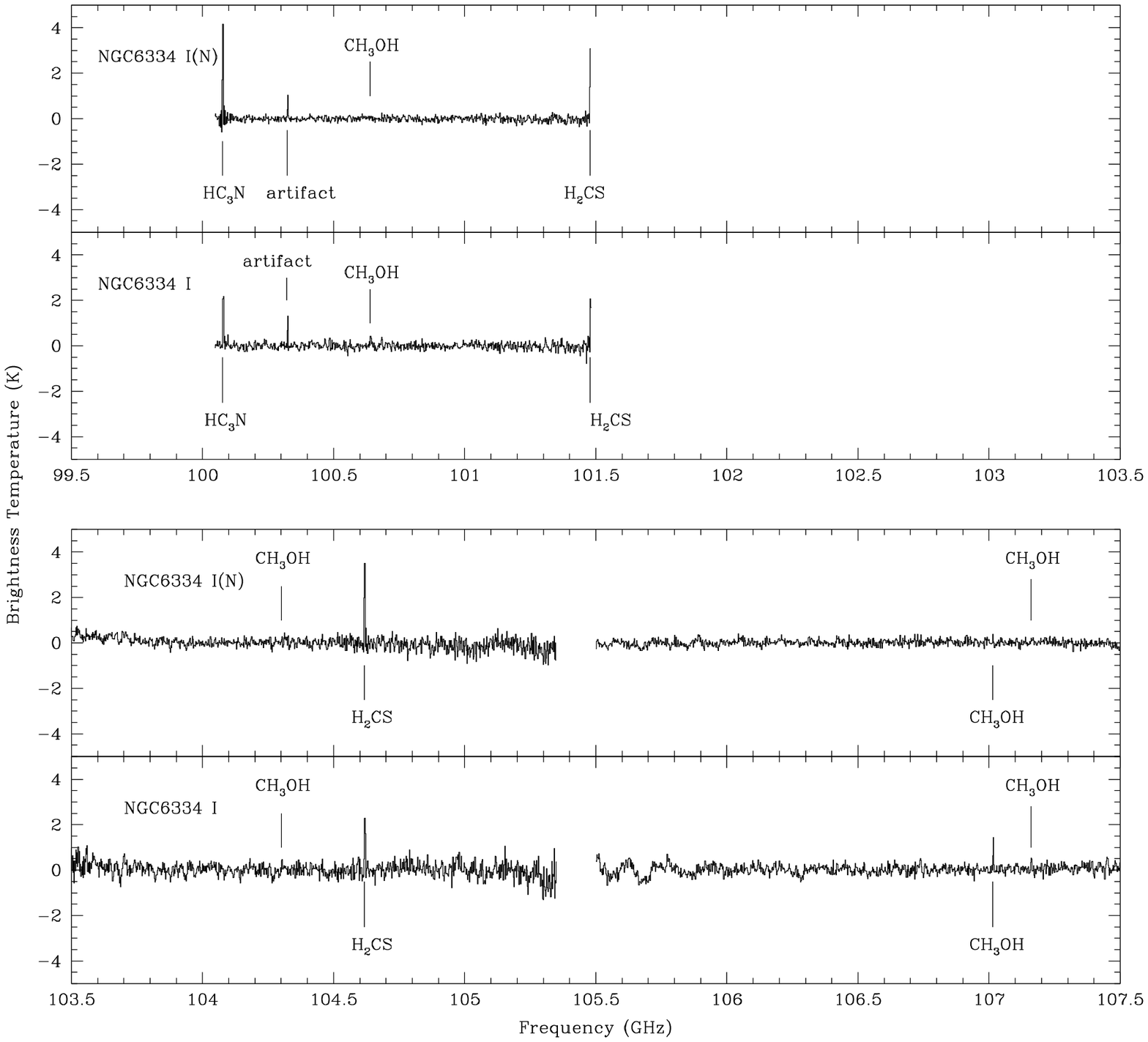}
\contcaption{... Frequency range is between
99.5 and 107.5\,GHz. The large blank areas shown on the upper plot are because the data
in these frequency ranges were unfortunately unusable. Note that one line appears at approximately
100.322\,GHz. This is a reflection of HNC at 90.664\,GHz, and not a real line.}
\end{figure*}

\begin{figure*}
\includegraphics[width=\textwidth]{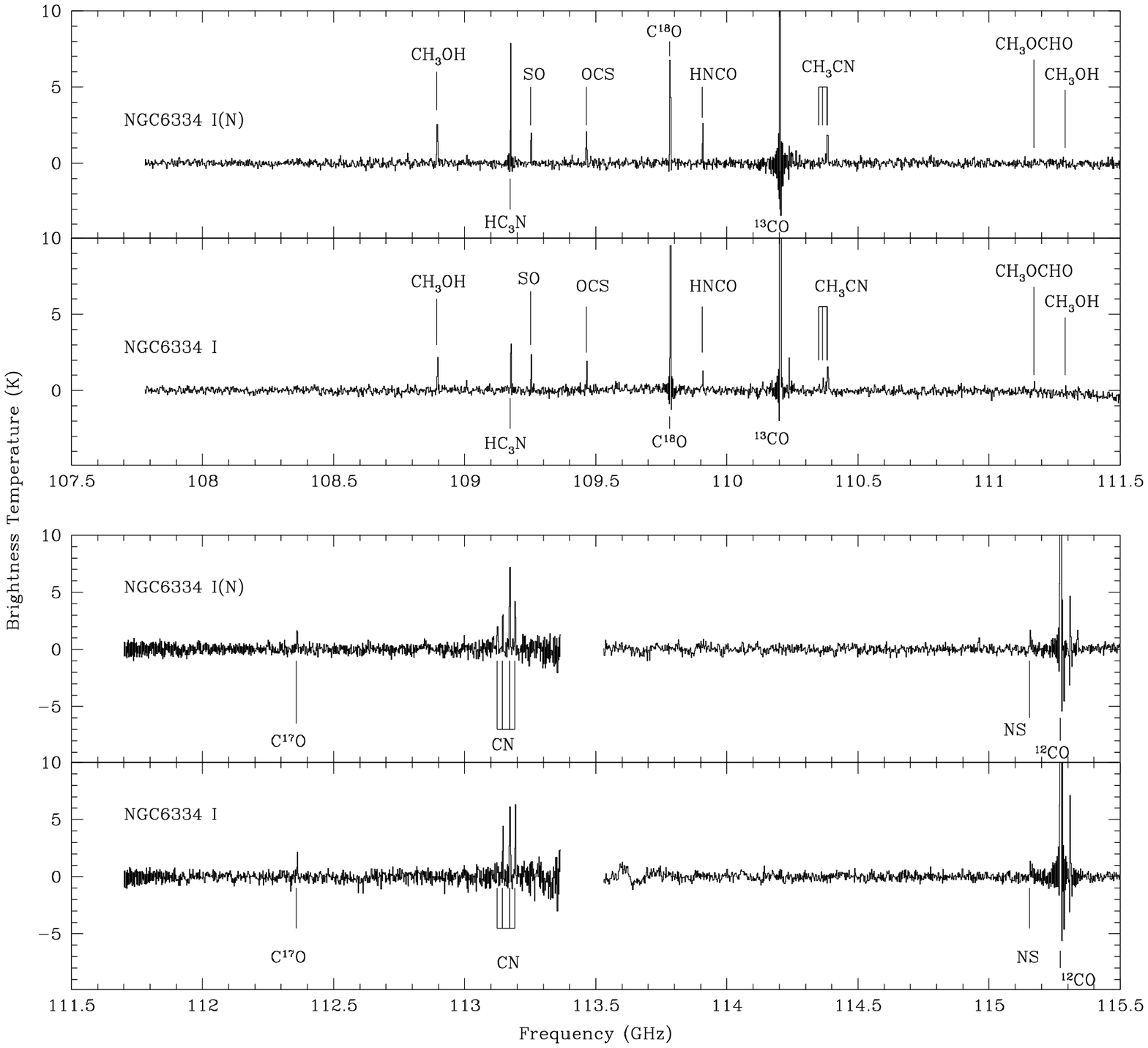}
\contcaption{... Frequency range is between
107.5 and 115.5\,GHz. The frequencies of identified lines in each
spectrum are indicated with a solid vertical line below the spectrum.}
\end{figure*}

\begin{figure*}
\begin{tabular}{cc}
\includegraphics[width=0.43\textwidth]{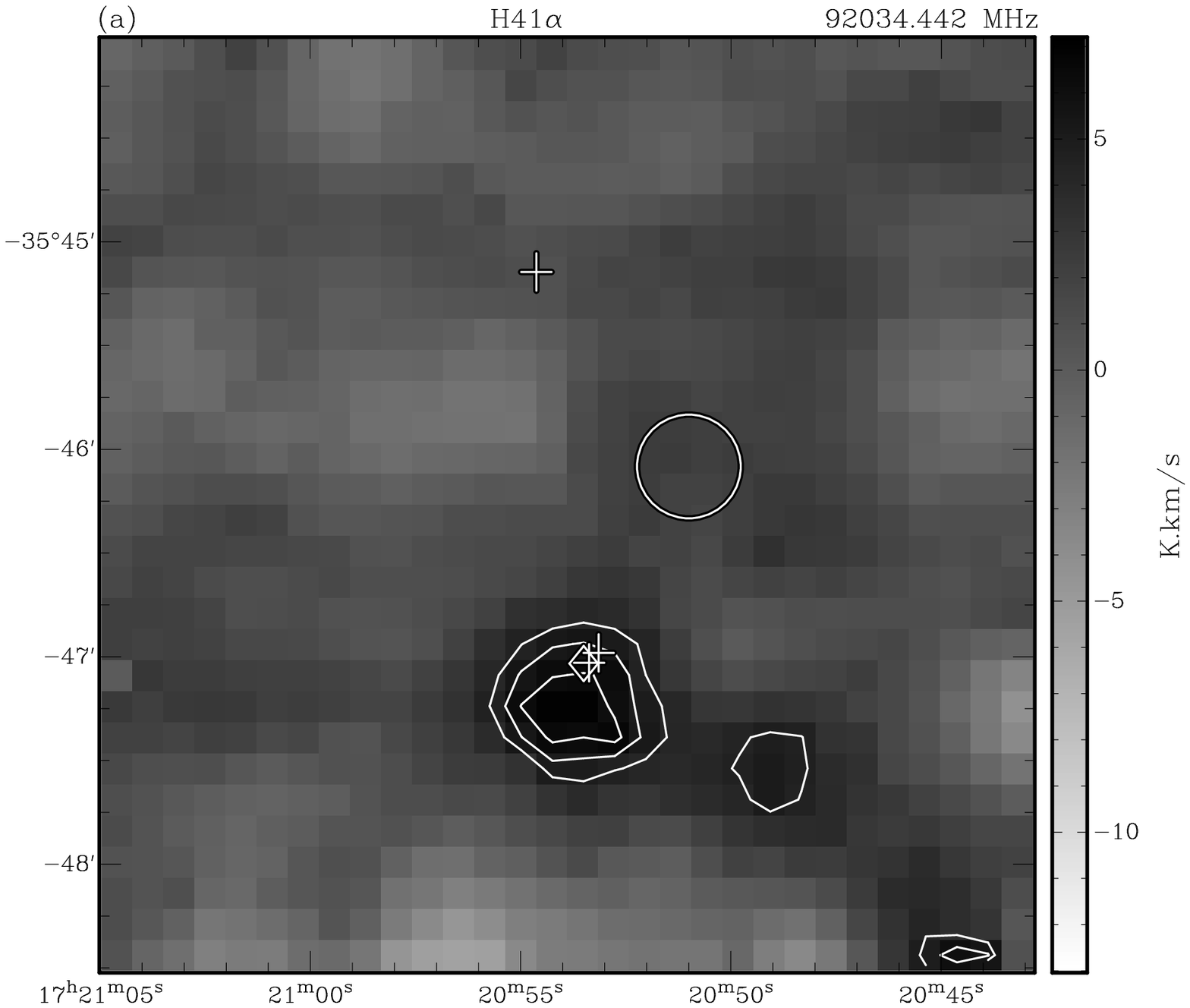}&
\includegraphics[width=0.43\textwidth]{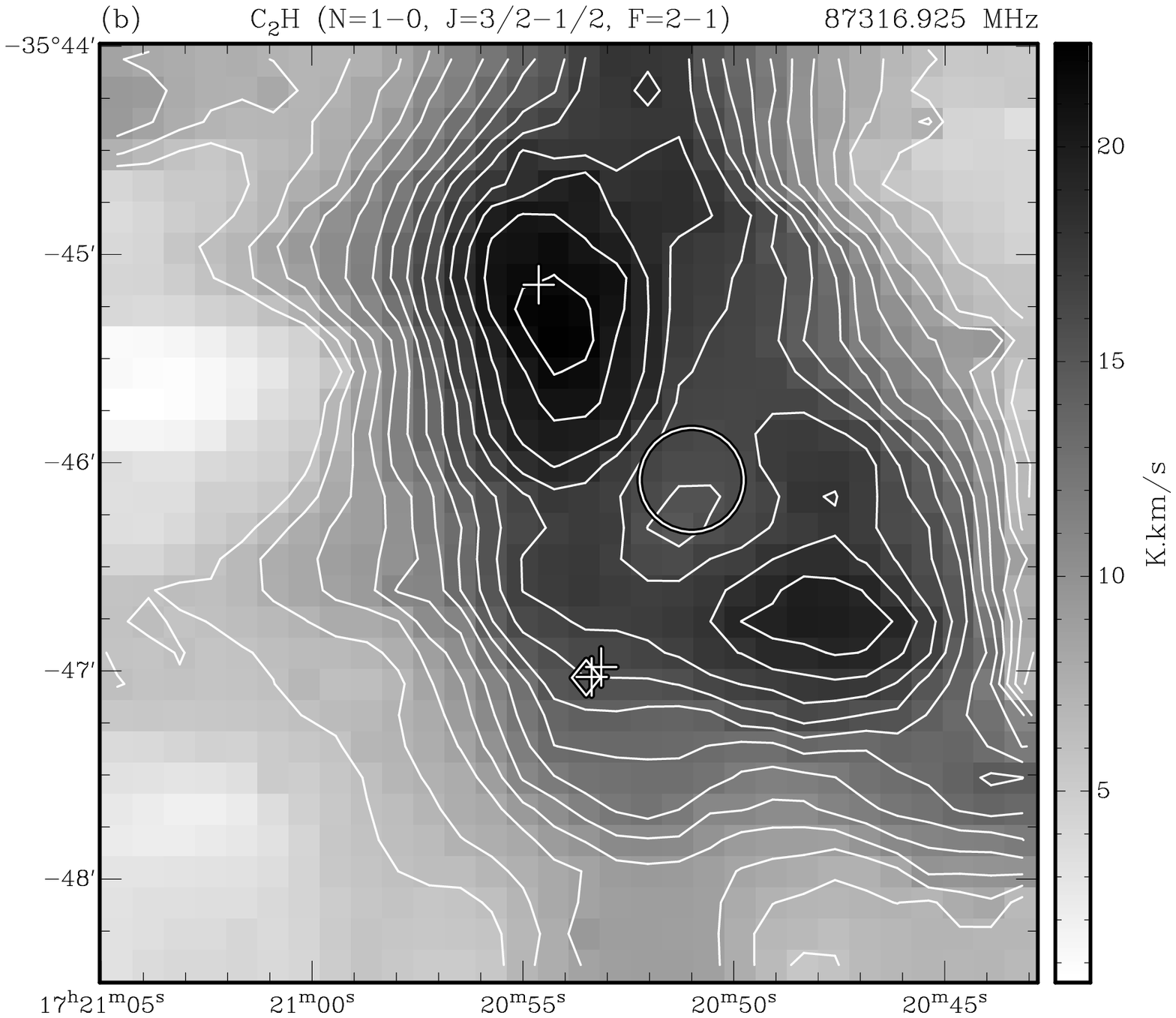}\\
\includegraphics[width=0.43\textwidth]{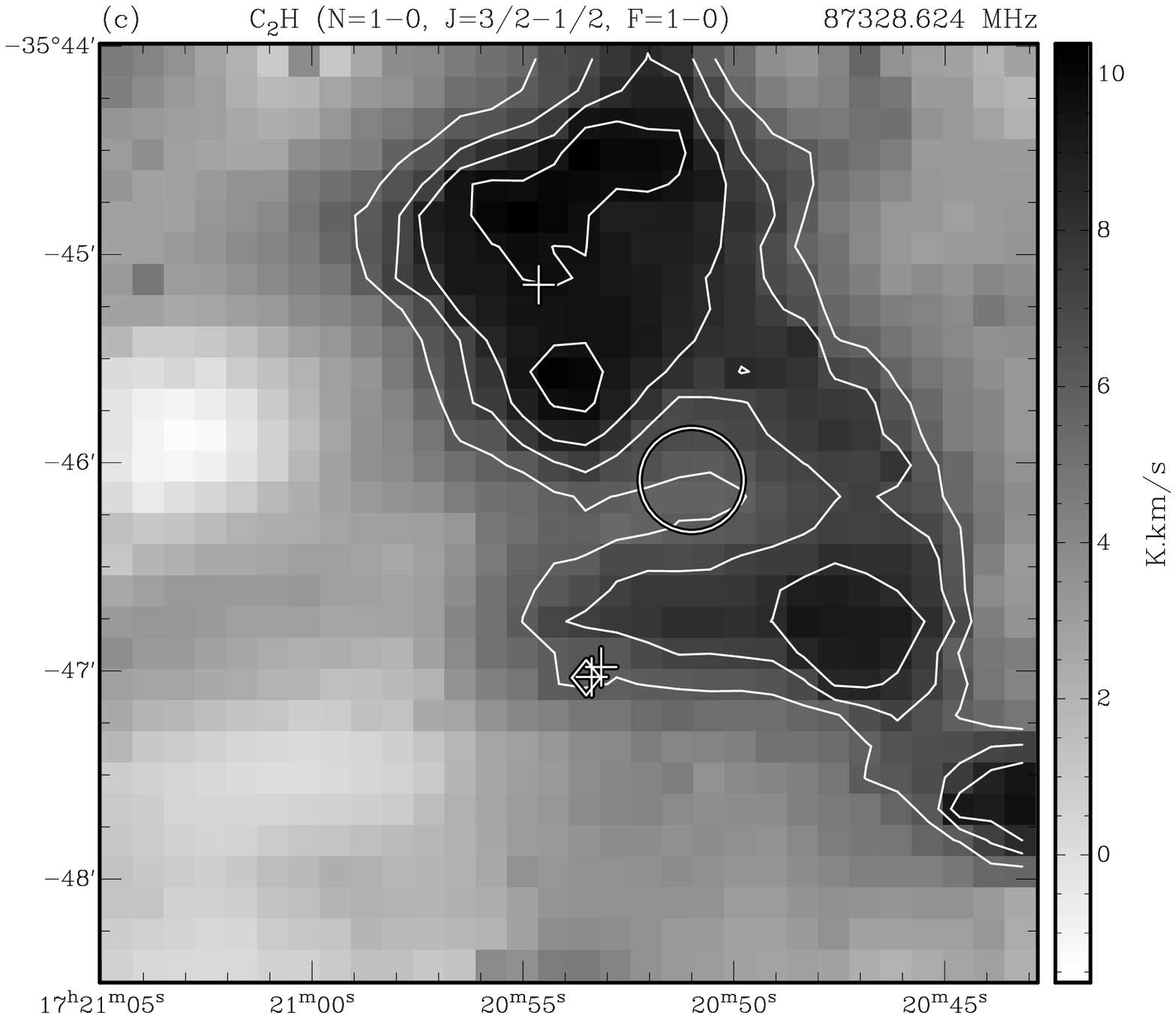}&
\includegraphics[width=0.43\textwidth]{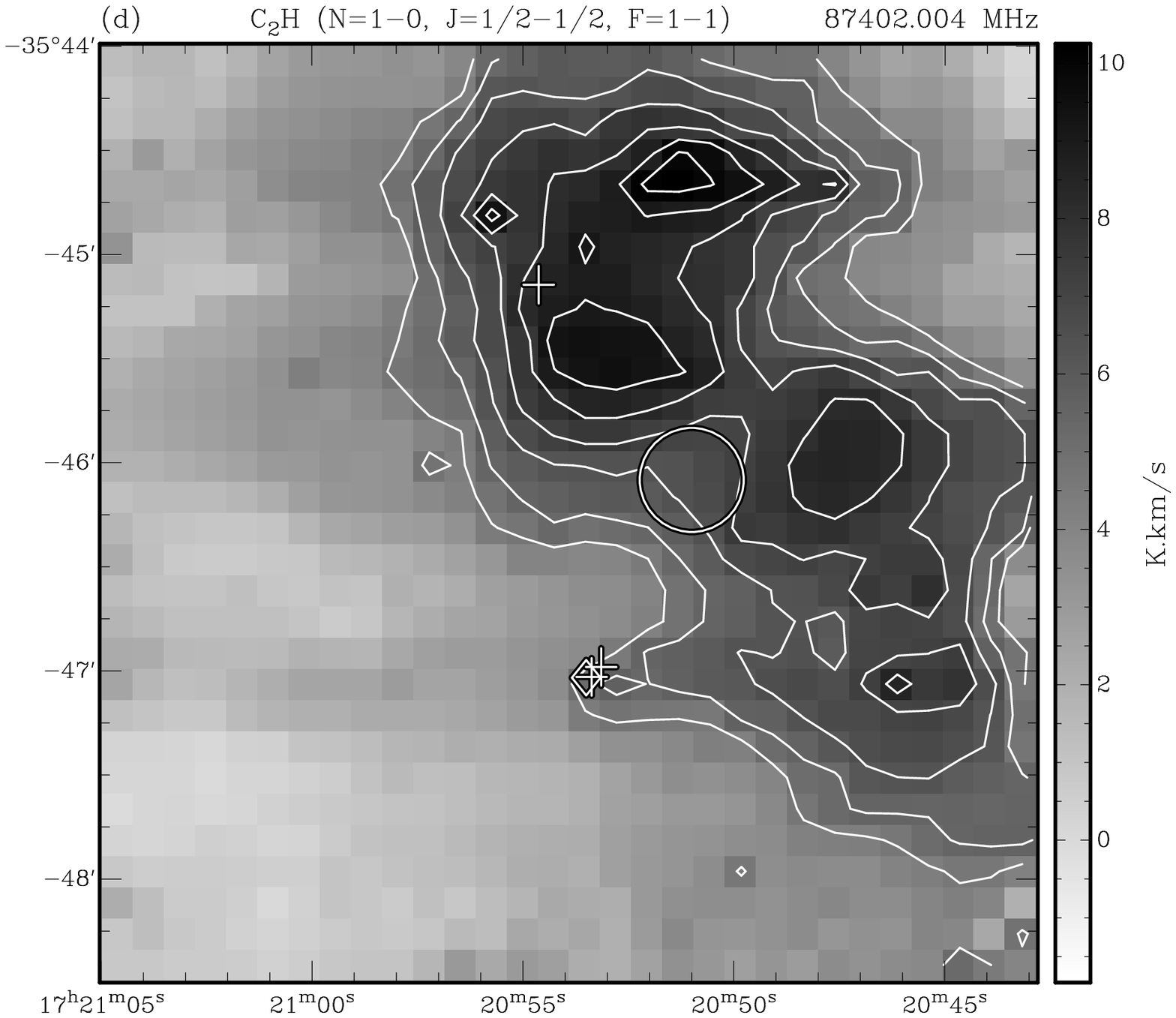}\\
\includegraphics[width=0.43\textwidth]{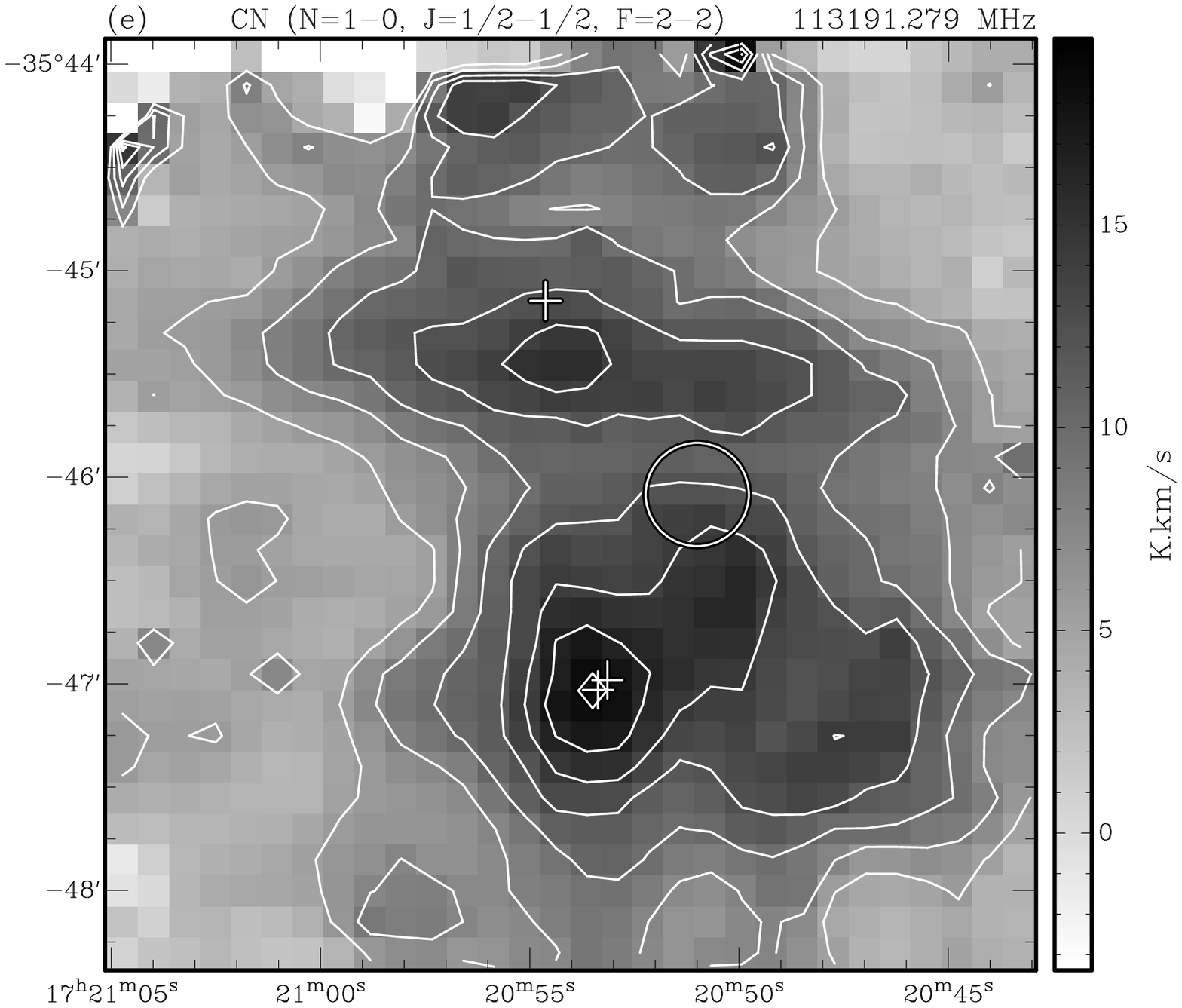}&
\includegraphics[width=0.43\textwidth]{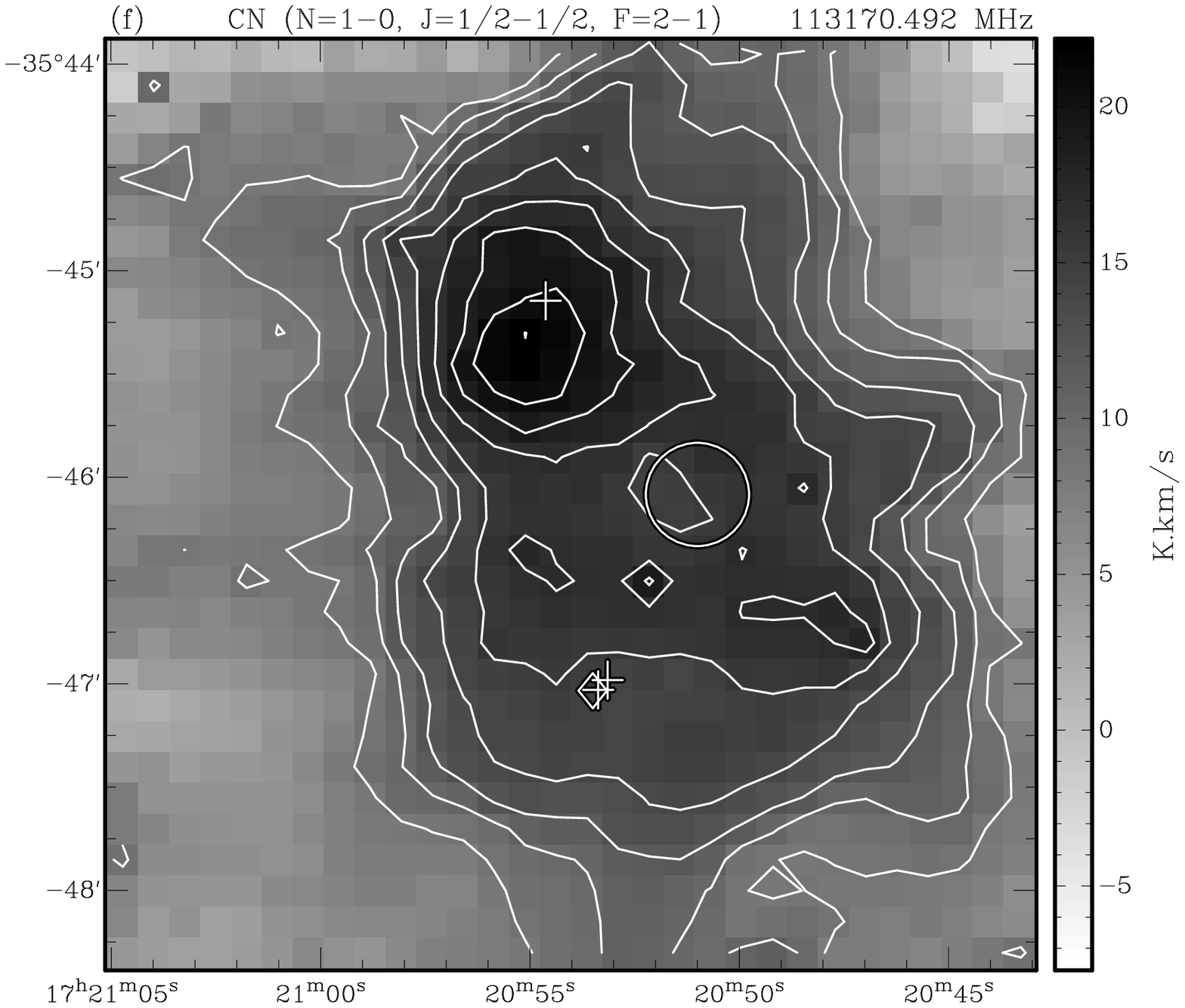}\\
\end{tabular}
\caption{Integrated intensity images and contours for detected transitions. The large circle
represents the approximate size and position of the \hii region NGC\,6334\,E \citep{carral02}.
The diamond represents the position of the \uchii region NGC\,6334\,F, also known as
NGC\,6334\,I. The two plus
symbols next to the diamond represent the position of two methanol maser sites
\citep{walsh98}. The single plus symbol in the northern half of each image represents
the position of NGC\,6334\,I(N), traced by a methanol maser site \citep{walsh98}.
Axes are labelled in coordinates of Ra and Dec (J2000). The intensity scale is in units
of brightness temperature multiplied by velocity.
{\bf (a)} H41$\alpha$ radio recombination line. Contours start at 5$\sigma$ and increase
in 1$\sigma$ steps, where 1$\sigma$ is 0.9\,K\,km\,s$^{-1}$. {\bf (b)} C$_2$H
(N=1--0, J=$3/2$--$1/2$, F=2--1) --- ethynyl radical. Contours start at 5$\sigma$ and increase
in 1$\sigma$ steps, where 1$\sigma$ is 1.2\,K\,km\,s$^{-1}$. {\bf (c)} C$_2$H
(N=1--0, J=$3/2$--$1/2$, F=1--0) --- ethynyl radical. Contours start at 5$\sigma$ and increase
in 1$\sigma$ steps, where 1$\sigma$ is 1.2\,K\,km\,s$^{-1}$. {\bf (d)} C$_2$H
(N=1--0, J=$1/2$--$1/2$, F=1--1) --- ethynyl radical. Contours start at 5$\sigma$ and increase
in 1$\sigma$ steps, where 1$\sigma$ is 0.9\,K\,km\,s$^{-1}$. {\bf (e)} CN
(N=1--0, J=$1/2$--$1/2$, F=2--2) --- cyanogen radical. Contours start at 5$\sigma$ and increase
in 2$\sigma$ steps, where 1$\sigma$ is 1.1\,K\,km\,s$^{-1}$. Note that this transition was
particularly sensitive to weather. Features in the northern quarter of
the image (north of I(N)) are almost certainly a result of poor weather and not real. {\bf (f)} CN
(N=1--0, J=$1/2$--$1/2$, F=2--1) --- cyanogen radical. Contours start at 5$\sigma$ and increase
in 1$\sigma$ steps, where 1$\sigma$ is 1.7\,K\,km\,s$^{-1}$.
}
\label{images}
\end{figure*}

\begin{figure*}
\begin{tabular}{cc}
\includegraphics[width=0.45\textwidth]{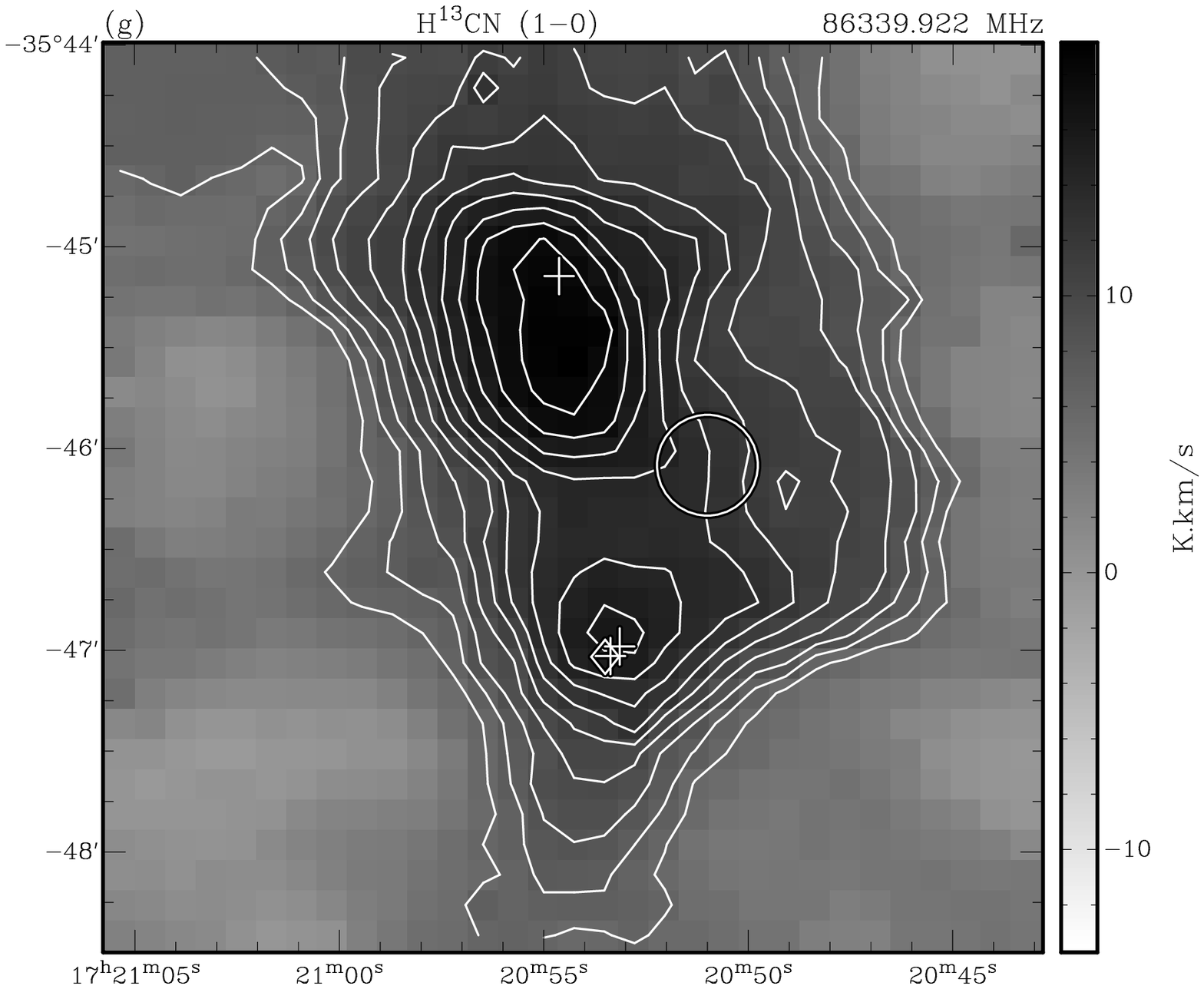}&
\includegraphics[width=0.45\textwidth]{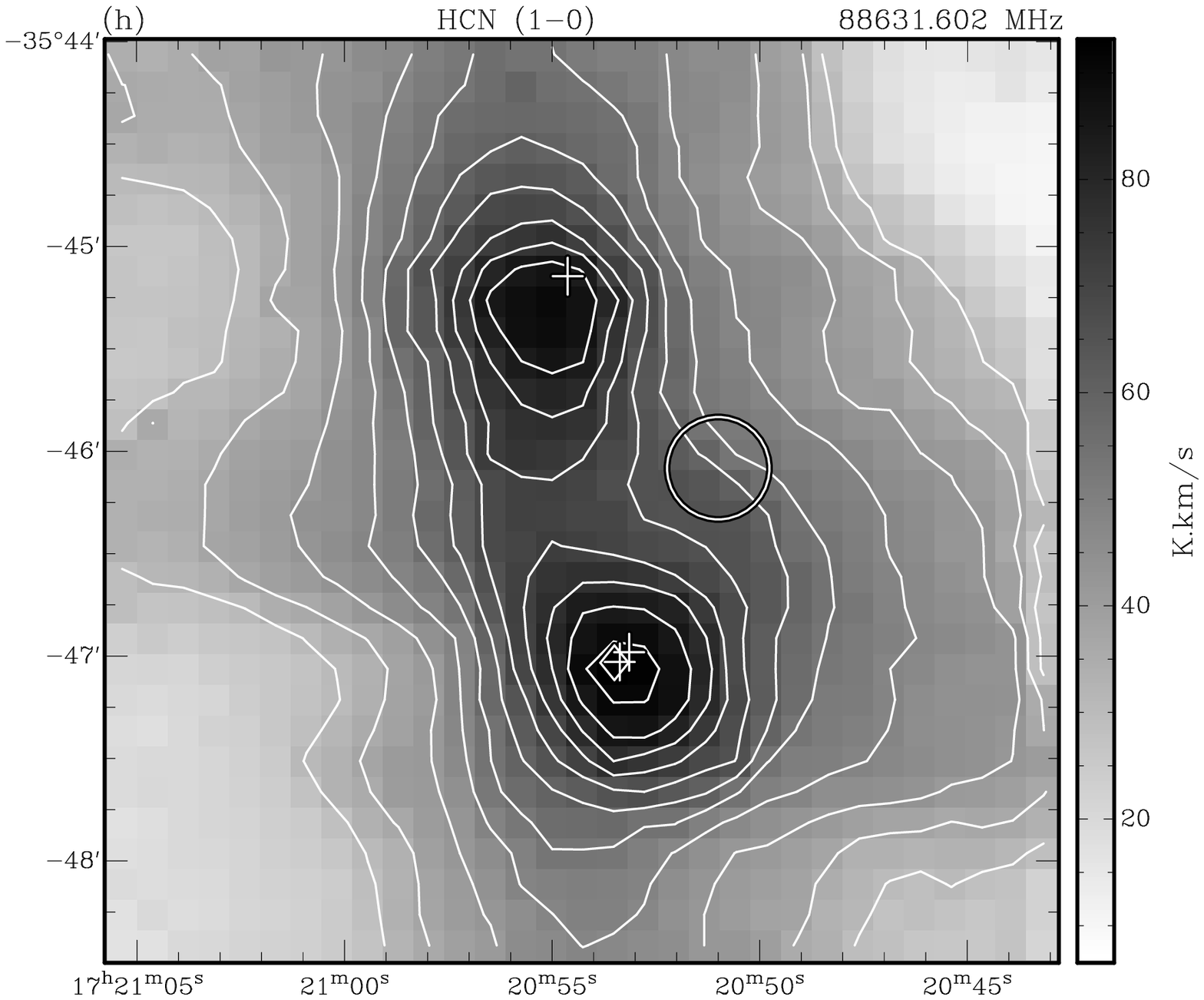}\\
\includegraphics[width=0.45\textwidth]{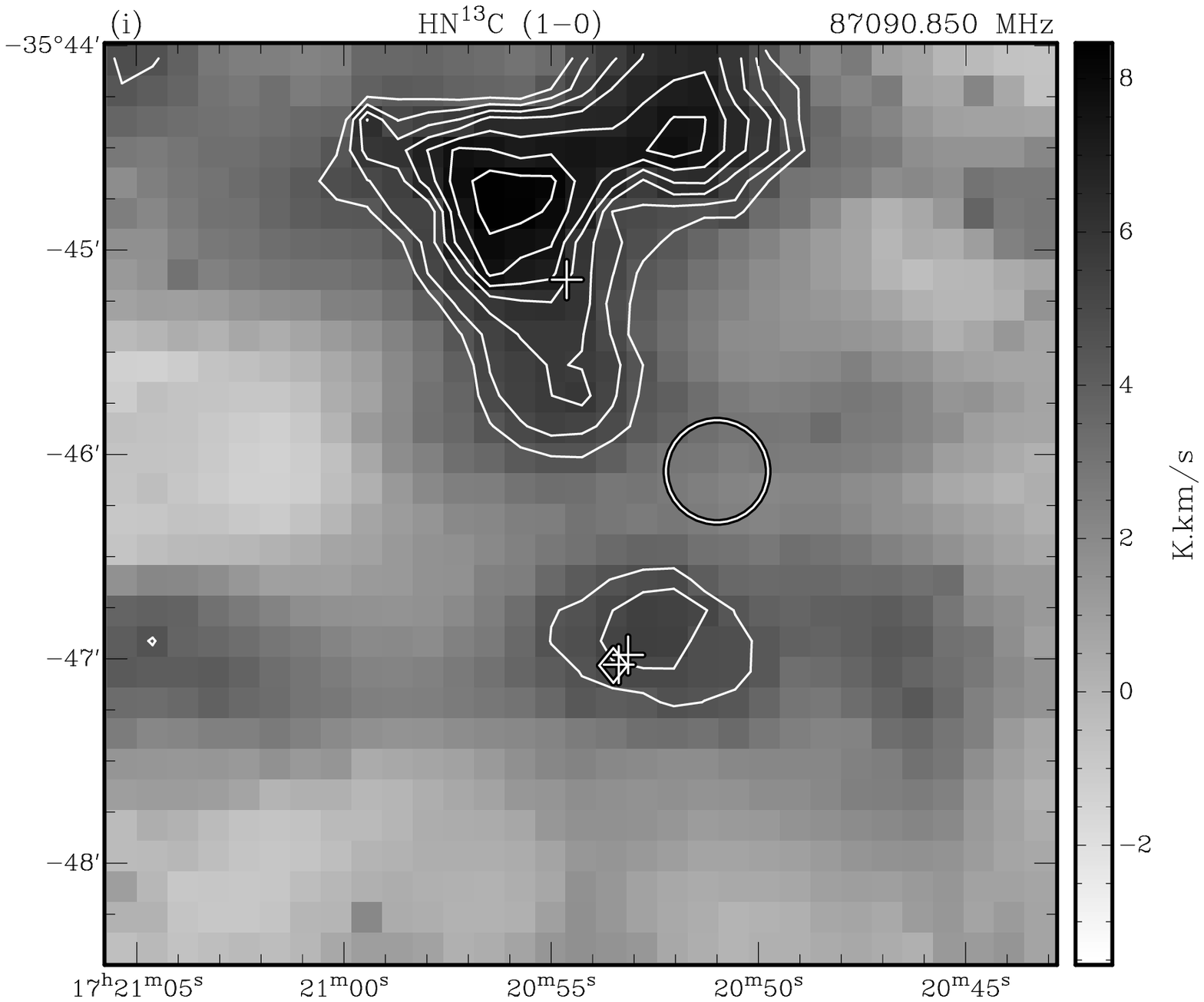}&
\includegraphics[width=0.45\textwidth]{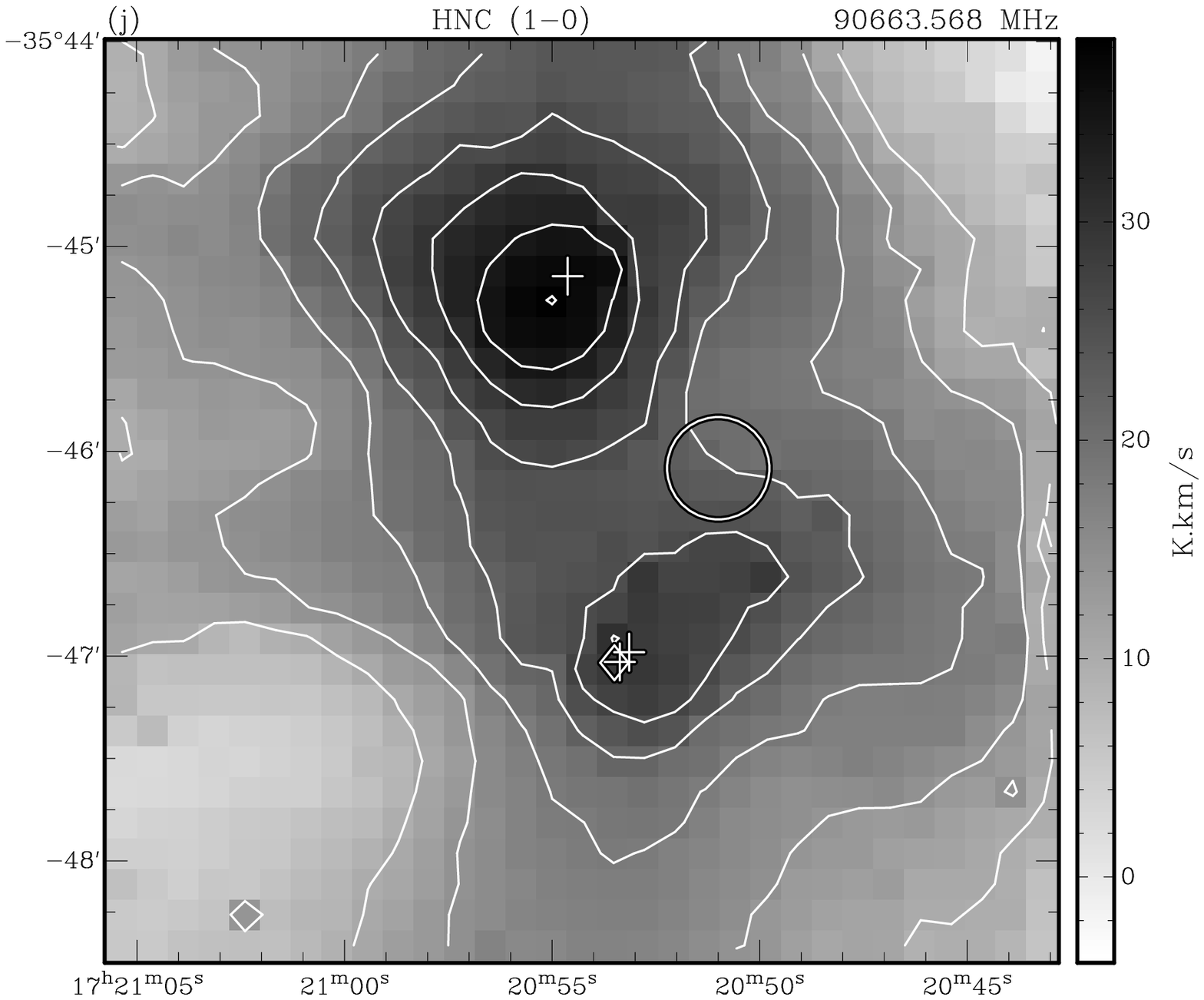}\\
\includegraphics[width=0.45\textwidth]{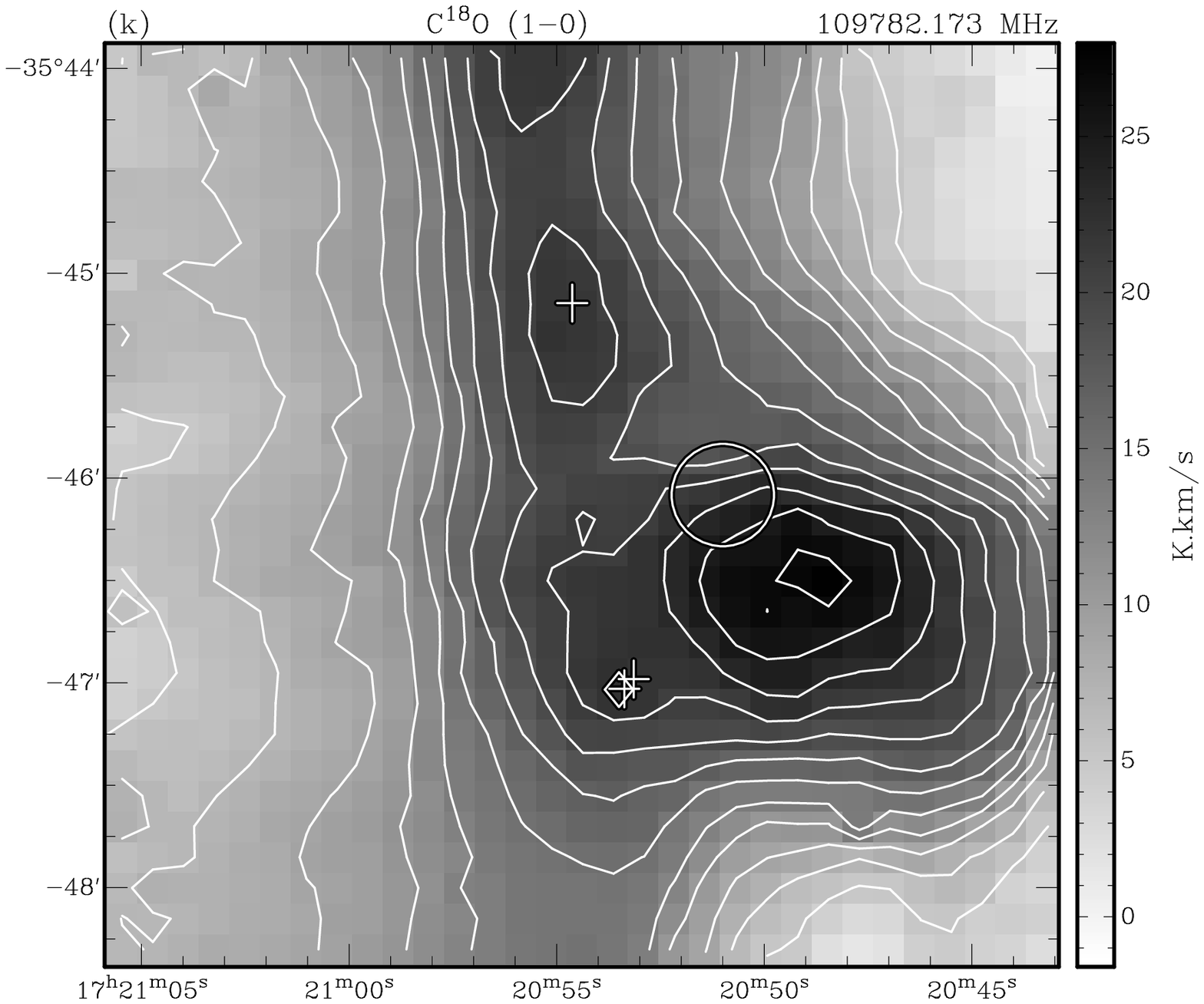}&
\includegraphics[width=0.45\textwidth]{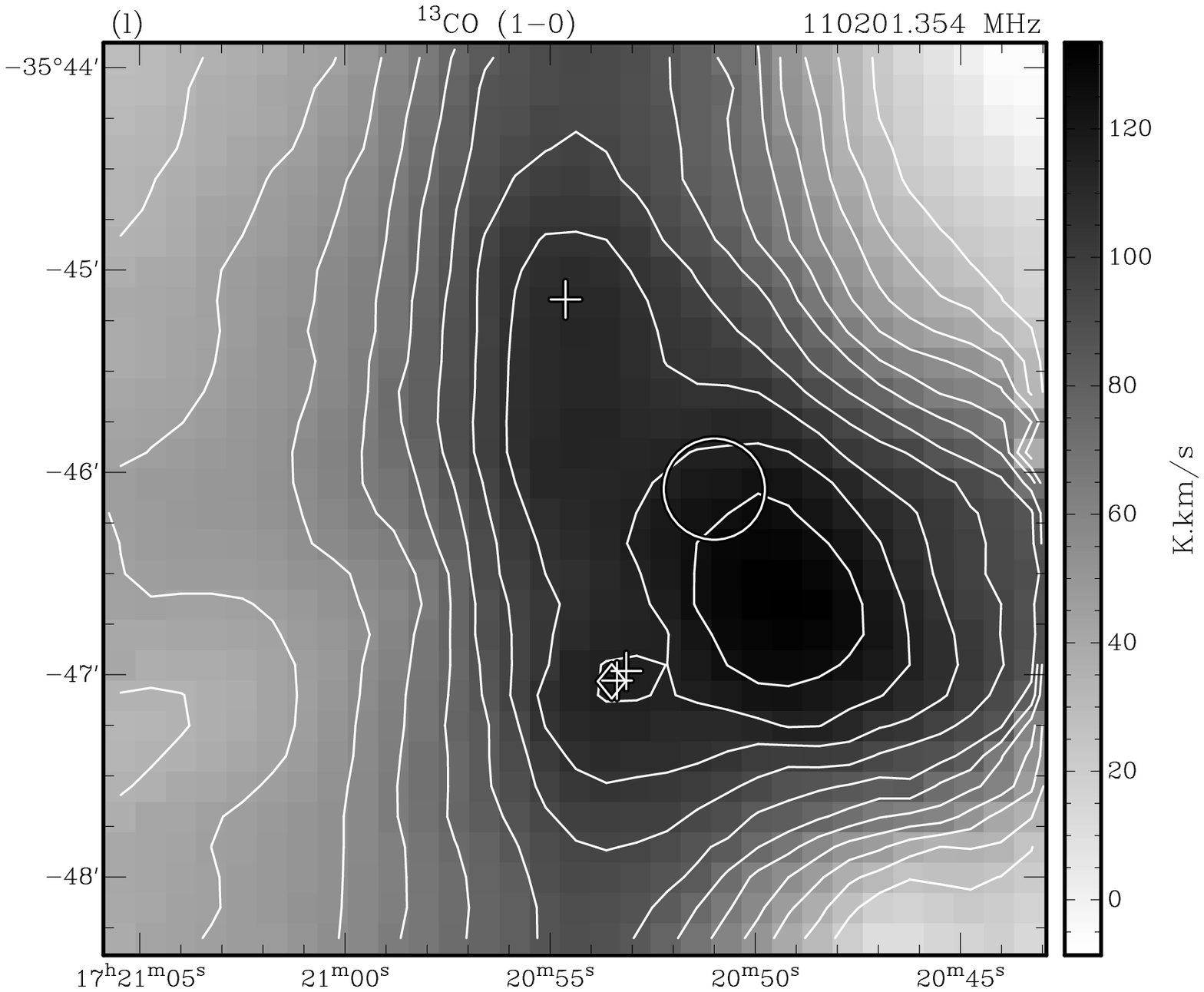}\\
\end{tabular}

\contcaption{{\bf (g)} H$^{13}$CN (1--0) --- hydrogen cyanide. Contours start
at 5$\sigma$ and increase in 1$\sigma$ steps, where 1$\sigma$
is 1.2\,K\,km\,s$^{-1}$.
{\bf (h)} HCN (1--0) --- hydrogen cyanide. Contours start
at 5$\sigma$ and increase in 1$\sigma$ steps, where 1$\sigma$
is 6.0\,K\,km\,s$^{-1}$.
{\bf (i)} HN$^{13}$C (1--0) --- hydrogen isocyanide. Contours start
at 5$\sigma$ and increase in 1$\sigma$ steps, where 1$\sigma$
is 0.6\,K\,km\,s$^{-1}$.
{\bf (j)} HNC (1--0) --- hydrogen isocyanide. Contours start
at 5$\sigma$ and increase in 2$\sigma$ steps, where 1$\sigma$
is 2.0\,K\,km\,s$^{-1}$.
{\bf (k)} C$^{18}$O (1--0) --- carbon monoxide. Contours start
at 5$\sigma$ and increase in 2$\sigma$ steps, where 1$\sigma$
is 1.1\,K\,km\,s$^{-1}$.
{\bf (l)} $^{13}$CO (1--0) --- carbon monoxide. Contours start
at 5$\sigma$ and increase in 2$\sigma$ steps, where 1$\sigma$
is 5.0\,K\,km\,s$^{-1}$.
}
\end{figure*}

\begin{figure*}
\begin{tabular}{cc}
\includegraphics[width=0.45\textwidth]{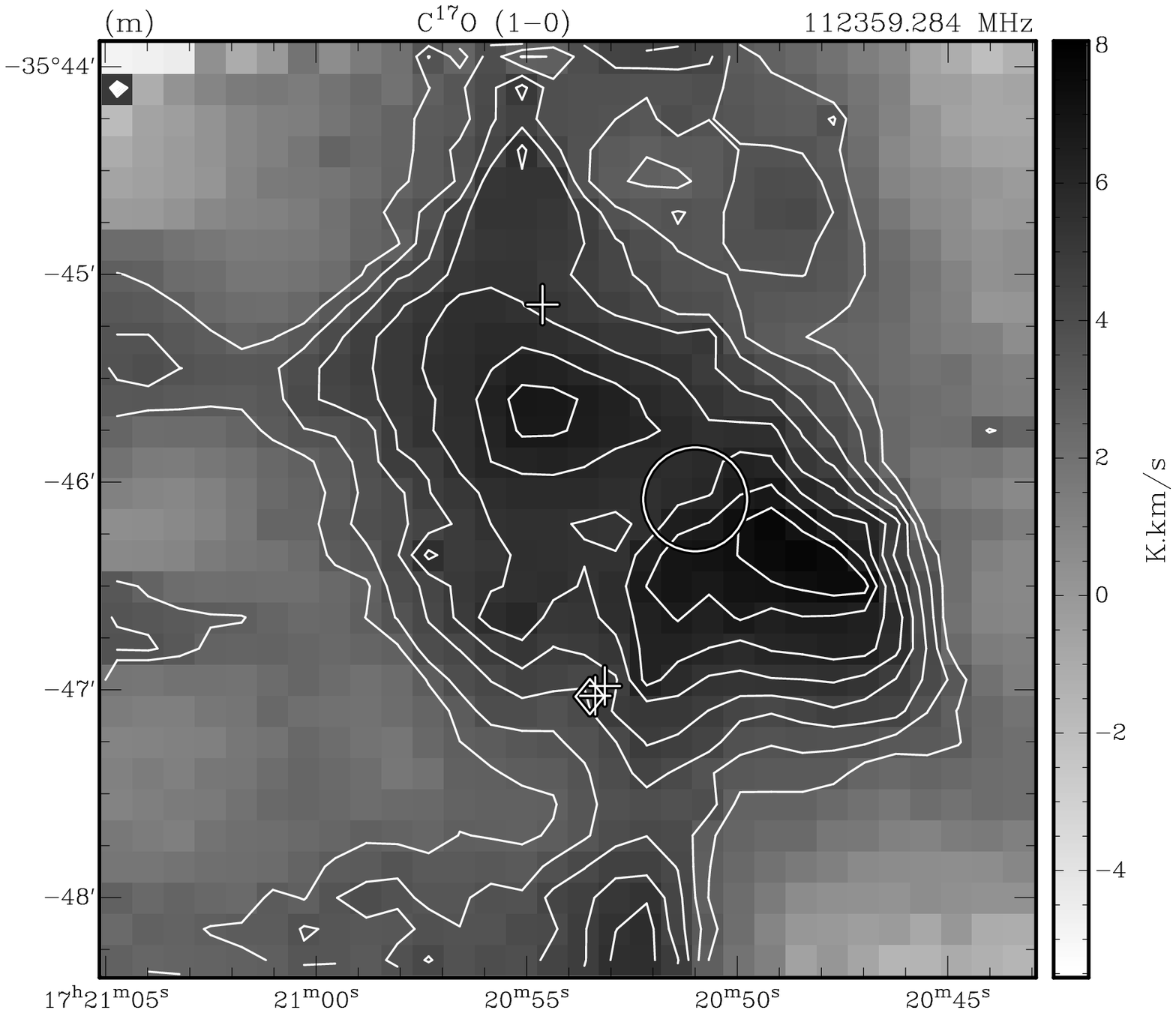}&
\includegraphics[width=0.45\textwidth]{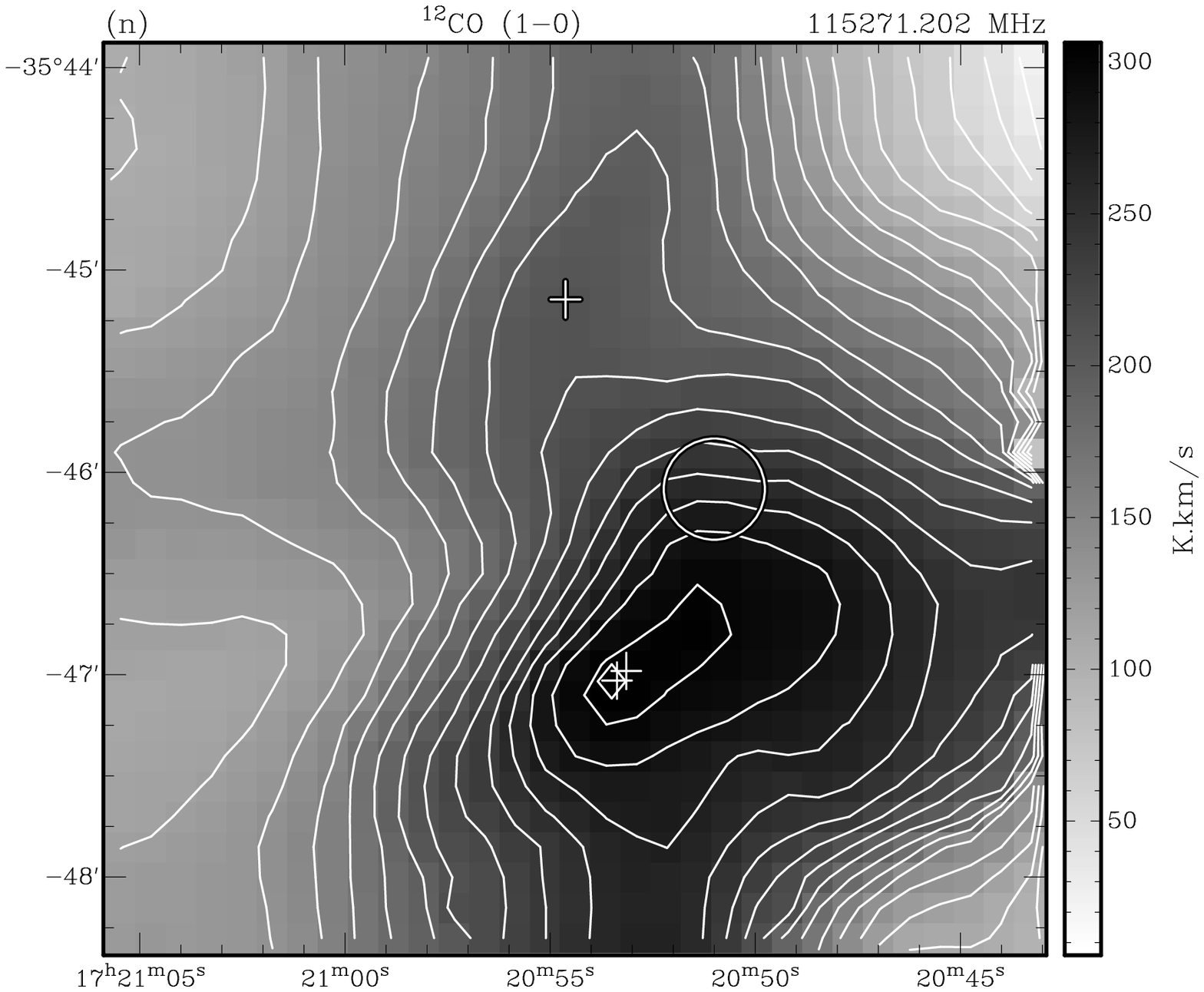}\\
\includegraphics[width=0.45\textwidth]{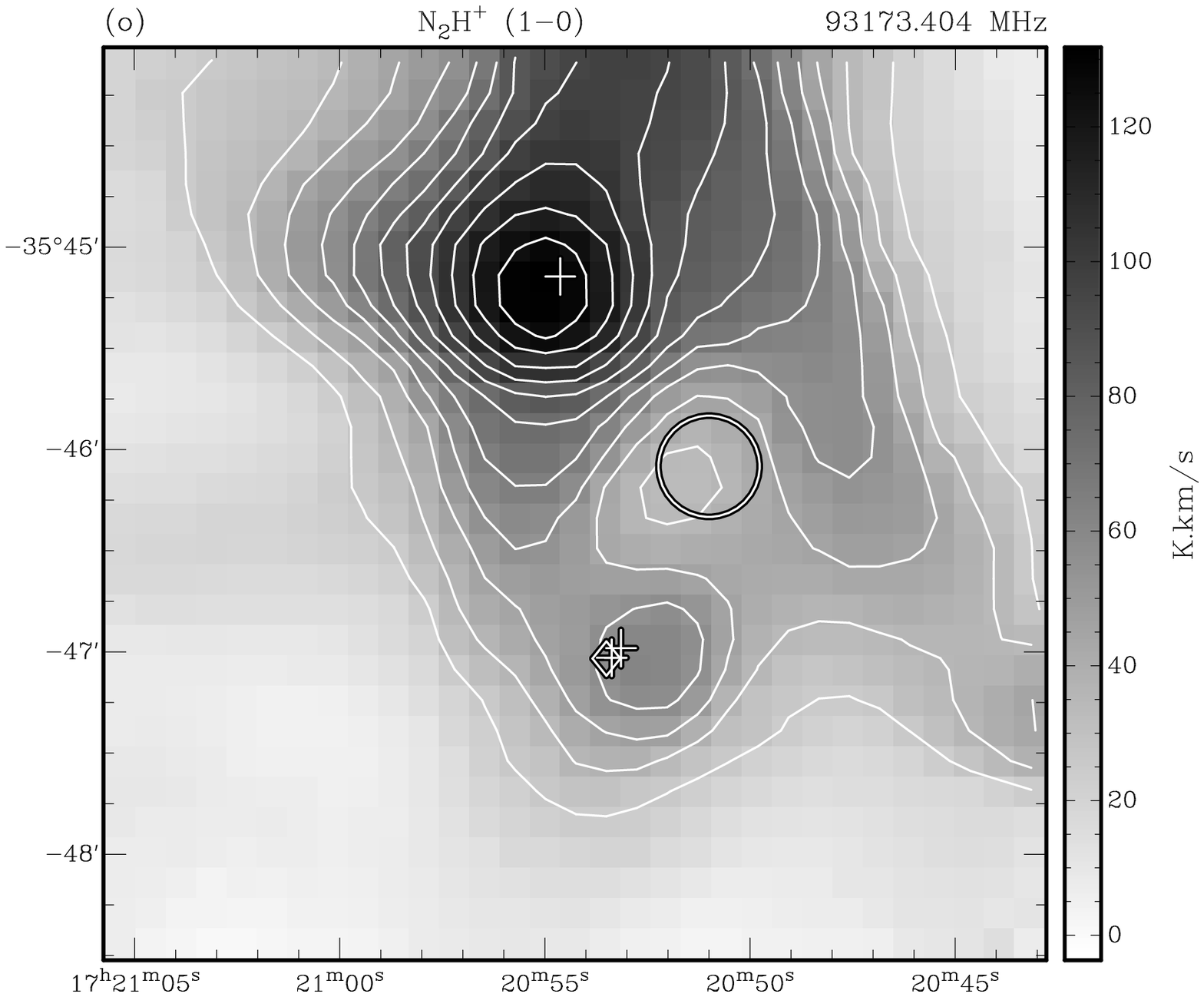}&
\includegraphics[width=0.45\textwidth]{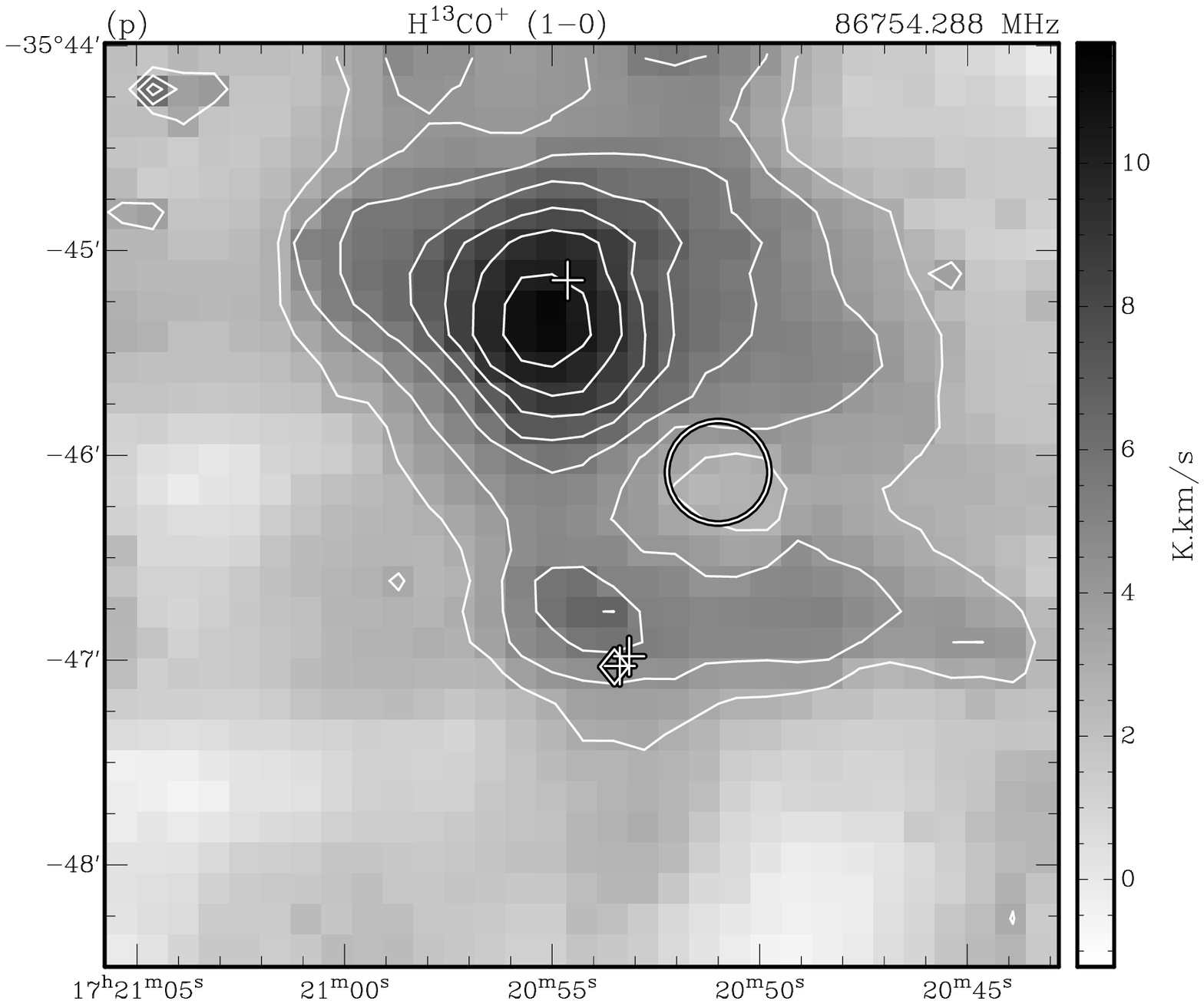}\\
\includegraphics[width=0.45\textwidth]{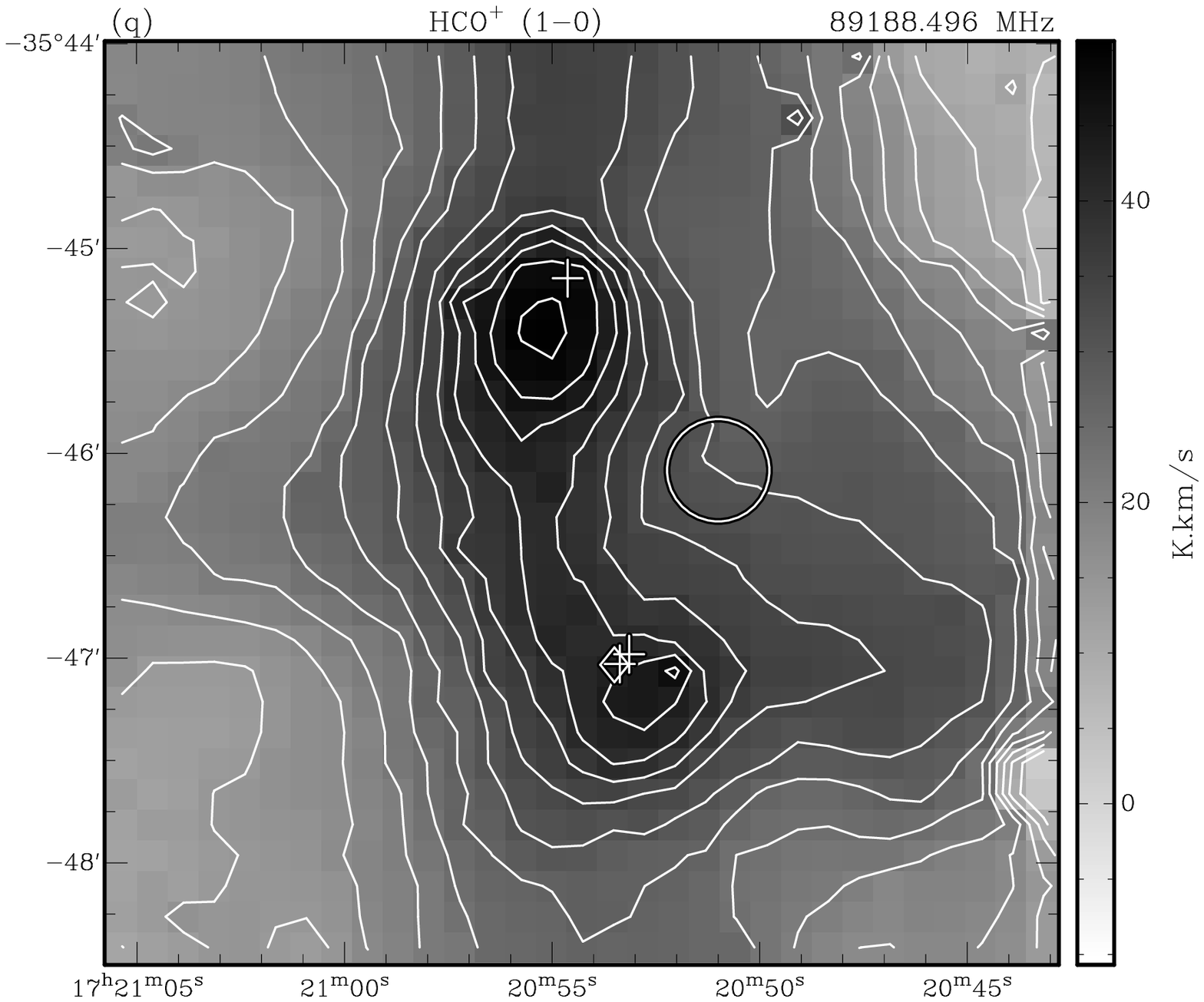}&
\includegraphics[width=0.45\textwidth]{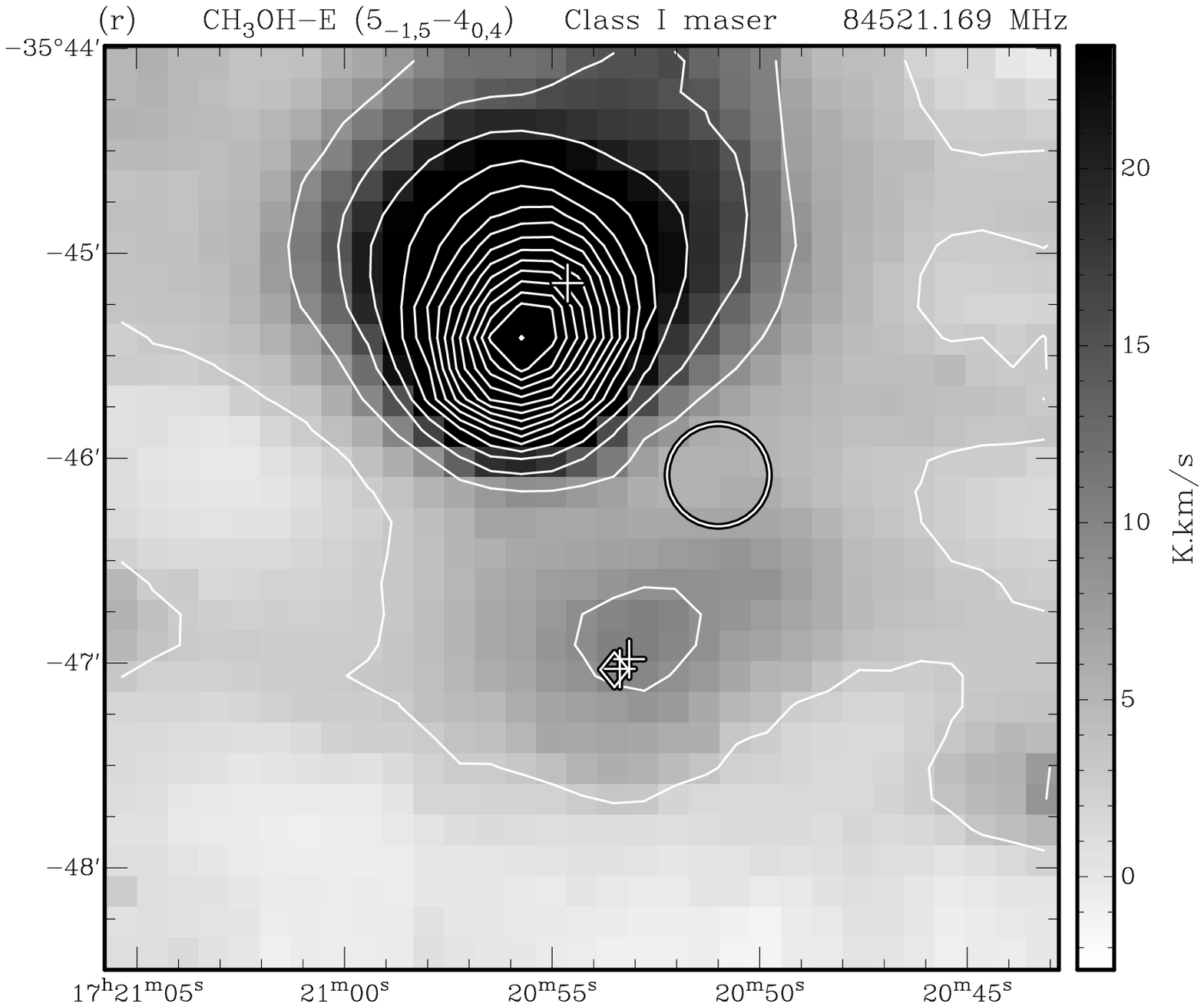}\\
\end{tabular}

\contcaption{{\bf (m)} C$^{17}$O (1--0) --- carbon monoxide. Contours start
at 5$\sigma$ and increase in 1$\sigma$ steps, where 1$\sigma$
is 0.6\,K\,km\,s$^{-1}$.
{\bf (n)} $^{12}$CO (1--0) --- carbon monoxide. Contours start
at 5$\sigma$ and increase in 3$\sigma$ steps, where 1$\sigma$
is 5.0\,K\,km\,s$^{-1}$.
{\bf (o)} N$_2$H$^+$ (1--0) --- diazenylium. Contours start
at 5$\sigma$ and increase in 2$\sigma$ steps, where 1$\sigma$
is 5.0\,K\,km\,s$^{-1}$.
{\bf (p)} H$^{13}$CO$^+$ (1--0) --- oxomethylium. Contours start
at 5$\sigma$ and increase in 2$\sigma$ steps, where 1$\sigma$
is 0.6\,K\,km\,s$^{-1}$.
{\bf (q)} HCO$^+$ (1--0) --- oxomethylium. Contours start
at 5$\sigma$ and increase in 2$\sigma$ steps, where 1$\sigma$
is 1.6\,K\,km\,s$^{-1}$.
{\bf (r)} CH$_3$OH-E ($5_{-1,5}$ -- $4_{0,4}$) --- methanol. Contours start
at 5$\sigma$ and increase in 10$\sigma$ steps, where 1$\sigma$
is 0.6\,K\,km\,s$^{-1}$. This methanol transition is a Class I maser
\citep{mueller04}.
}
\end{figure*}

\begin{figure*}
\begin{tabular}{cc}
\includegraphics[width=0.44\textwidth]{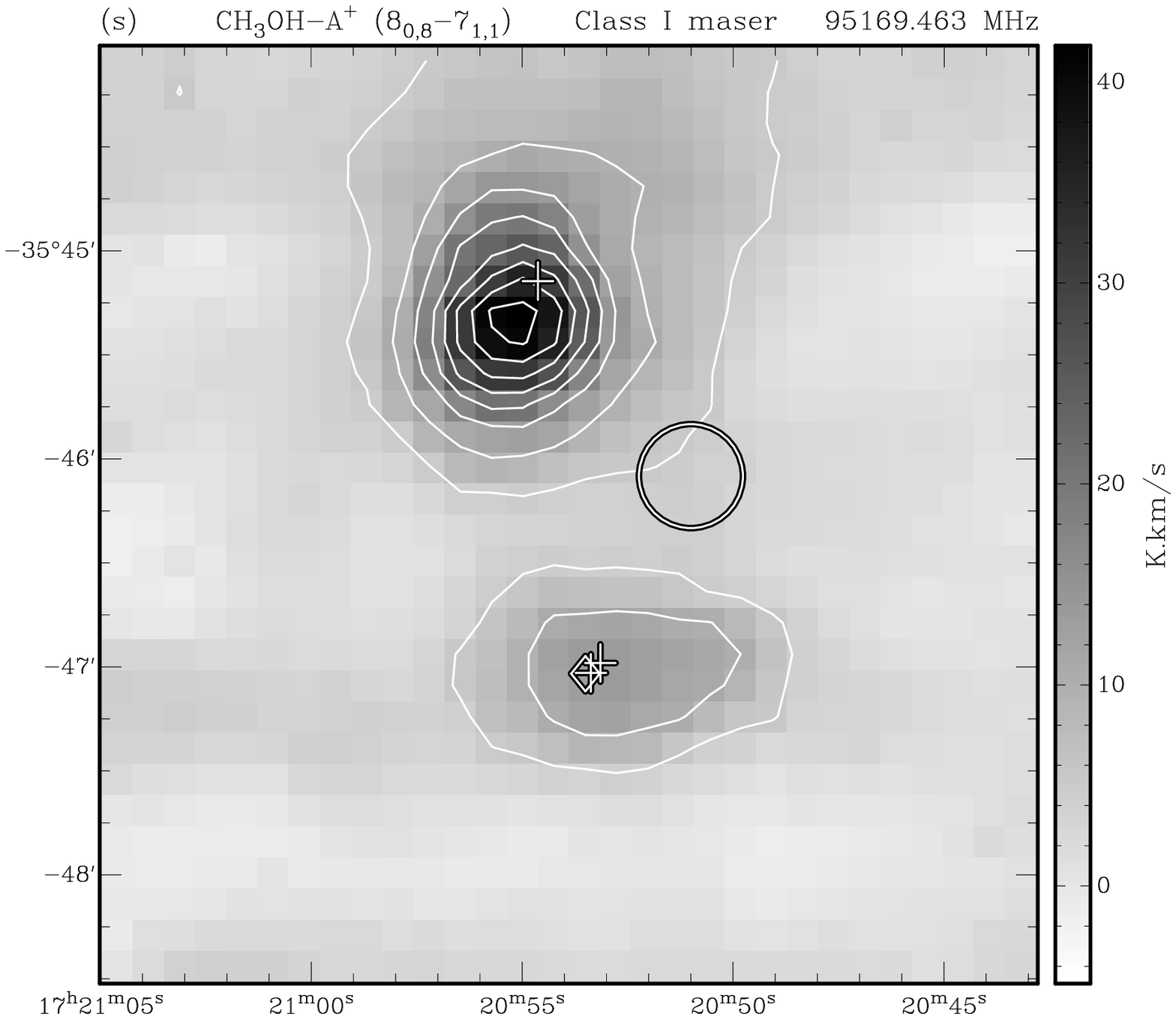}&
\includegraphics[width=0.44\textwidth]{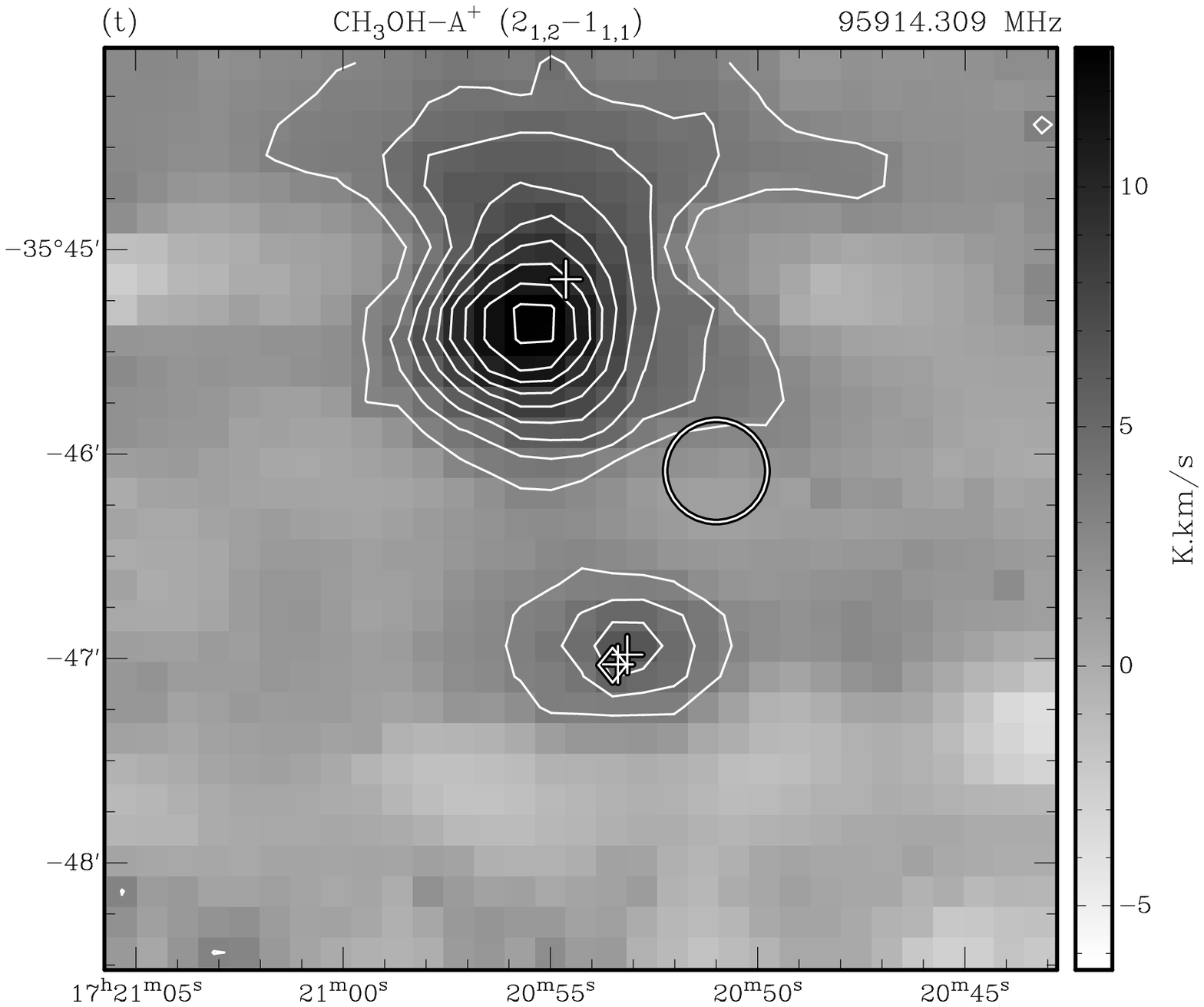}\\
\includegraphics[width=0.44\textwidth]{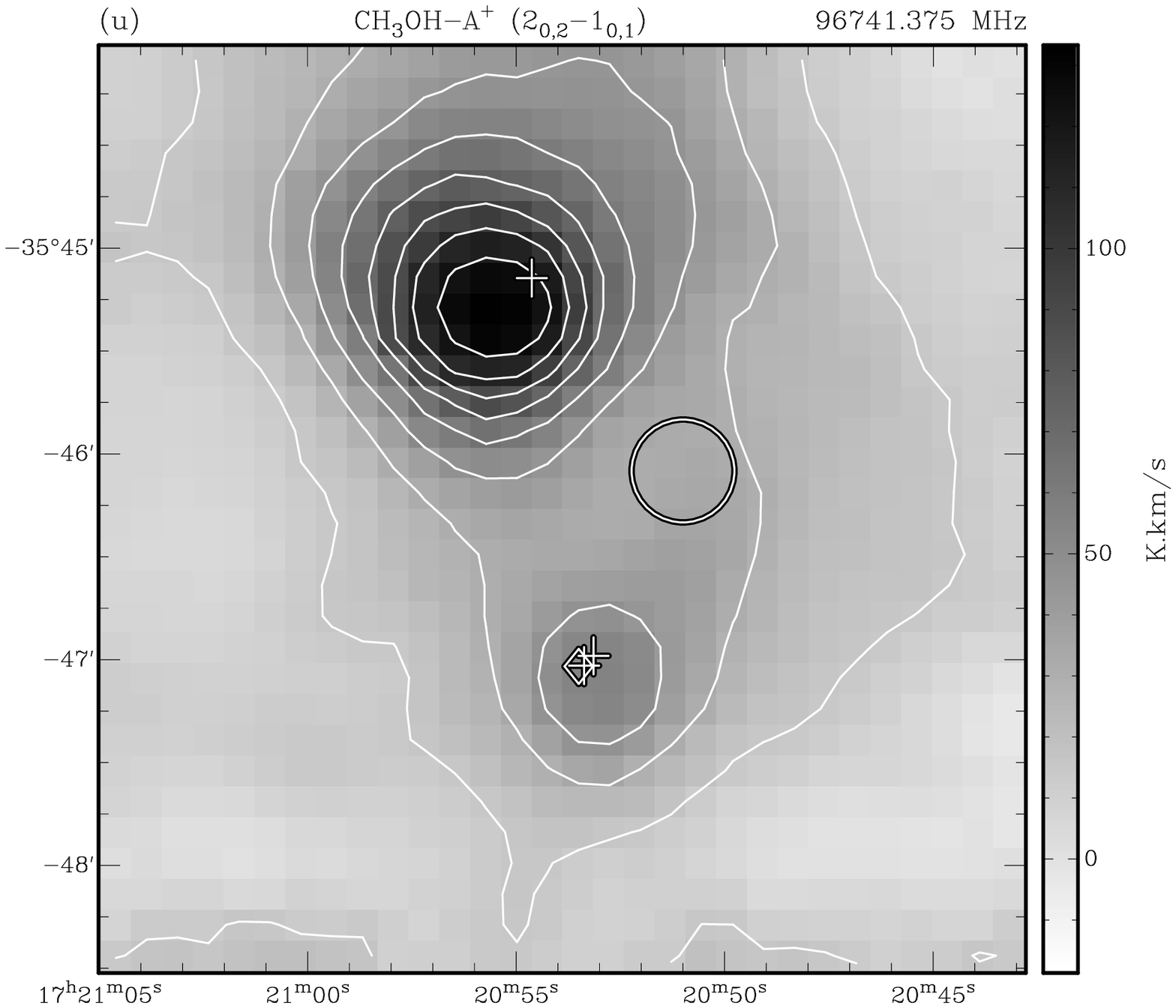}&
\includegraphics[width=0.44\textwidth]{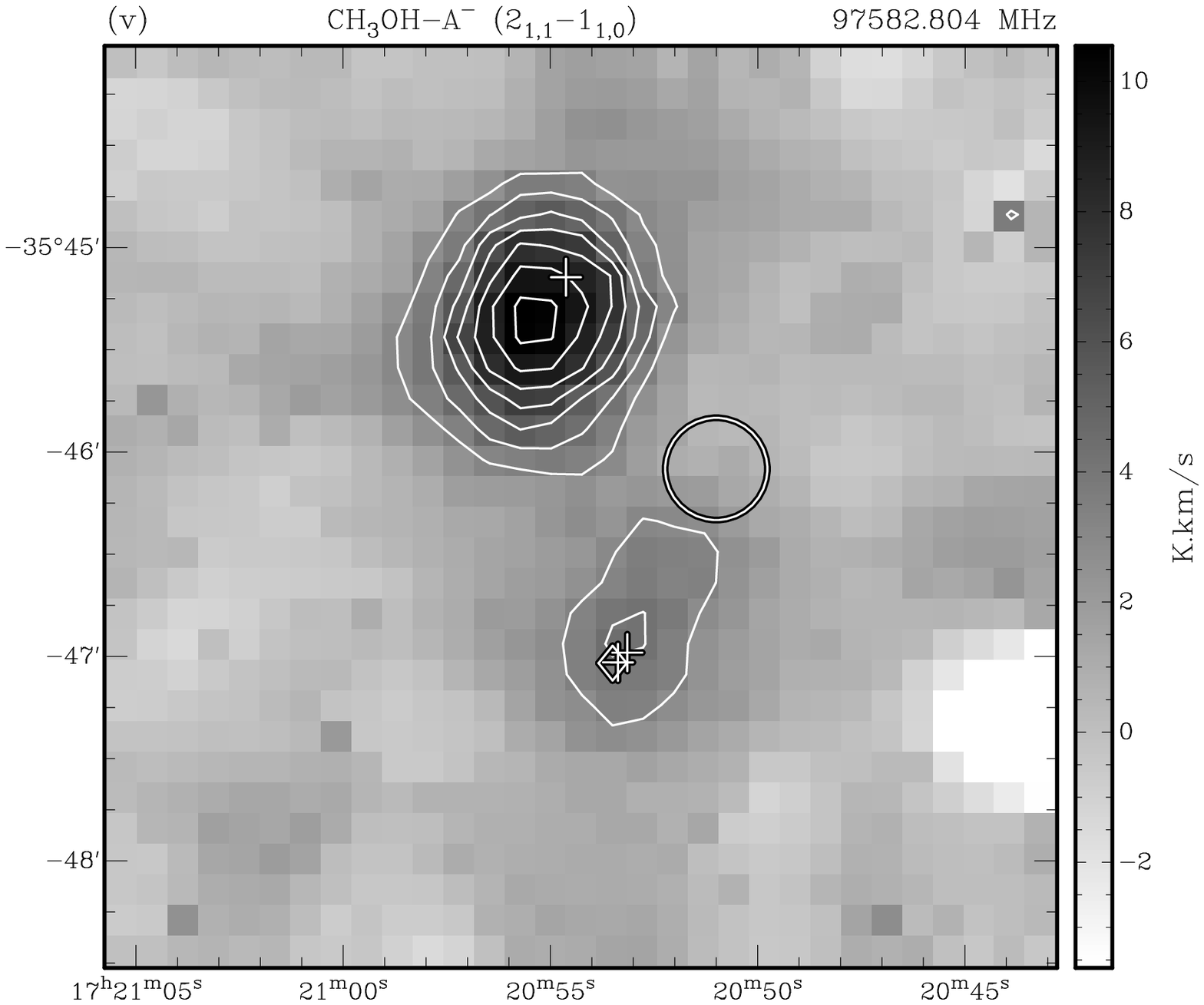}\\
\includegraphics[width=0.44\textwidth]{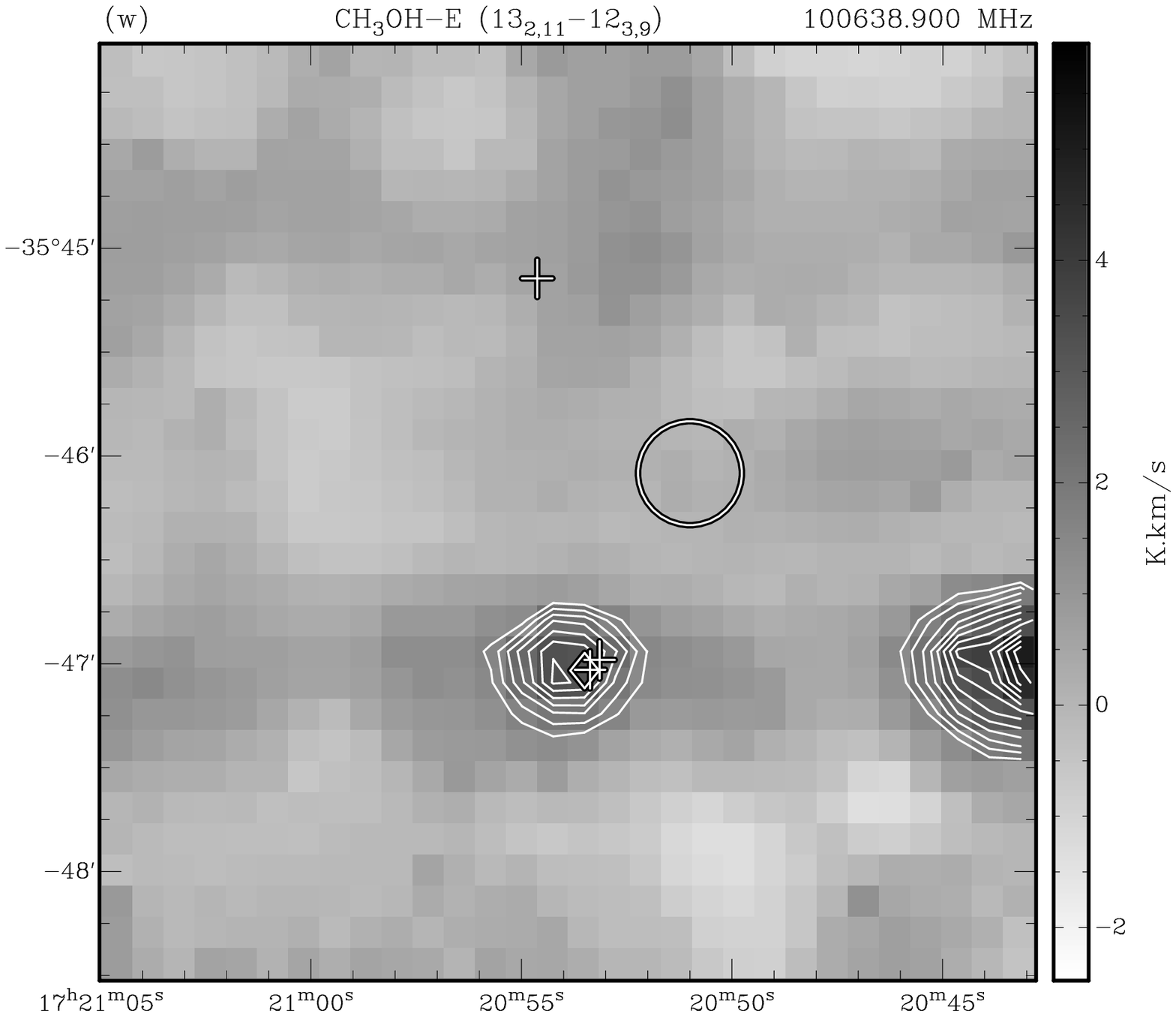}&
\includegraphics[width=0.44\textwidth]{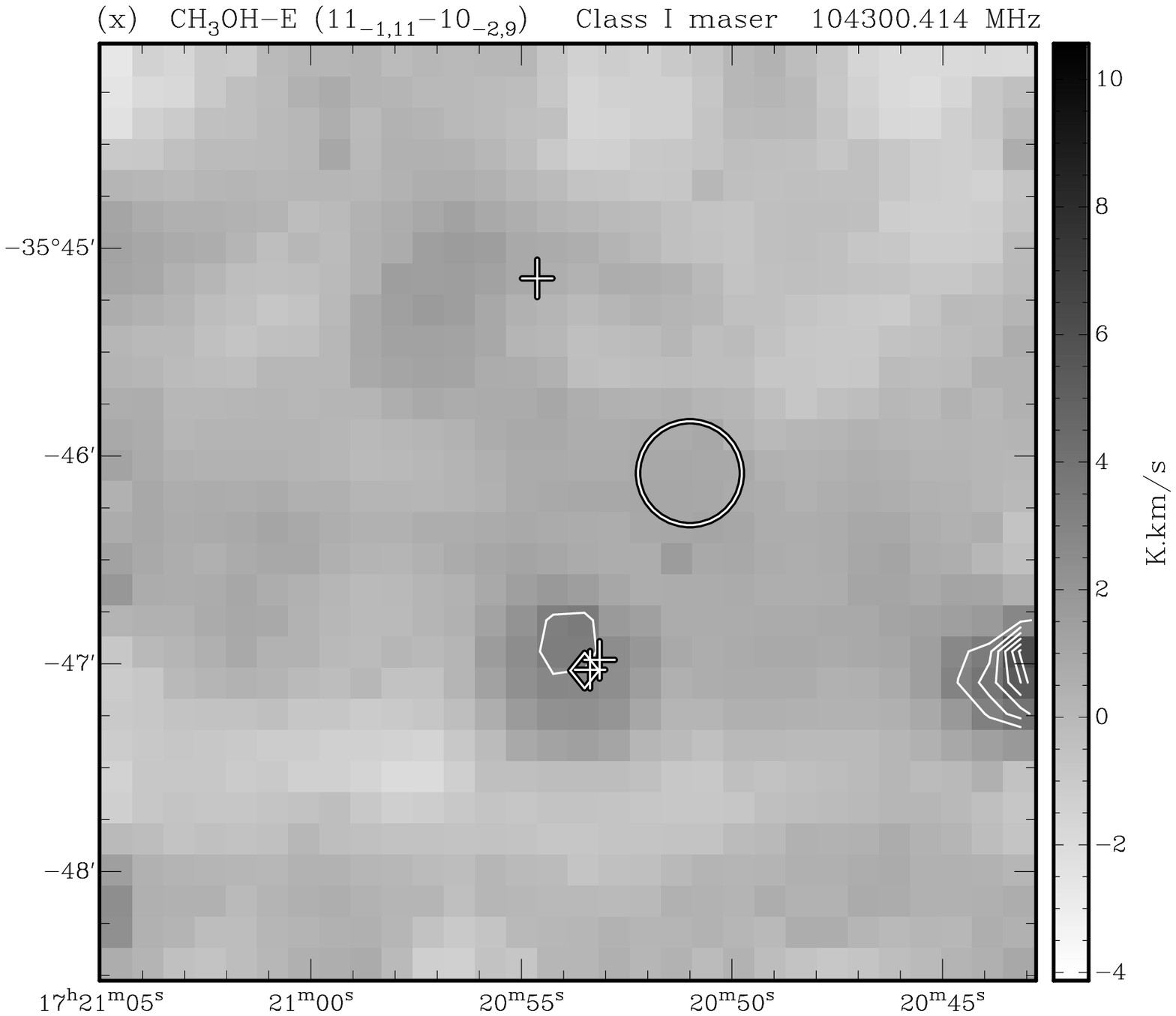}\\
\end{tabular}

\contcaption{{\bf (s)} CH$_3$OH-A$^+$ (8$_{0,8}$--7$_{1,7}$) --- methanol. Contours start
at 5$\sigma$ and increase in 5$\sigma$ steps, where 1$\sigma$
is 1.0\,K\,km\,s$^{-1}$. This methanol transition is a Class I maser
\citep{mueller04}.
{\bf (t)} CH$_3$OH-A$^+$ (2$_{1,2}$--1$_{1,1}$) --- methanol. Contours start
at 5$\sigma$ and increase in 2$\sigma$ steps, where 1$\sigma$
is 0.6\,K\,km\,s$^{-1}$.
{\bf (u)} CH$_3$OH-A$^+$ (2$_{0,2}$--1$_{0,1}$) --- methanol. Contours start
at 5$\sigma$ and increase in 5$\sigma$ steps, where 1$\sigma$
is 3.0\,K\,km\,s$^{-1}$.
{\bf (v)} CH$_3$OH-A$^-$ (2$_{1,1}$--1$_{1,0}$) --- methanol. Contours start
at 5$\sigma$ and increase in 2$\sigma$ steps, where 1$\sigma$
is 0.6\,K\,km\,s$^{-1}$.
{\bf (w)} CH$_3$OH-E (13$_{2,11}$--12$_{3,9}$) --- methanol. Contours start
at 5$\sigma$ and increase in 1$\sigma$ steps, where 1$\sigma$
is 0.3\,K\,km\,s$^{-1}$. Strong ``emission'' is seen on the right hand edge of the image,
at a declination of approximately $-35^\circ 47^\prime$, which 
is due to a bad data point, and is not real emission.
{\bf (x)} CH$_3$OH-E (11$_{-1,11}$--10$_{-2,9}$) --- methanol. Contours start
at 5$\sigma$ and increase in 1$\sigma$ steps, where 1$\sigma$
is 0.6\,K\,km\,s$^{-1}$. This methanol transition is a Class I maser
\citep{mueller04}. ``Emission'' is seen on the right hand edge of the image,
at a declination of approximately $-35^\circ 47^\prime$, which 
is due to a bad data point, and is not real emission.
}
\end{figure*}

\begin{figure*}
\begin{tabular}{cc}
\includegraphics[width=0.43\textwidth]{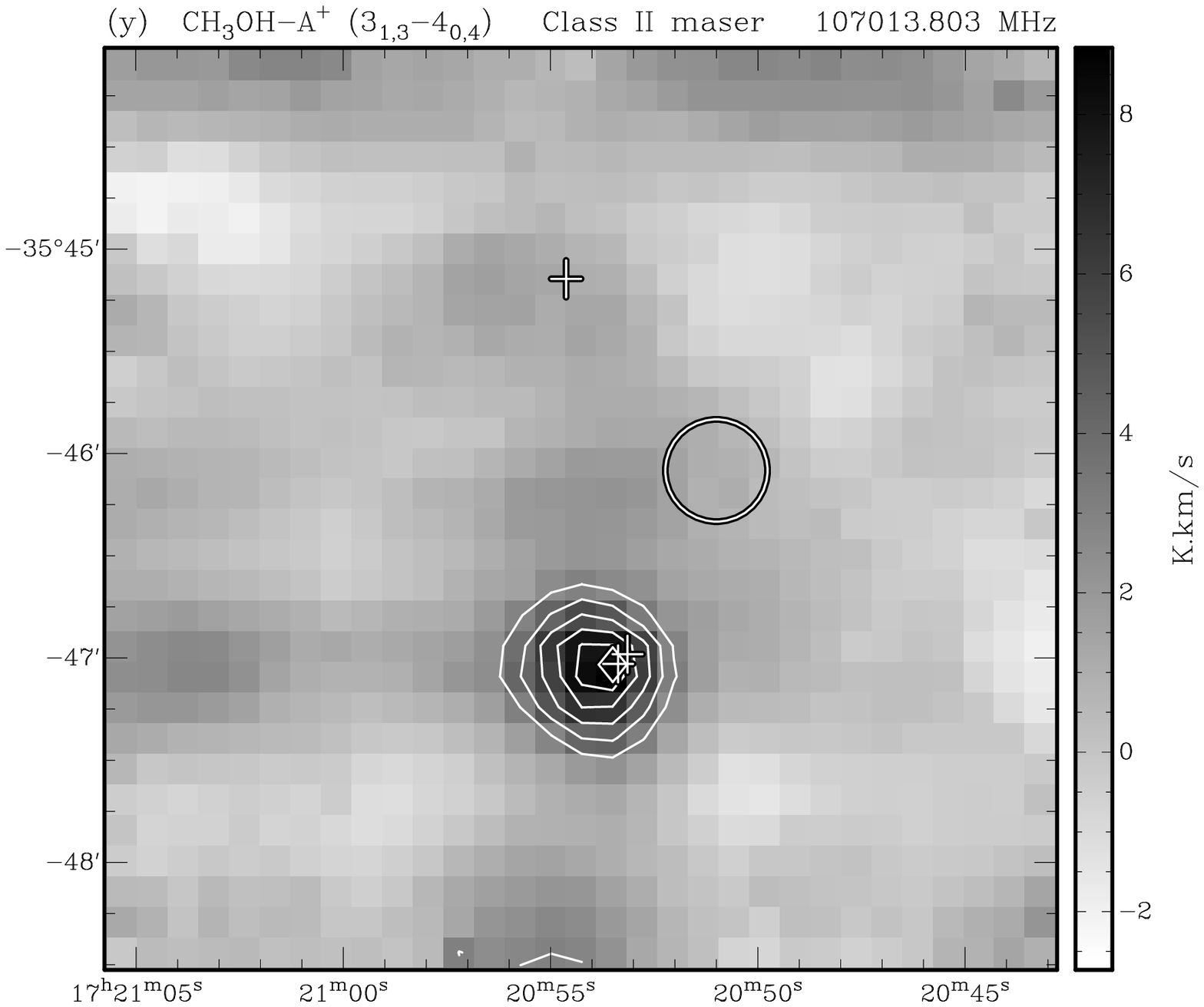}&
\includegraphics[width=0.43\textwidth]{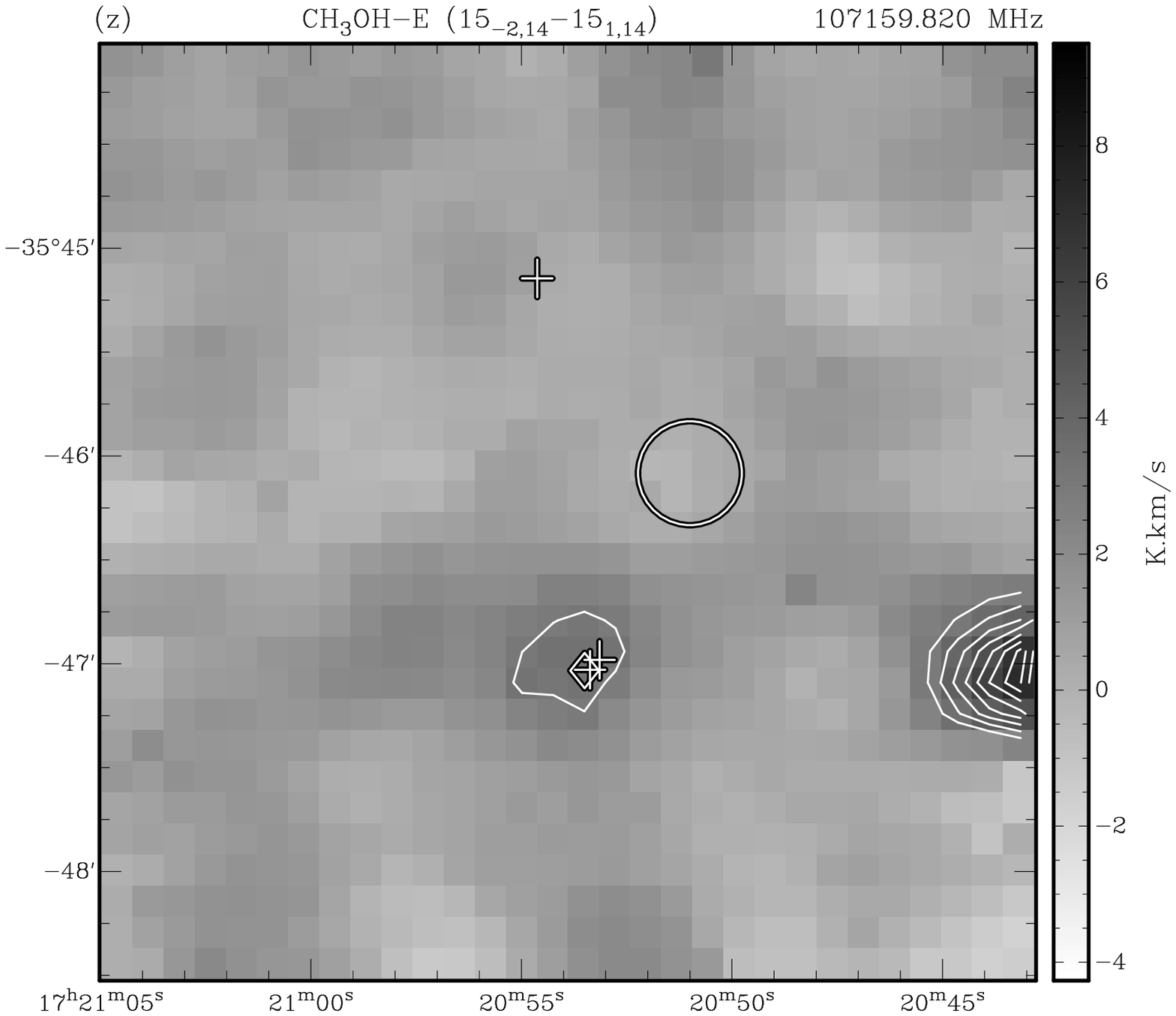}\\
\includegraphics[width=0.43\textwidth]{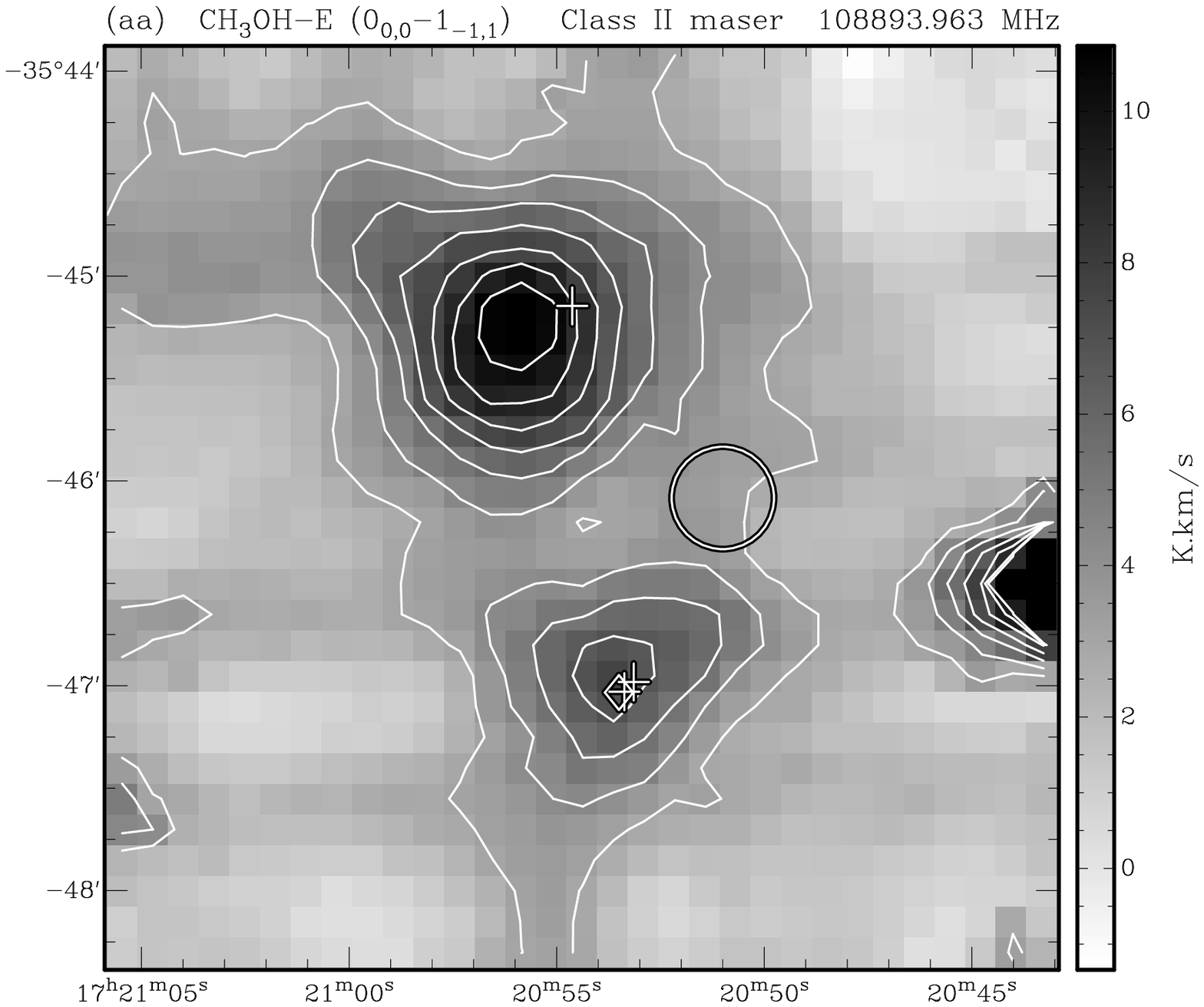}&
\includegraphics[width=0.43\textwidth]{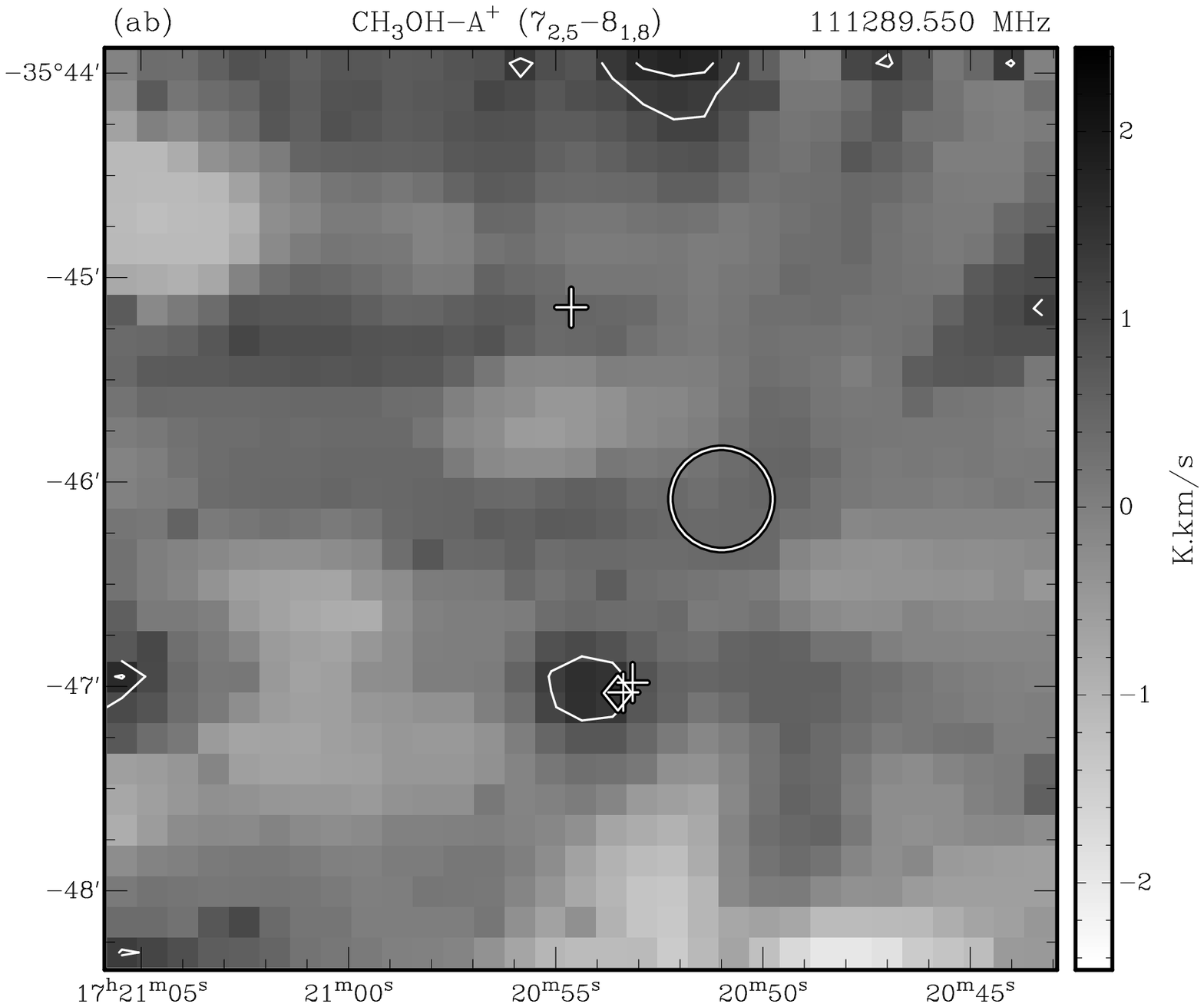}\\
\includegraphics[width=0.43\textwidth]{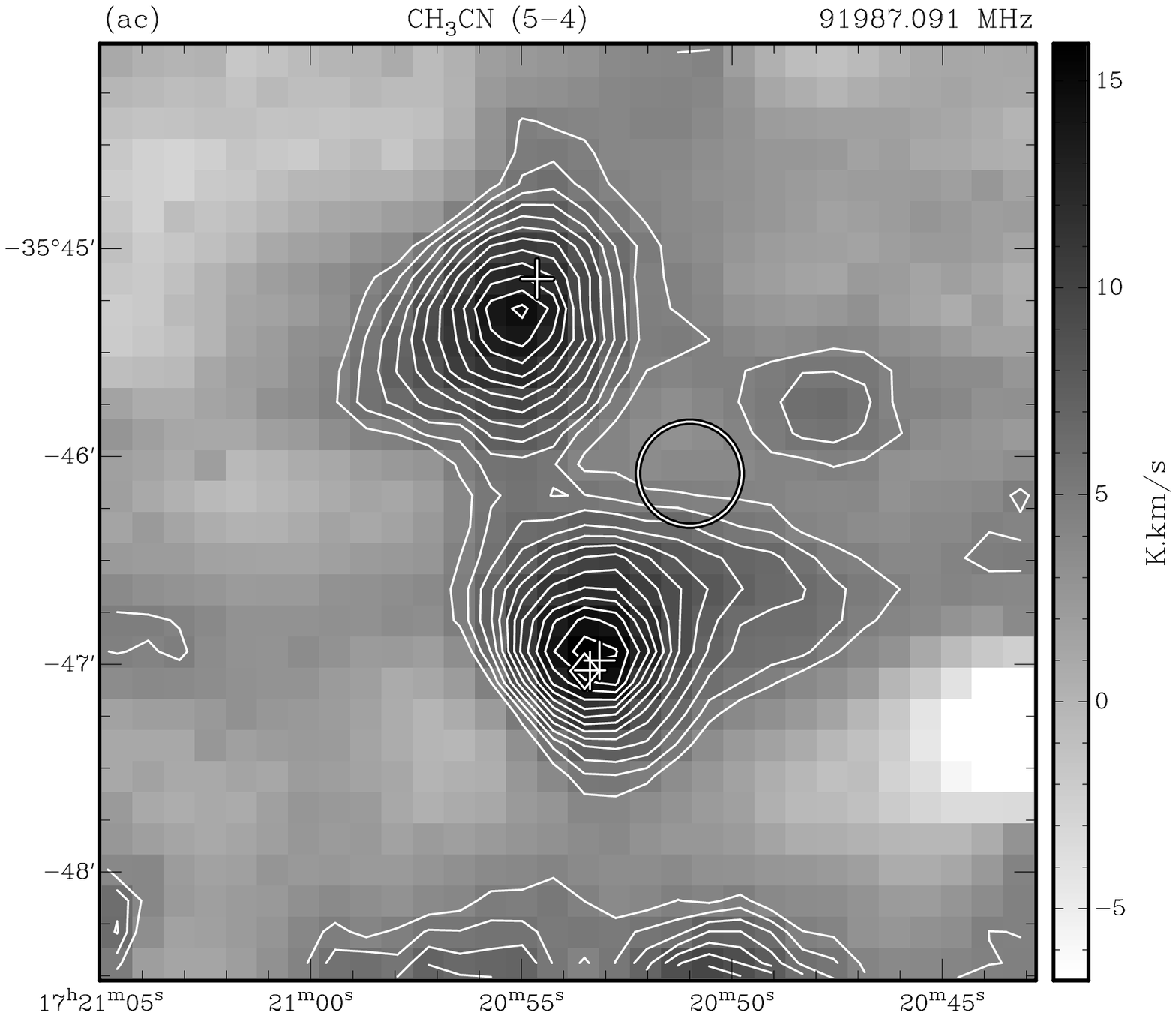}&
\includegraphics[width=0.43\textwidth]{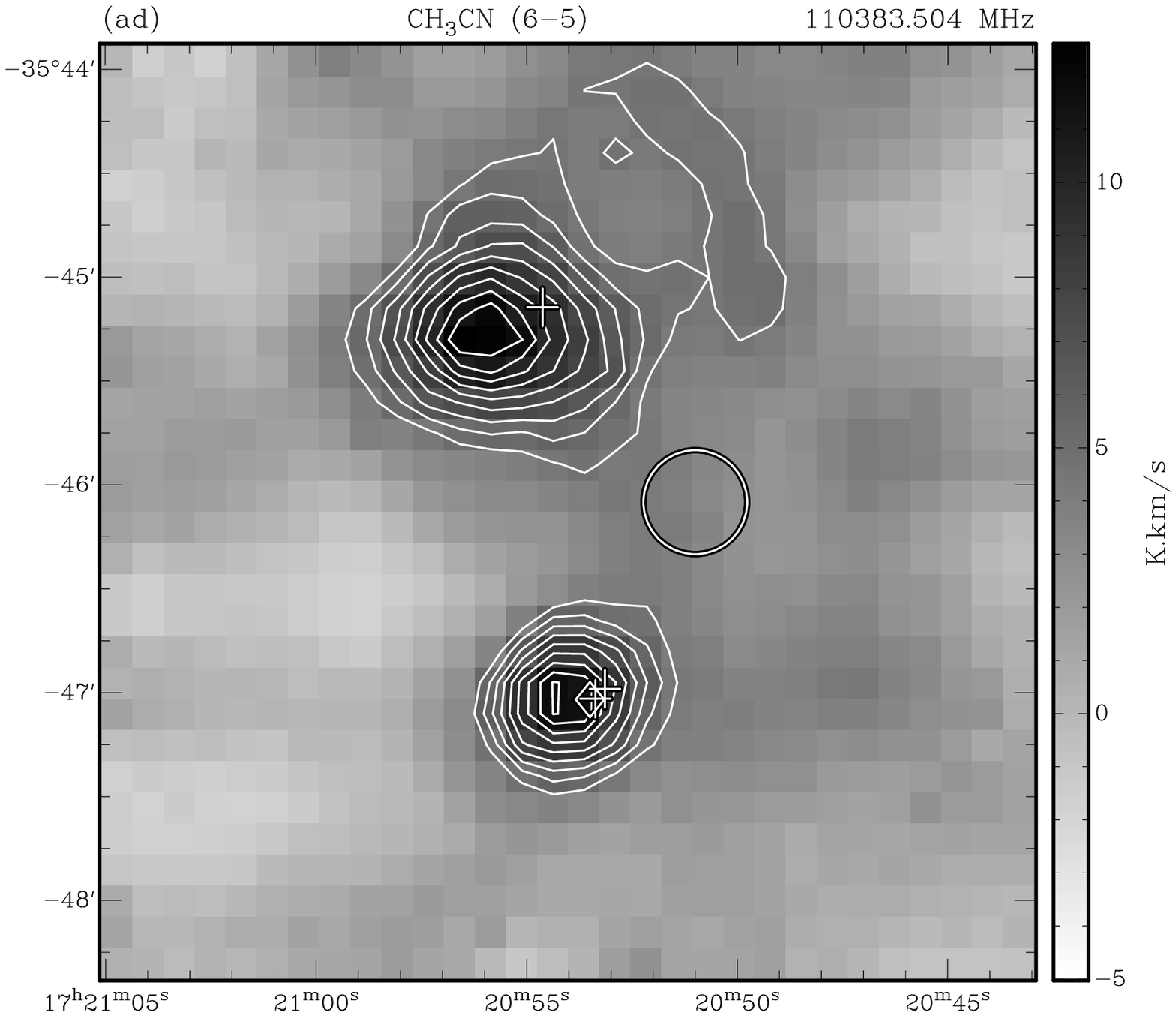}\\
\end{tabular}

\contcaption{{\bf (y)} CH$_3$OH-A$^+$ (3$_{1,3}$--4$_{0,4}$) --- methanol. Contours start
at 5$\sigma$ and increase in 2$\sigma$ steps, where 1$\sigma$
is 0.6\,K\,km\,s$^{-1}$. This methanol transition is a Class II maser
\citep{mueller04}.
{\bf (z)} CH$_3$OH-E (15$_{-2,14}$--15$_{1,14}$) --- methanol. Contours start
at 5$\sigma$ and increase in 1$\sigma$ steps, where 1$\sigma$
is 0.6\,K\,km\,s$^{-1}$. Strong ``emission'' is seen on the right hand edge of the image,
at a declination of approximately $-35^\circ 47^\prime$, which 
is due to a bad data point, and is not real emission.
{\bf (aa)} CH$_3$OH-E (0$_{0,0}$--1$_{-1,1}$) --- methanol. Contours start
at 5$\sigma$ and increase in 2$\sigma$ steps, where 1$\sigma$
is 0.6\,K\,km\,s$^{-1}$. This methanol transition is a Class II maser
\citep{mueller04}. Strong ``emission'' is seen on the right hand edge of the image,
at a declination of approximately $-35^\circ 46^\prime 30^{\prime\prime}$, which 
is due to a bad data point, and is not real emission.
{\bf (ab)} CH$_3$OH-A$^+$ (7$_{2,5}$--8$_{1,8}$) --- methanol. Contours start
at 3$\sigma$ and increase in 1$\sigma$ steps, where 1$\sigma$
is 0.5\,K\,km\,s$^{-1}$.
{\bf (ac)} CH$_3$CN (5--4) --- methyl cyanide. The emission includes
all transitions in the K ladder from 0 to 3, inclusive. Contours start
at 5$\sigma$ and increase in 1$\sigma$ steps, where 1$\sigma$
is 0.9\,K\,km\,s$^{-1}$. We attribute the ``emission'' seen south of
$-35^\circ 48^\prime$ to be due to poor weather and is thus not real.
{\bf (ad)} CH$_3$CN (6--5) --- methyl cyanide. The emission includes
all transitions in the K ladder from 0 to 3, inclusive. Contours start
at 5$\sigma$ and increase in 1$\sigma$ steps, where 1$\sigma$
is 0.9\,K\,km\,s$^{-1}$.
}
\end{figure*}

\begin{figure*}
\begin{tabular}{cc}
\includegraphics[width=0.45\textwidth]{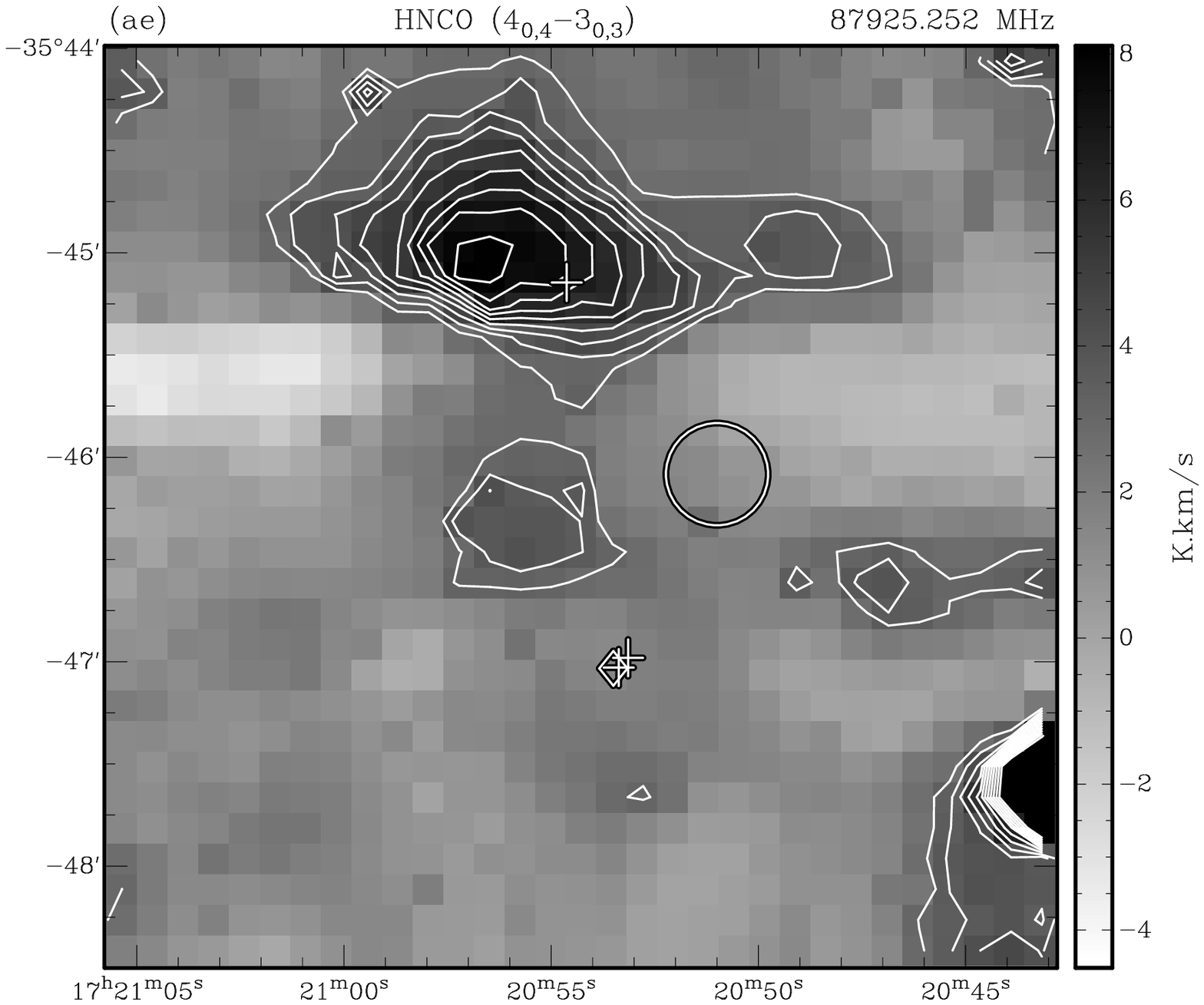}&
\includegraphics[width=0.45\textwidth]{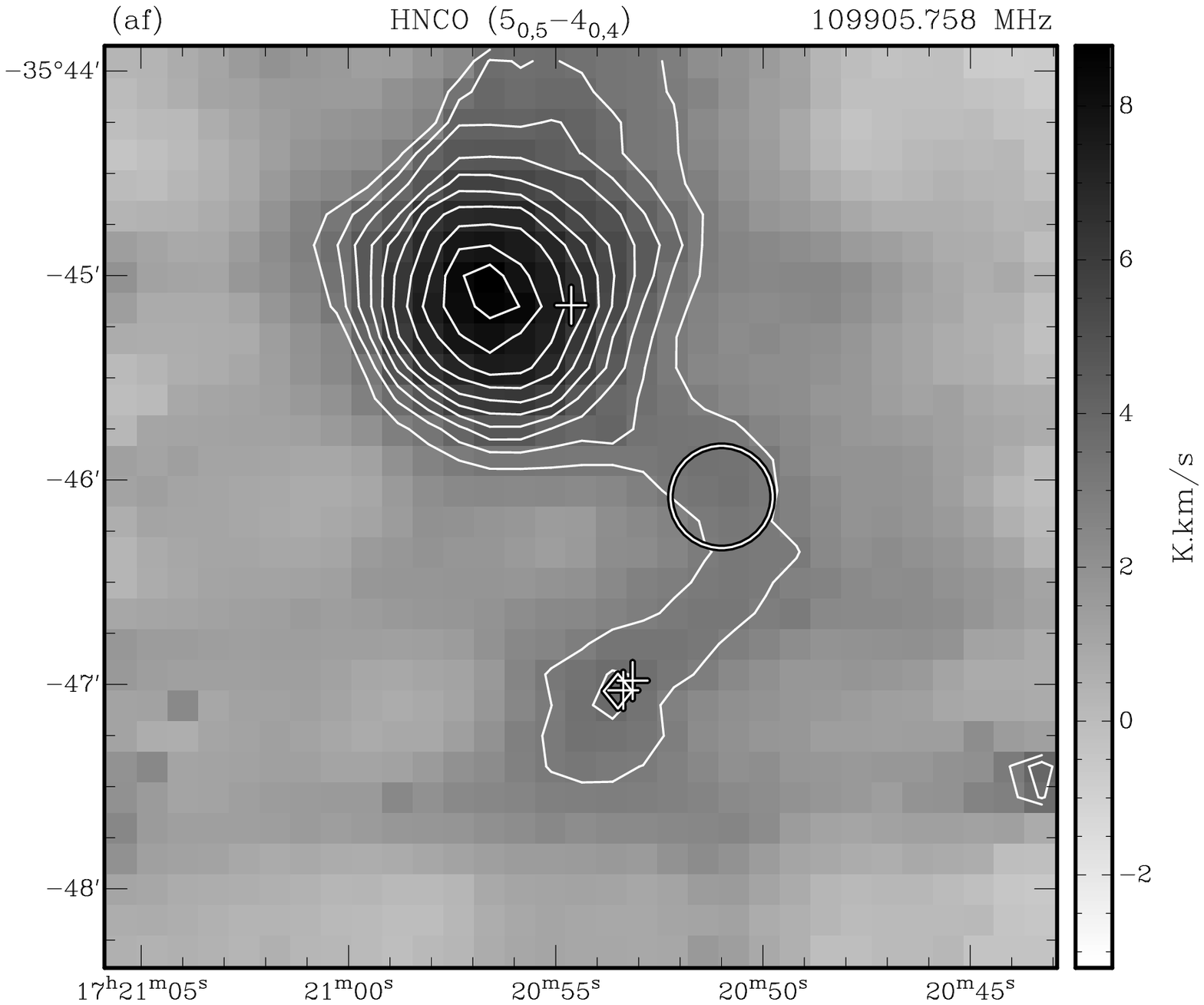}\\
\includegraphics[width=0.45\textwidth]{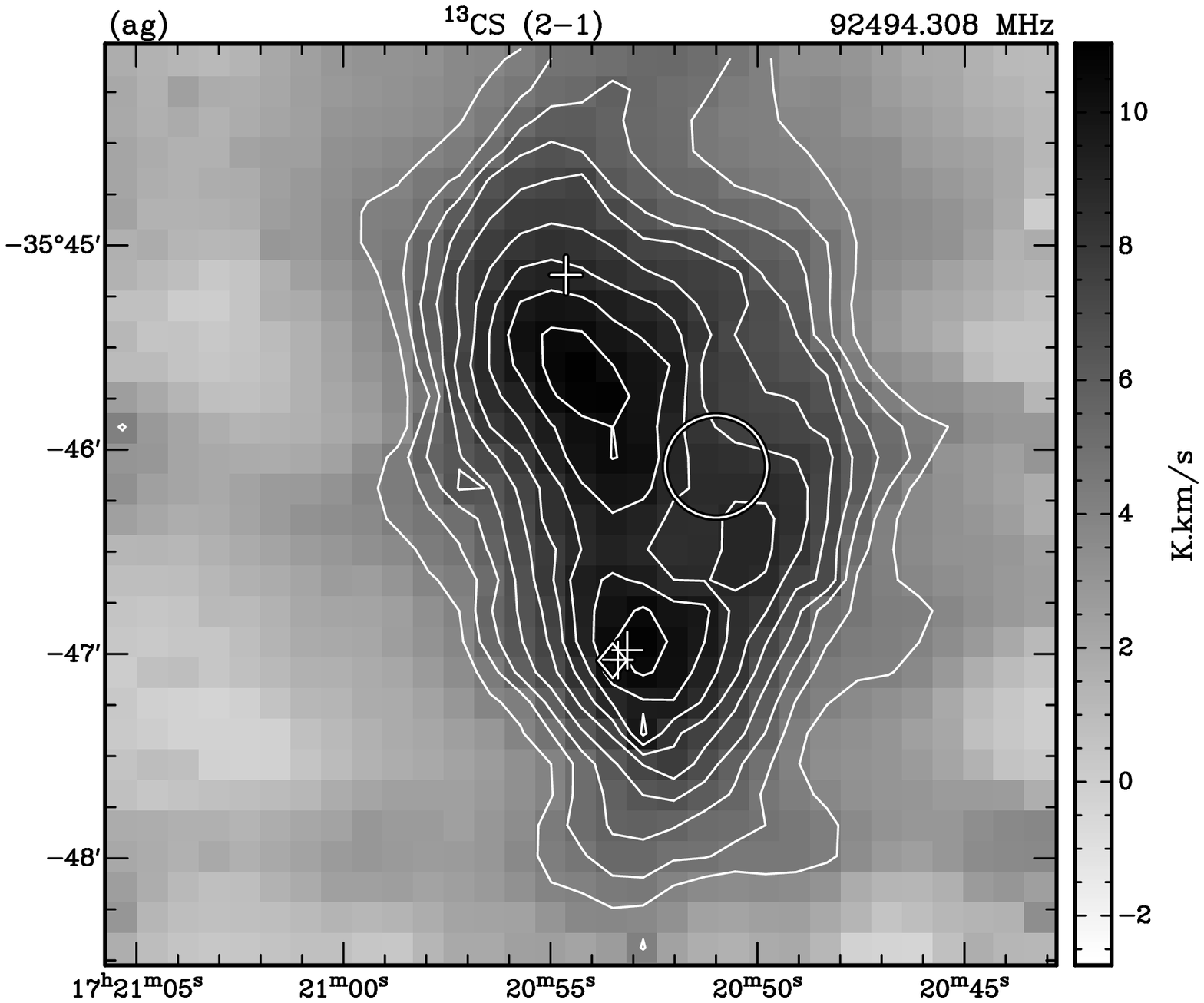}&
\includegraphics[width=0.45\textwidth]{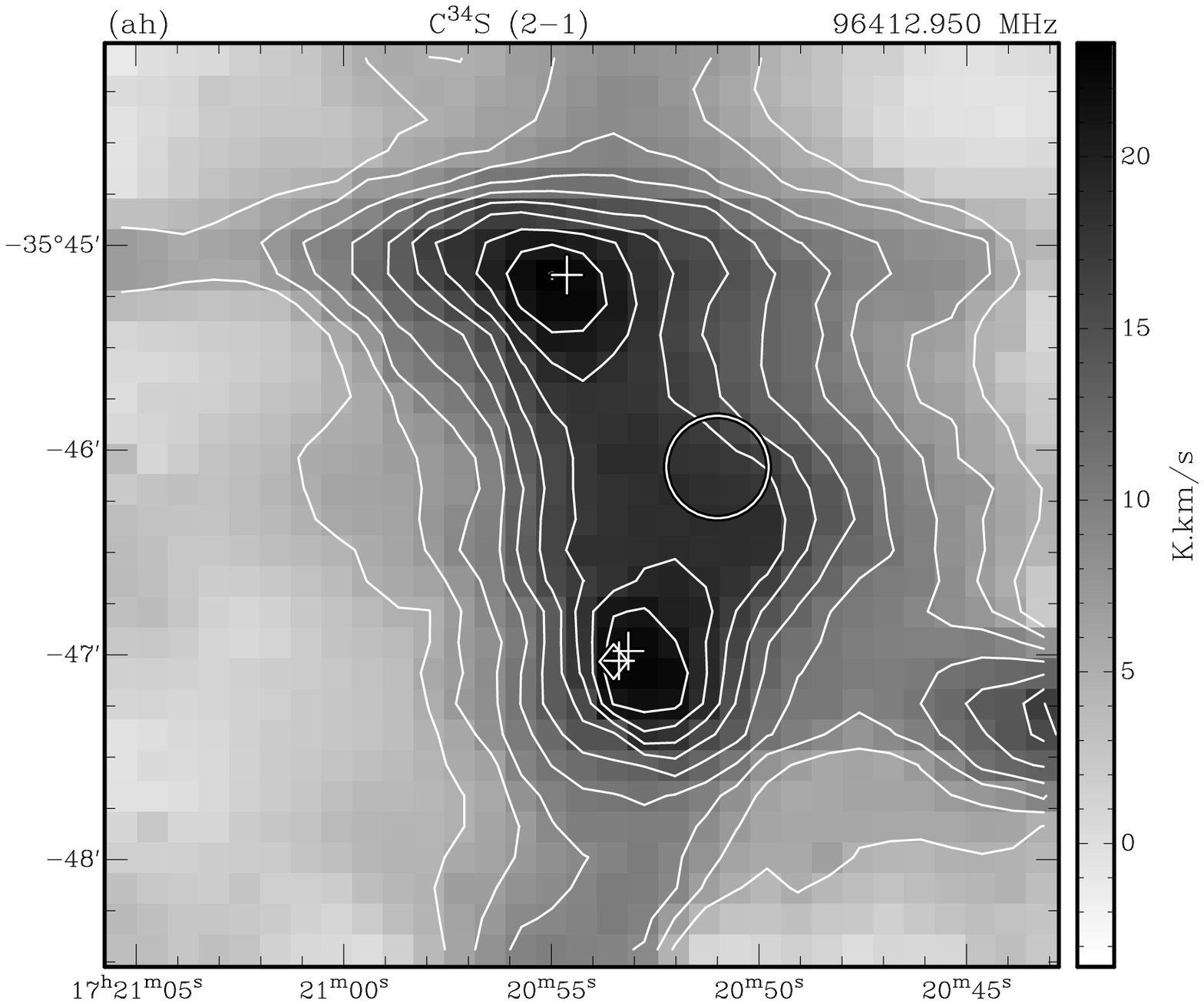}\\
\includegraphics[width=0.45\textwidth]{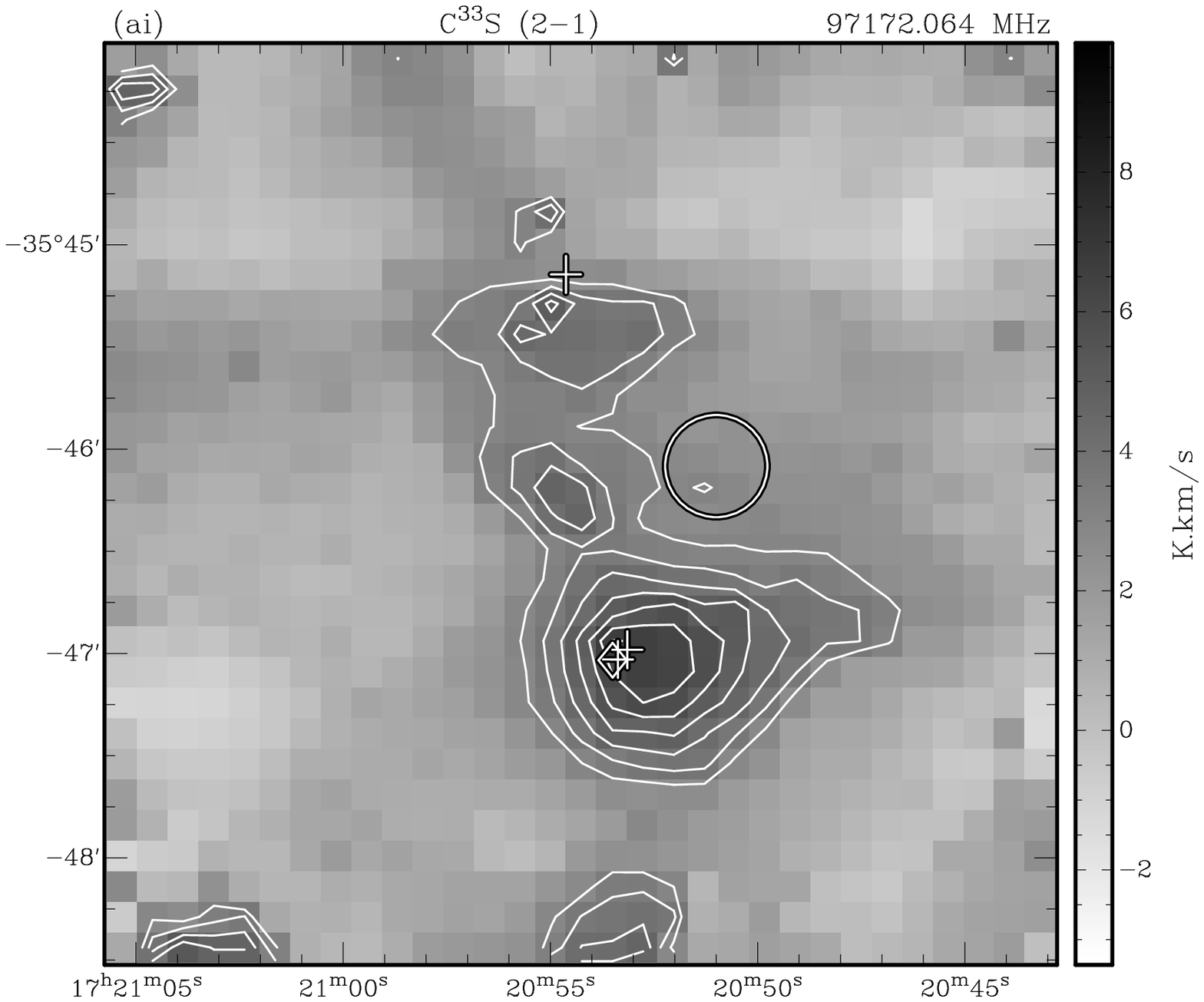}&
\includegraphics[width=0.45\textwidth]{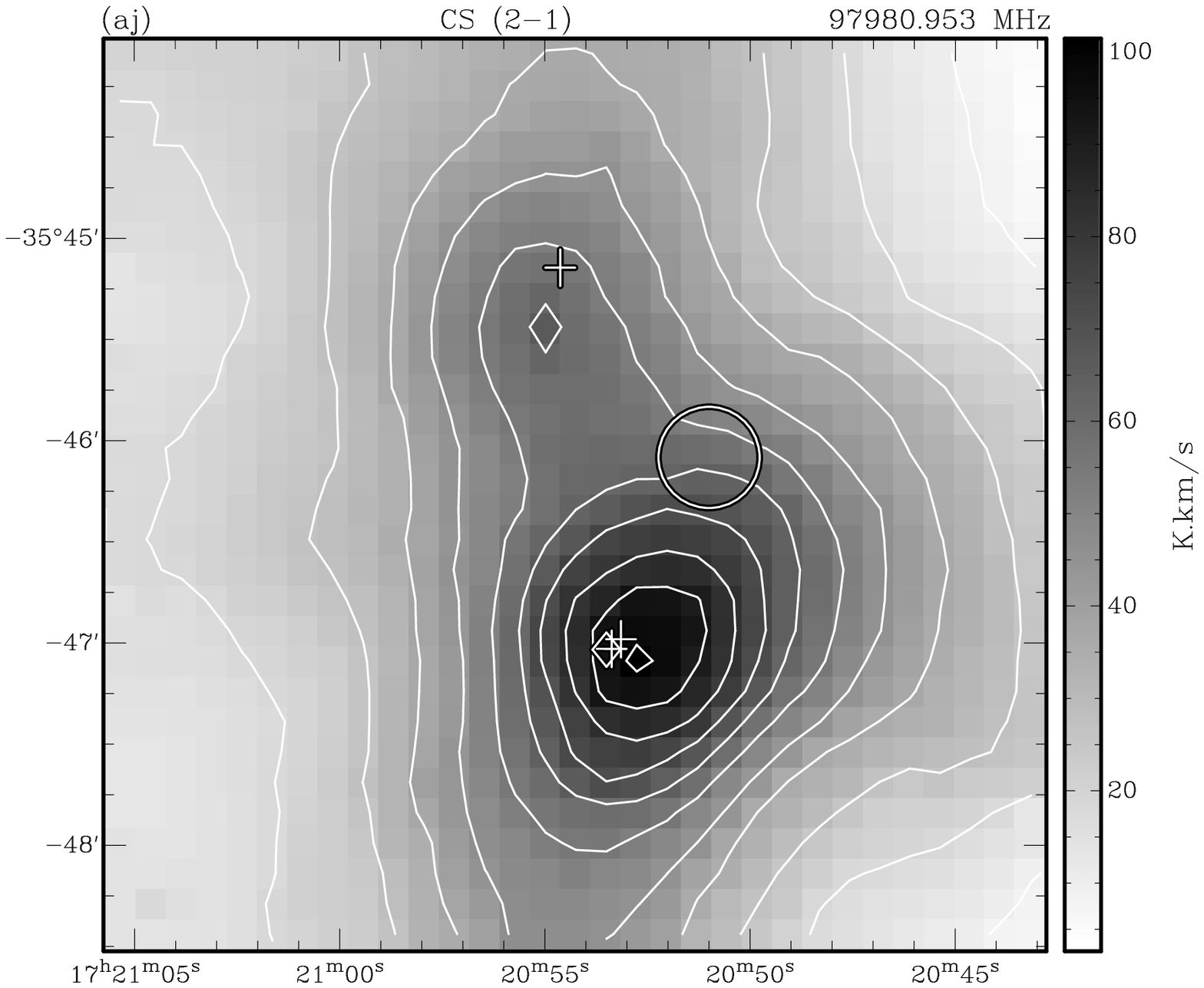}\\
\end{tabular}

\contcaption{{\bf (ae)} HNCO (4$_{0,4}$--3$_{0,3}$) --- isocyanic acid. Contours start
at 5$\sigma$ and increase in 1$\sigma$ steps, where 1$\sigma$
is 0.6\,K\,km\,s$^{-1}$. Strong ``emission'' is seen on the right hand edge of the image,
at a declination of approximately $-35^\circ 47^\prime 30^{\prime\prime}$, which
is due to a bad data point, and is not real emission.
{\bf (af)} HNCO (5$_{0,5}$--4$_{0,4}$)  --- isocyanic acid. Contours start
at 5$\sigma$ and increase in 1$\sigma$ steps, where 1$\sigma$
is 0.6\,K\,km\,s$^{-1}$.
{\bf (ag)} $^{13}$CS (2--1) --- carbon sulfide. Contours start
at 5$\sigma$ and increase in 1$\sigma$ steps, where 1$\sigma$
is 0.7\,K\,km\,s$^{-1}$.
{\bf (ah)} C$^{34}$S (2--1) --- carbon sulfide. Contours start
at 5$\sigma$ and increase in 2$\sigma$ steps, where 1$\sigma$
is 1.0\,K\,km\,s$^{-1}$. Strong ``emission'' is seen on the right hand edge of the image,
at a declination of approximately $-35^\circ 47^\prime 15^{\prime\prime}$, which
is due to a bad data point, and is not real emission.
{\bf (ai)} C$^{33}$S (2--1) --- carbon sulfide. Contours start
at 5$\sigma$ and increase in 1$\sigma$ steps, where 1$\sigma$
is 0.6\,K\,km\,s$^{-1}$.
{\bf (aj)} CS (2--1) --- carbon sulfide. Contours start
at 5$\sigma$ and increase in 5$\sigma$ steps, where 1$\sigma$
is 1.8\,K\,km\,s$^{-1}$.
}
\end{figure*}

\begin{figure*}
\begin{tabular}{cc}
\includegraphics[width=0.45\textwidth]{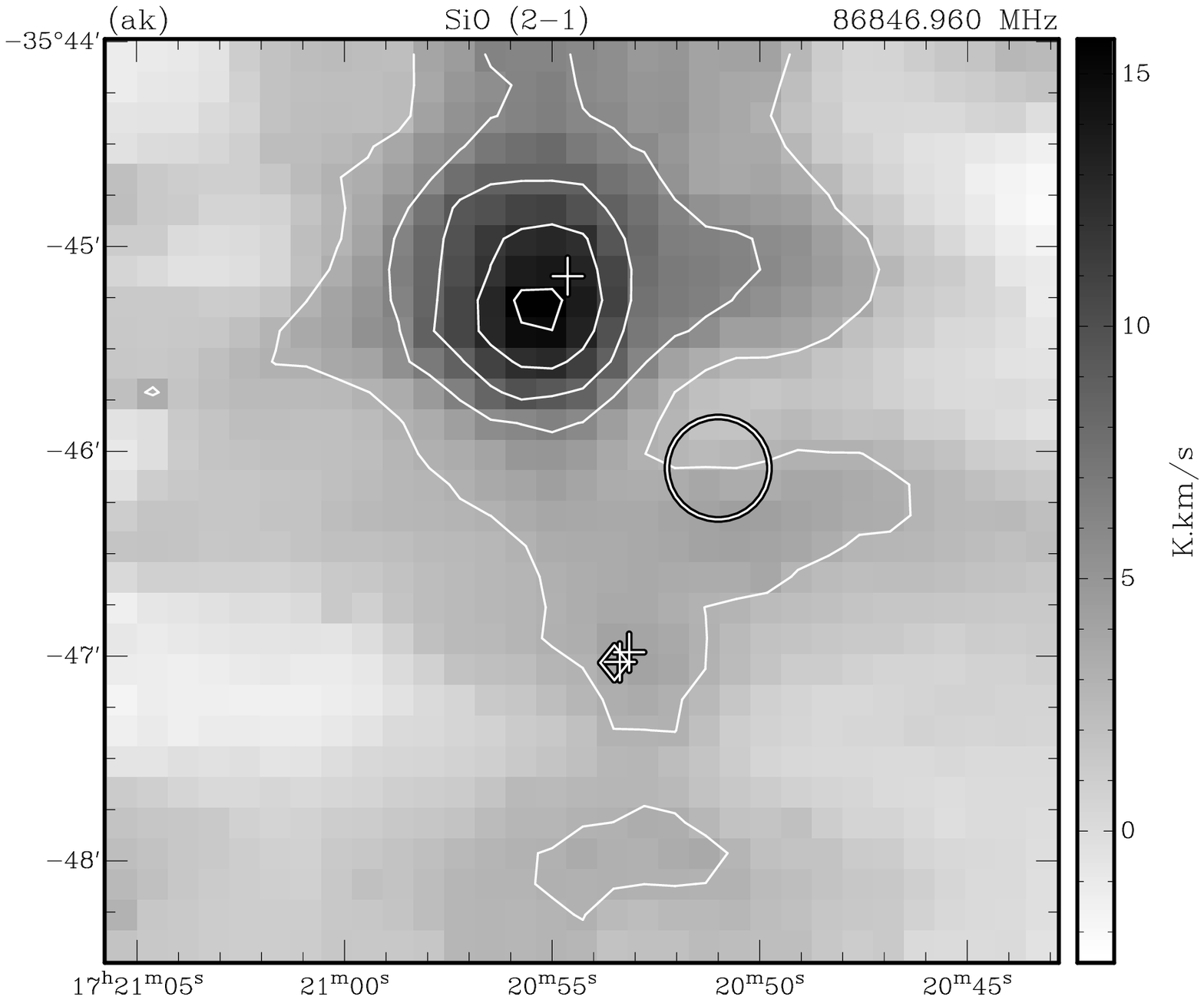}&
\includegraphics[width=0.45\textwidth]{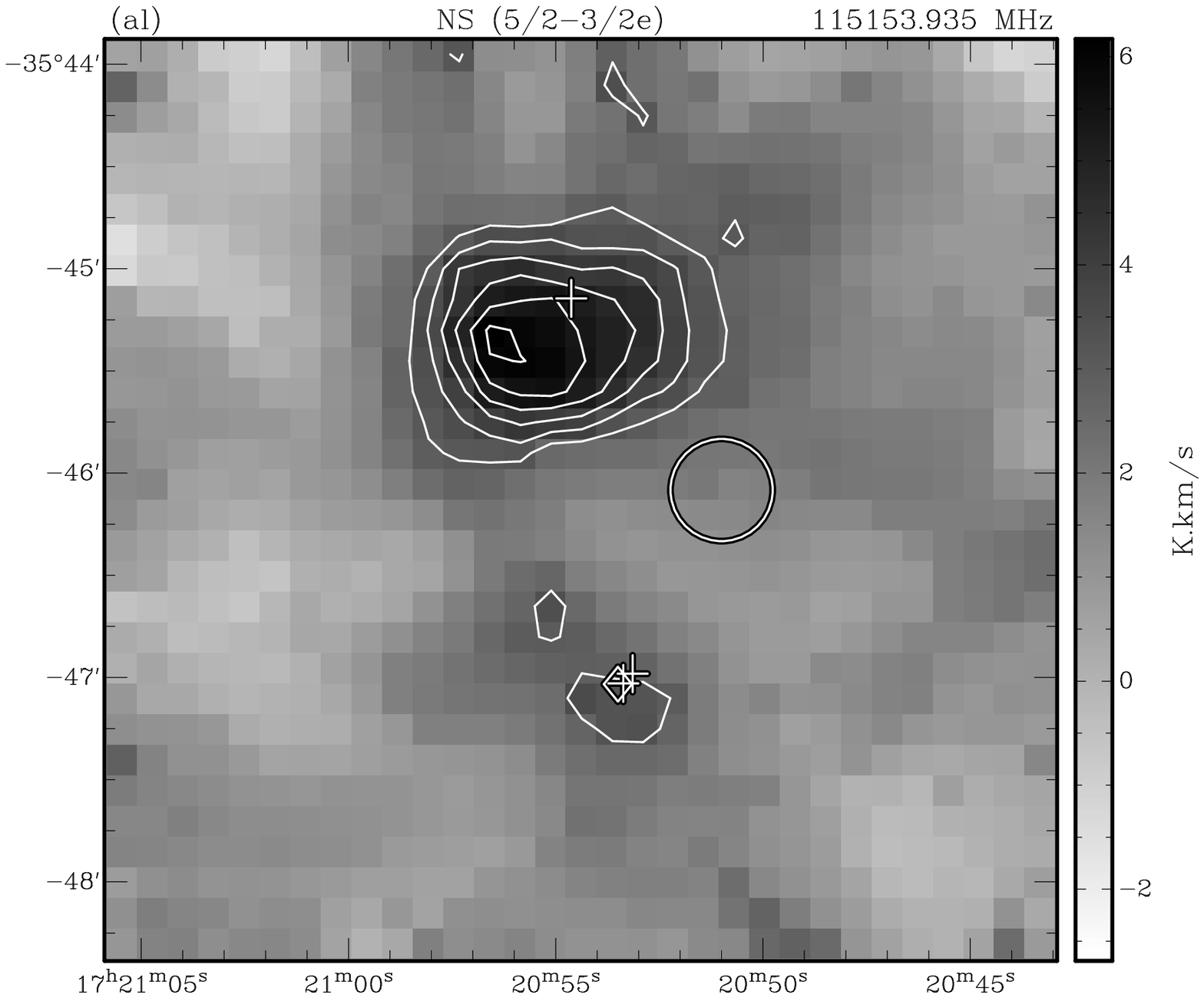}\\
\includegraphics[width=0.45\textwidth]{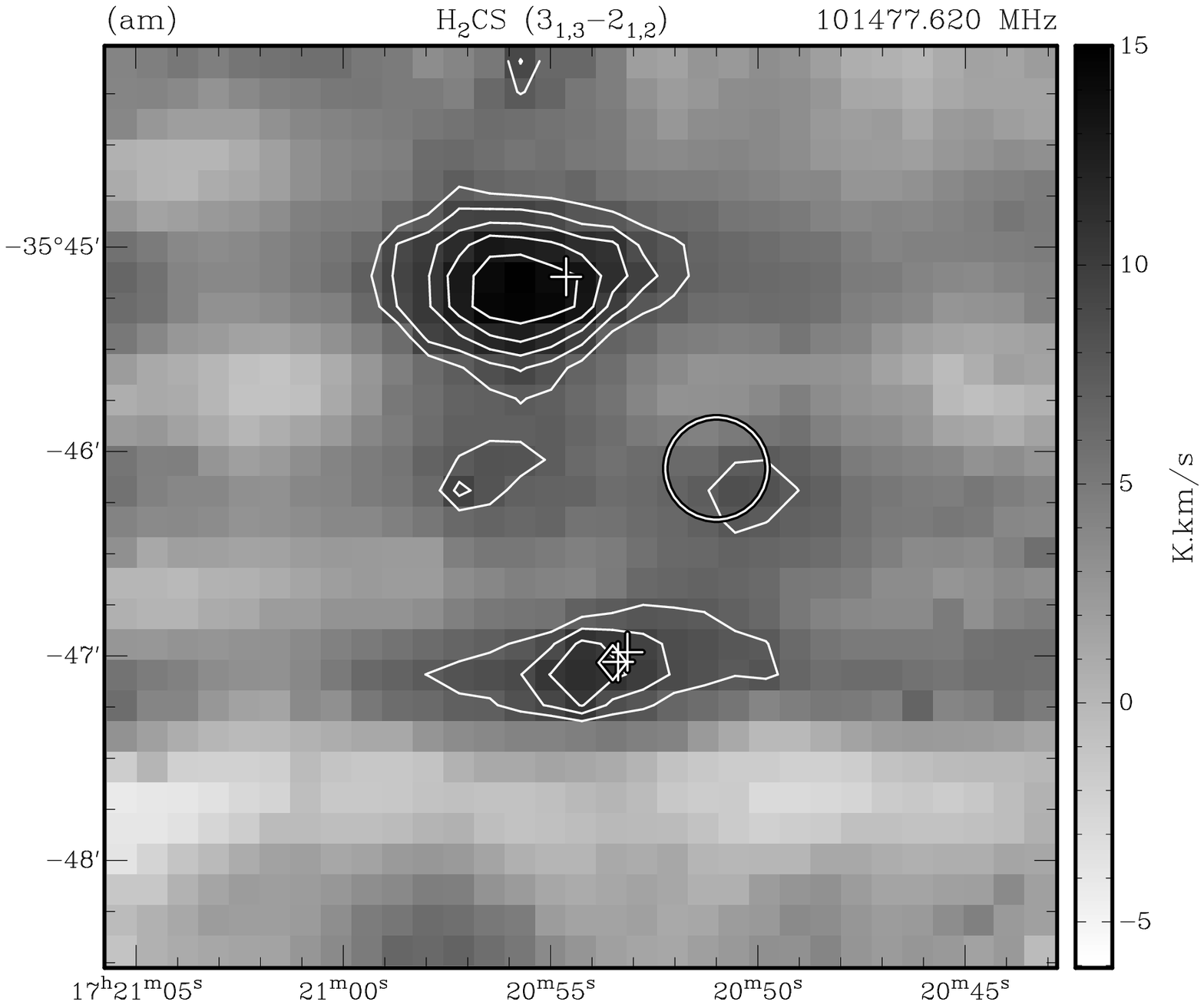}&
\includegraphics[width=0.45\textwidth]{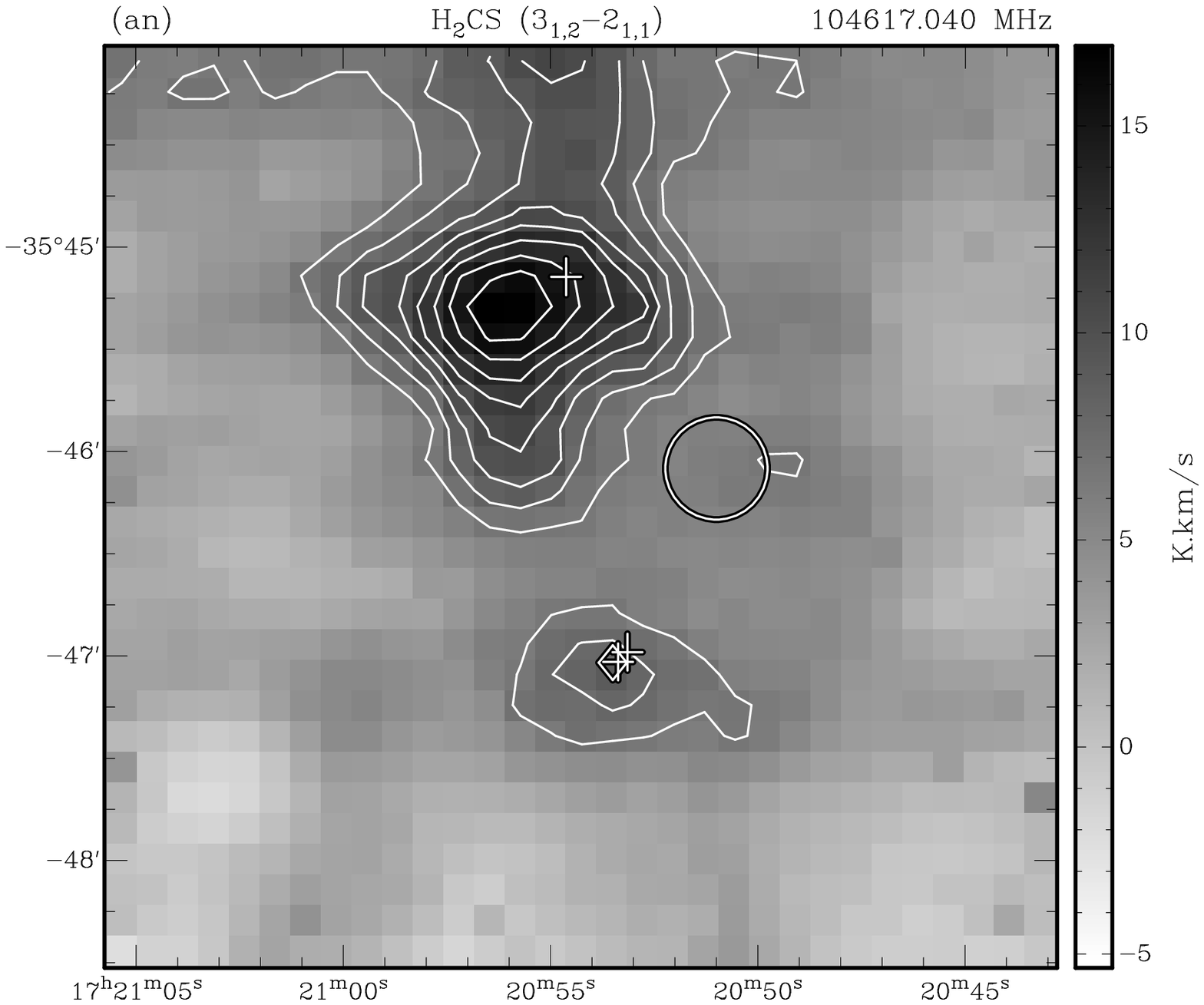}\\
\includegraphics[width=0.45\textwidth]{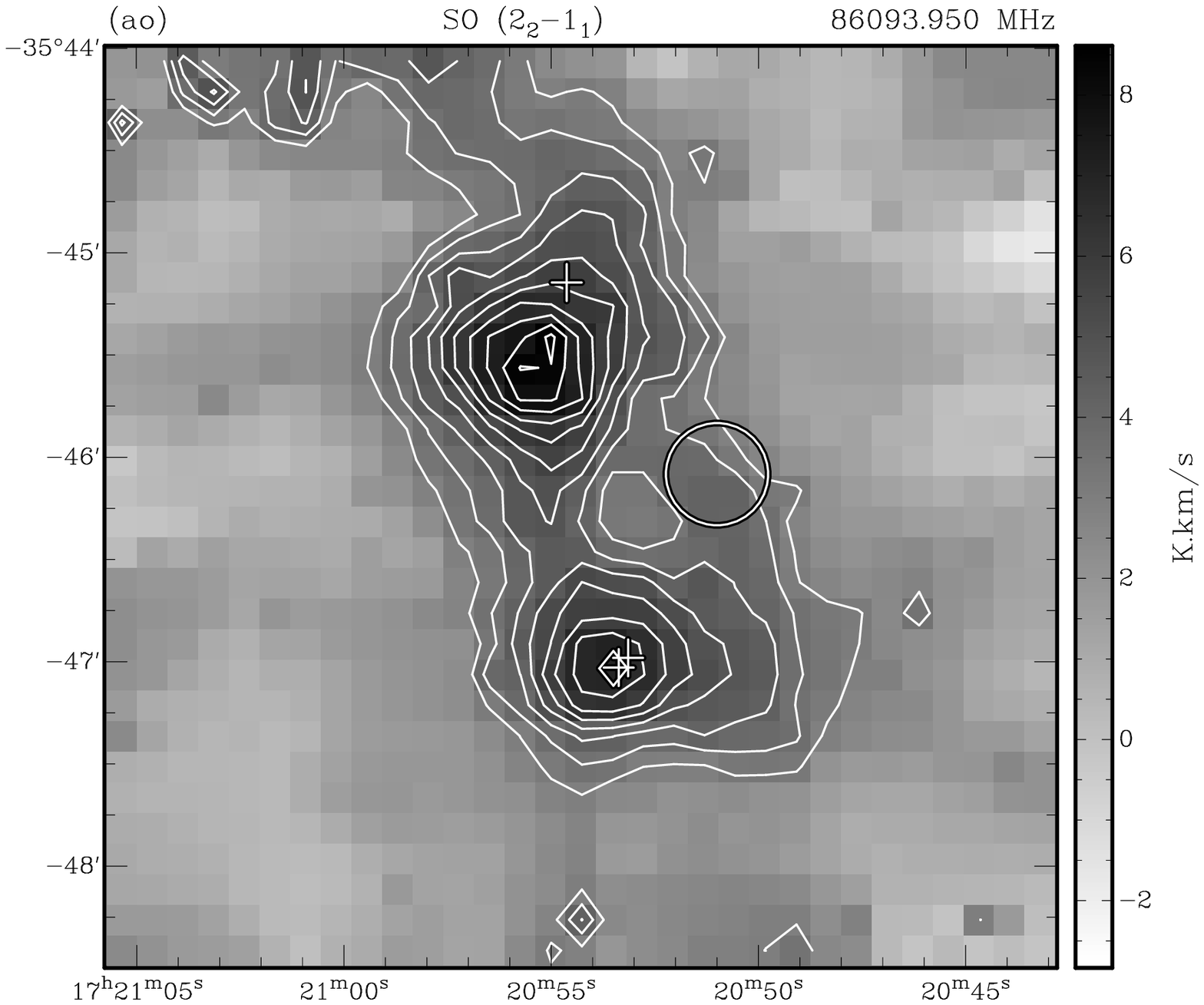}&
\includegraphics[width=0.45\textwidth]{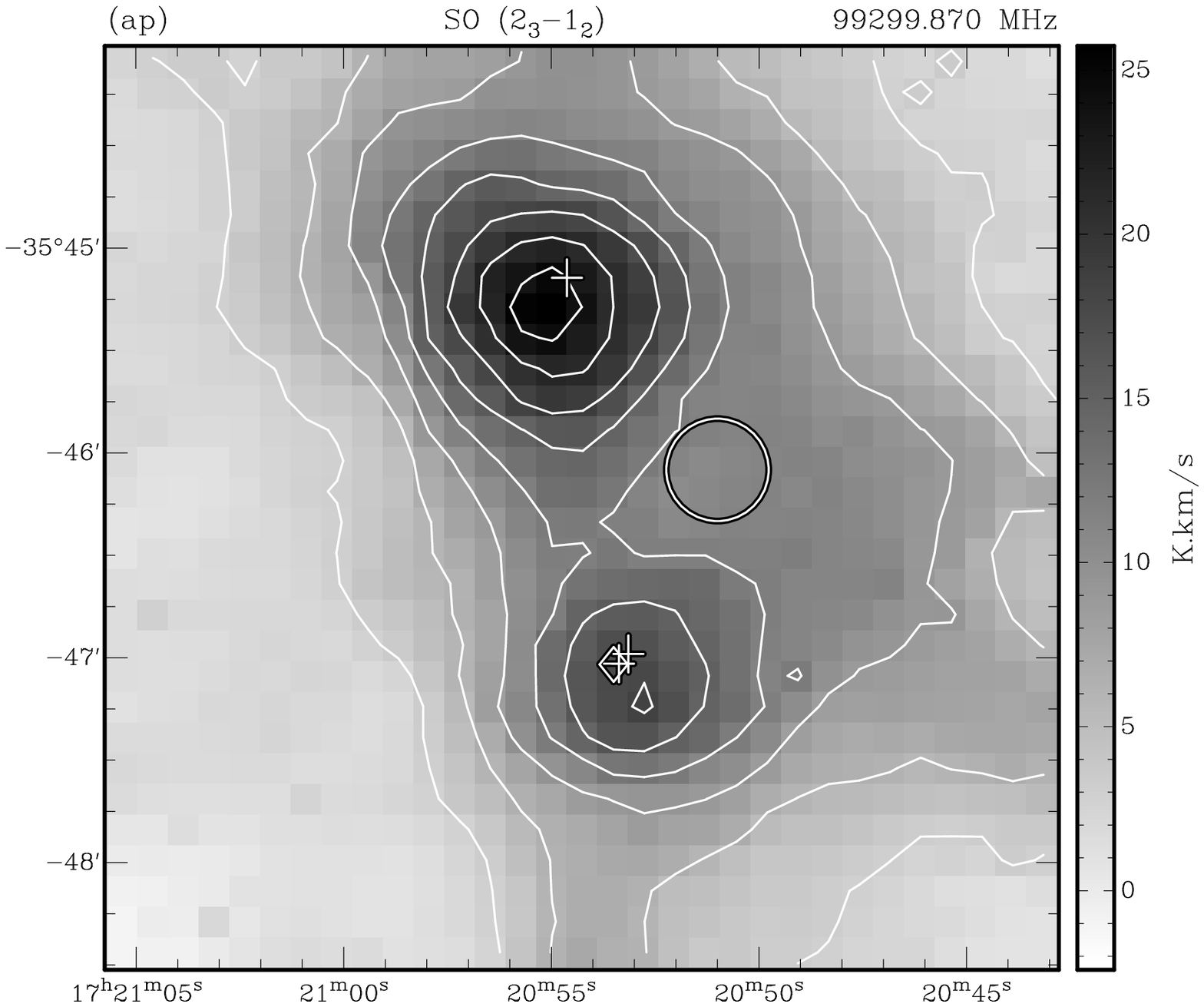}\\
\end{tabular}

\contcaption{{\bf (ak)} SiO (2--1) --- silicon monoxide. Contours start
at 5$\sigma$ and increase in 5$\sigma$ steps, where 1$\sigma$
is 0.6\,K\,km\,s$^{-1}$.
{\bf (al)} NS ($5/2$--$3/2$e) --- nitrogen sulfide. Contours start
at 5$\sigma$ and increase in 1$\sigma$ steps, where 1$\sigma$
is 0.6\,K\,km\,s$^{-1}$.
{\bf (am)} H$_2$CS (3$_{1,3}$--2$_{1,2}$) --- thioformaldehyde. Contours start
at 5$\sigma$ and increase in 1$\sigma$ steps, where 1$\sigma$
is 1.5\,K\,km\,s$^{-1}$.
{\bf (an)} H$_2$CS (3$_{1,2}$--2$_{1,1}$) --- thioformaldehyde. Contours start
at 5$\sigma$ and increase in 1$\sigma$ steps, where 1$\sigma$
is 1.3\,K\,km\,s$^{-1}$.
{\bf (ao)} SO (2$_2$--1$_1$) --- sulfur monoxide. Contours start
at 5$\sigma$ and increase in 1$\sigma$ steps, where 1$\sigma$
is 0.6\,K\,km\,s$^{-1}$.
{\bf (ap)} SO (2$_3$-1$_2$) --- sulfur monoxide. Contours start
at 5$\sigma$ and increase in 5$\sigma$ steps, where 1$\sigma$
is 0.6\,K\,km\,s$^{-1}$.
}
\end{figure*}

\begin{figure*}
\begin{tabular}{cc}
\includegraphics[width=0.45\textwidth]{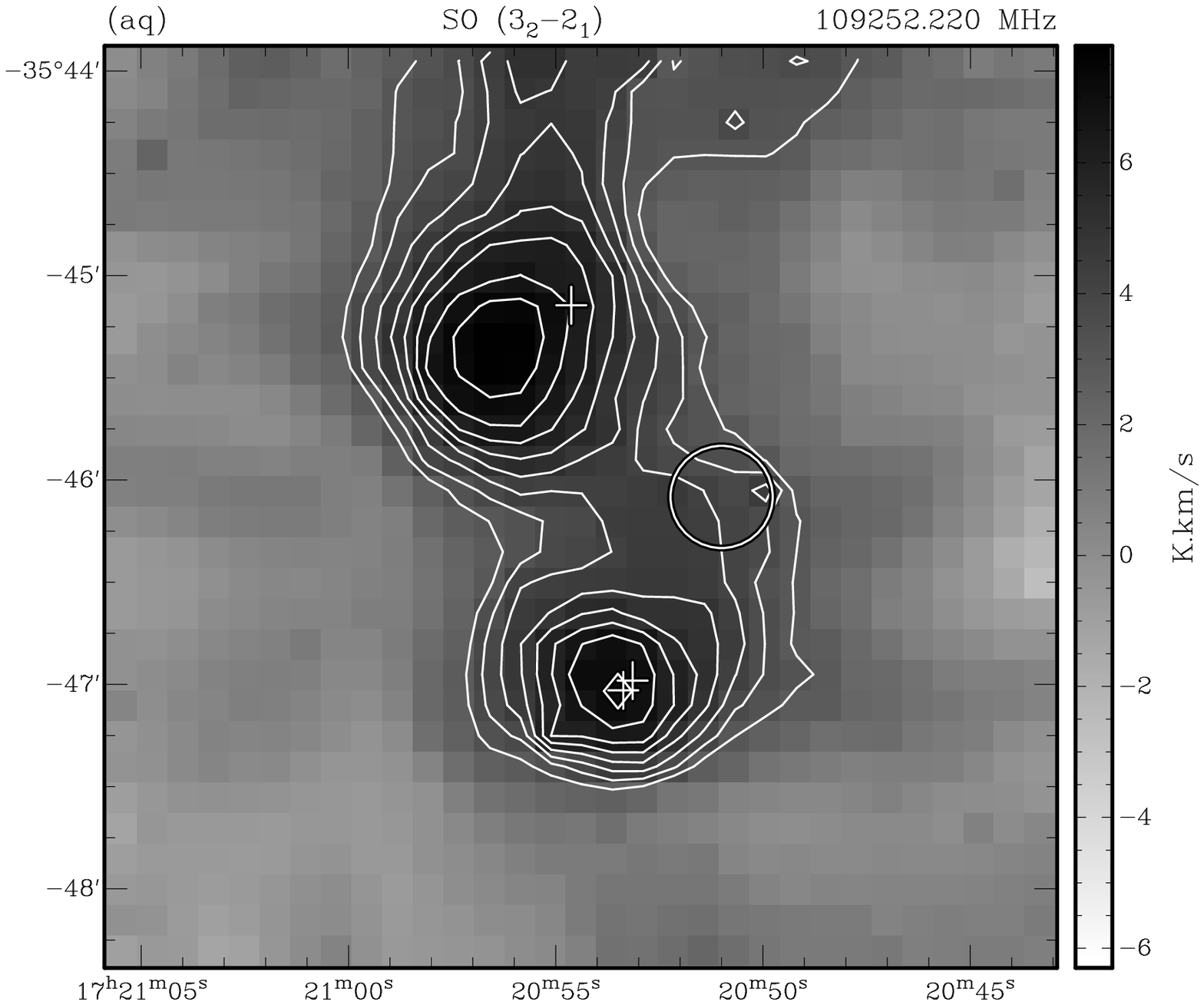}&
\includegraphics[width=0.45\textwidth]{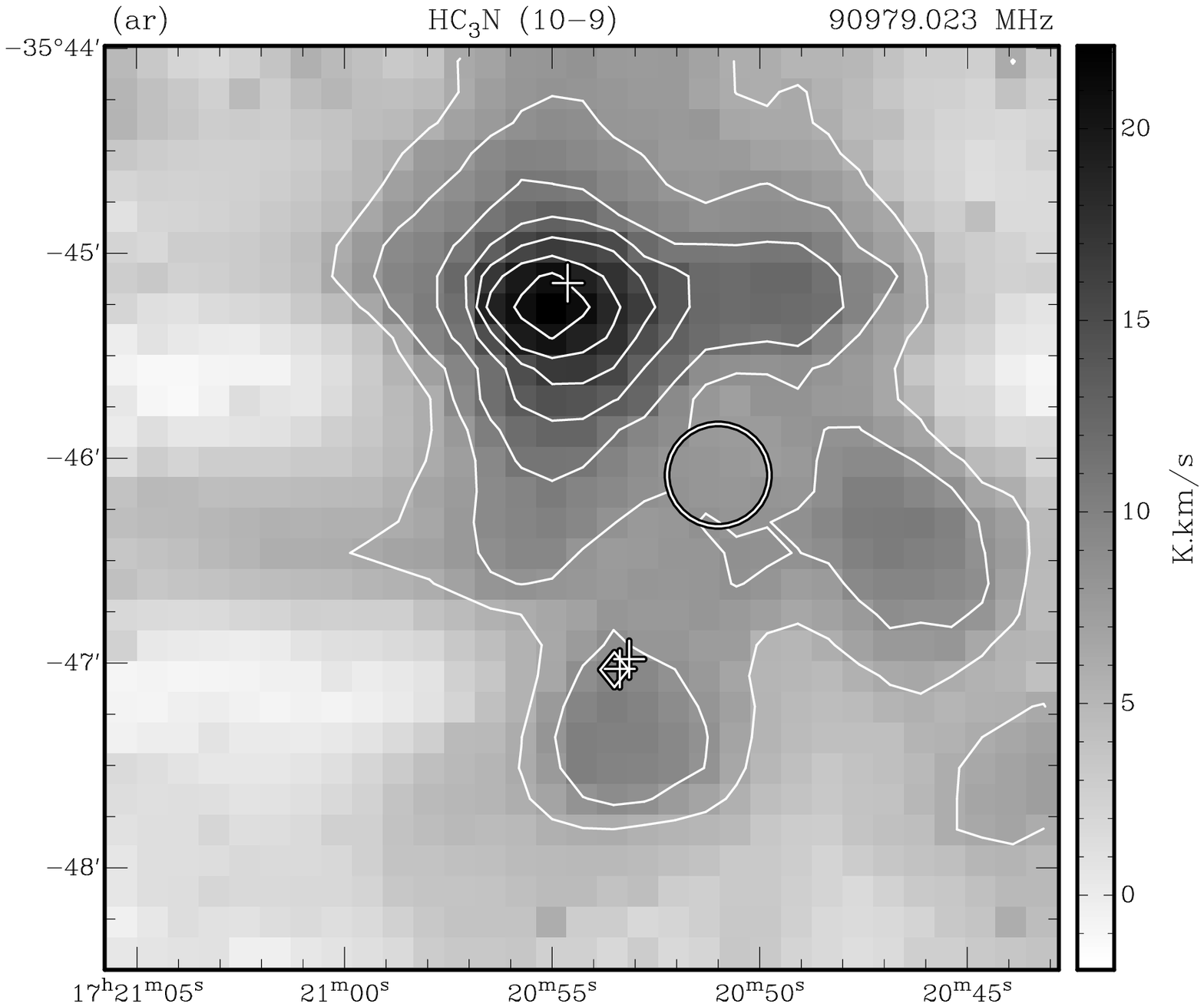}\\
\includegraphics[width=0.45\textwidth]{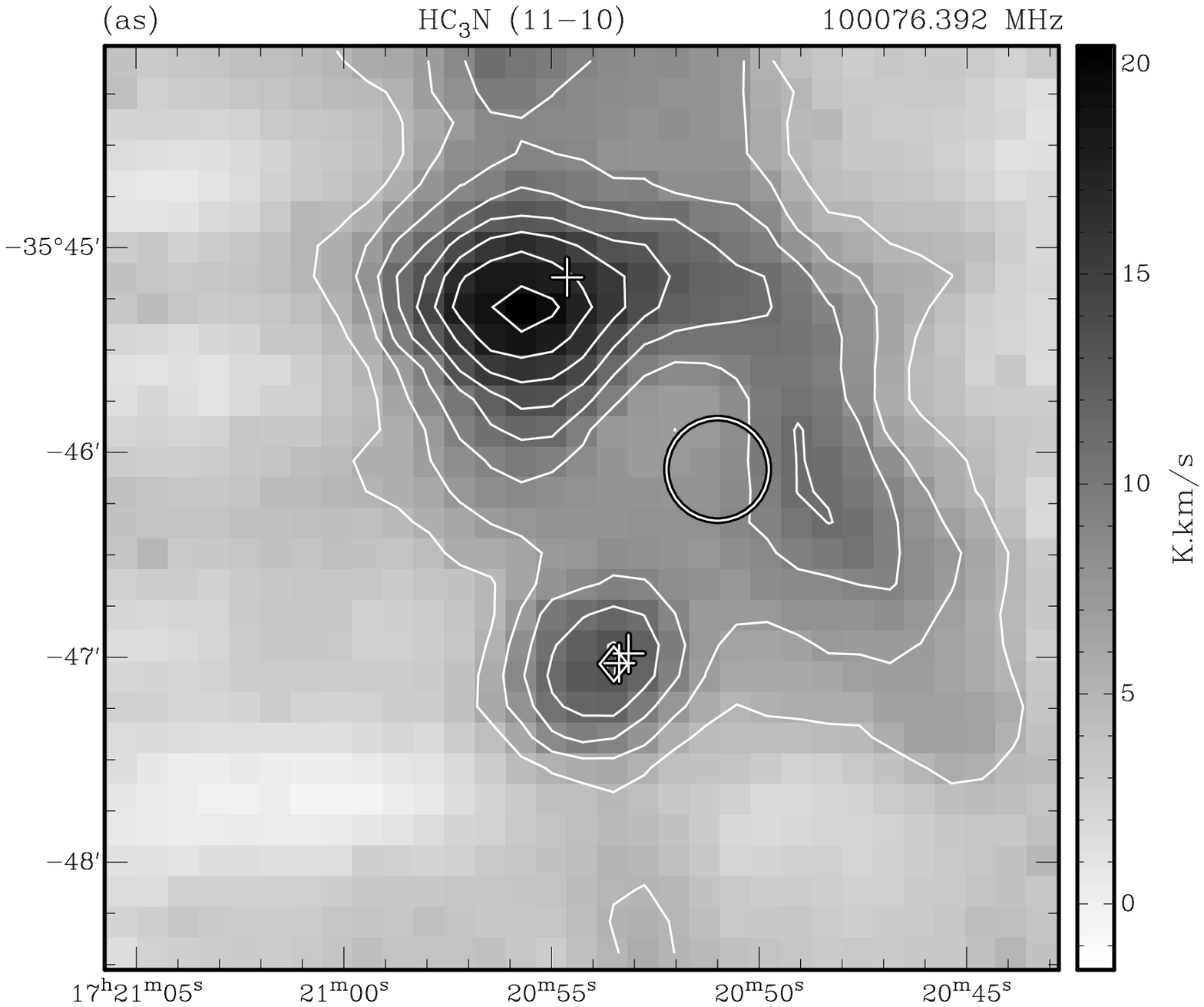}&
\includegraphics[width=0.45\textwidth]{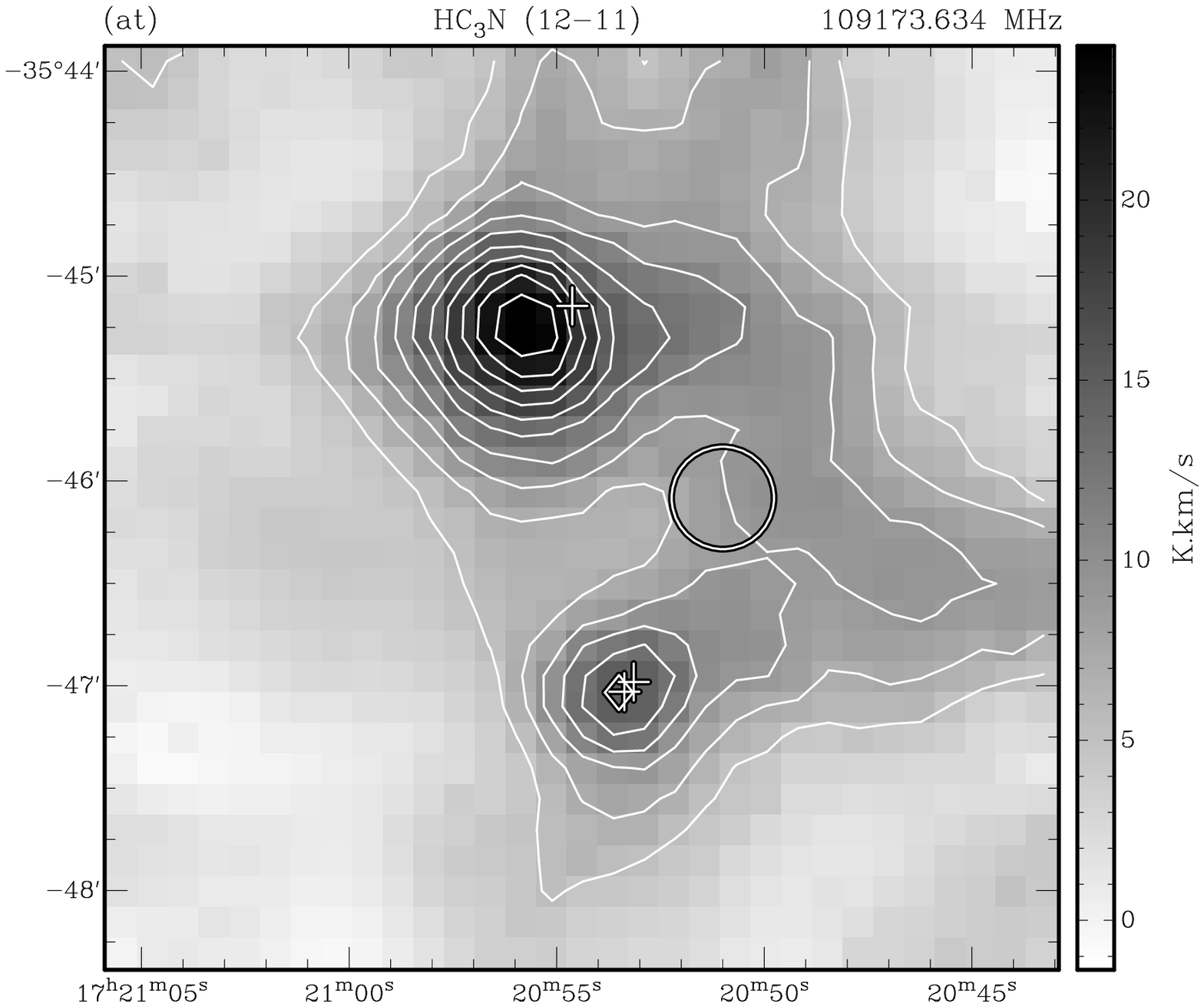}\\
\includegraphics[width=0.45\textwidth]{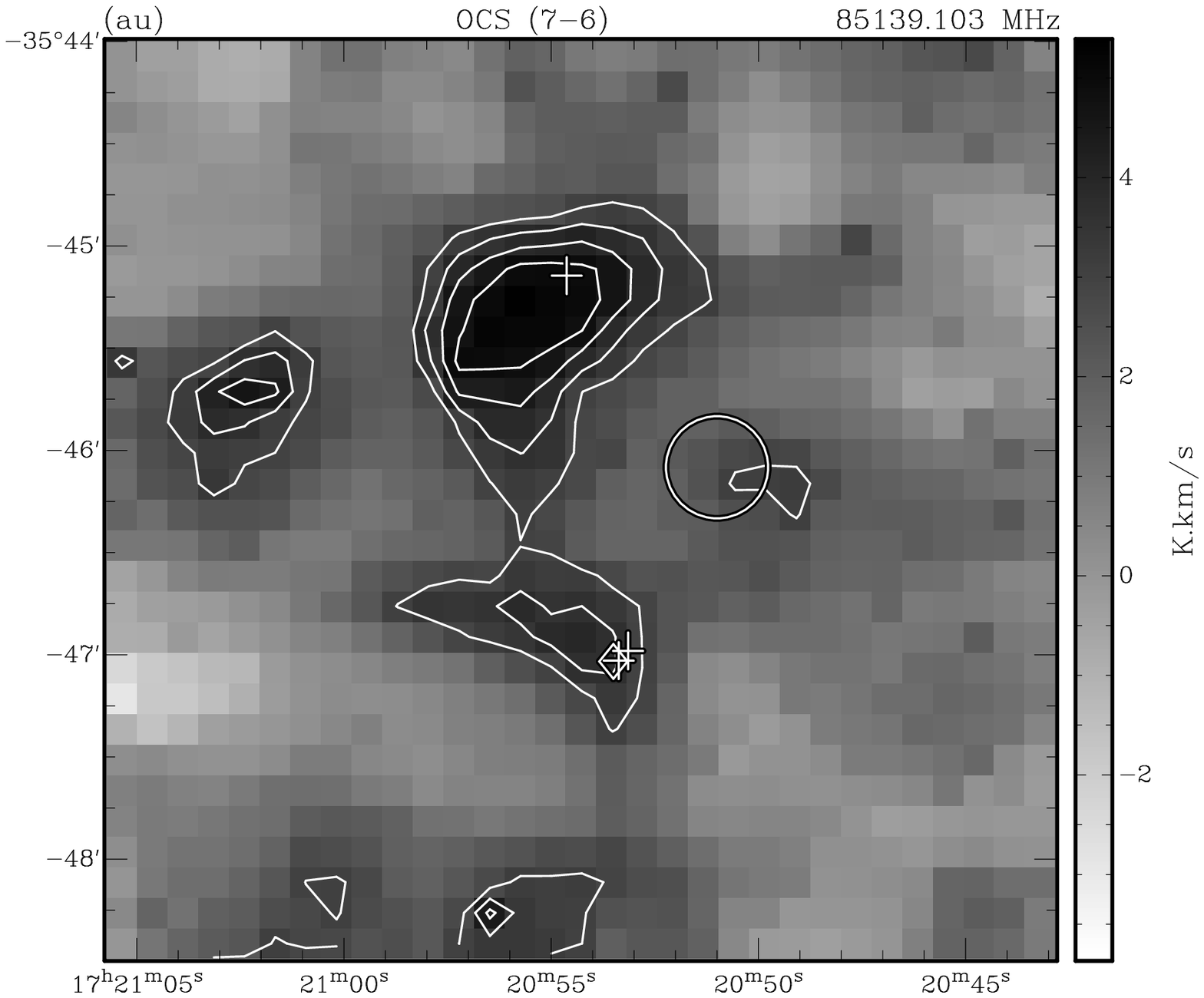}&
\includegraphics[width=0.45\textwidth]{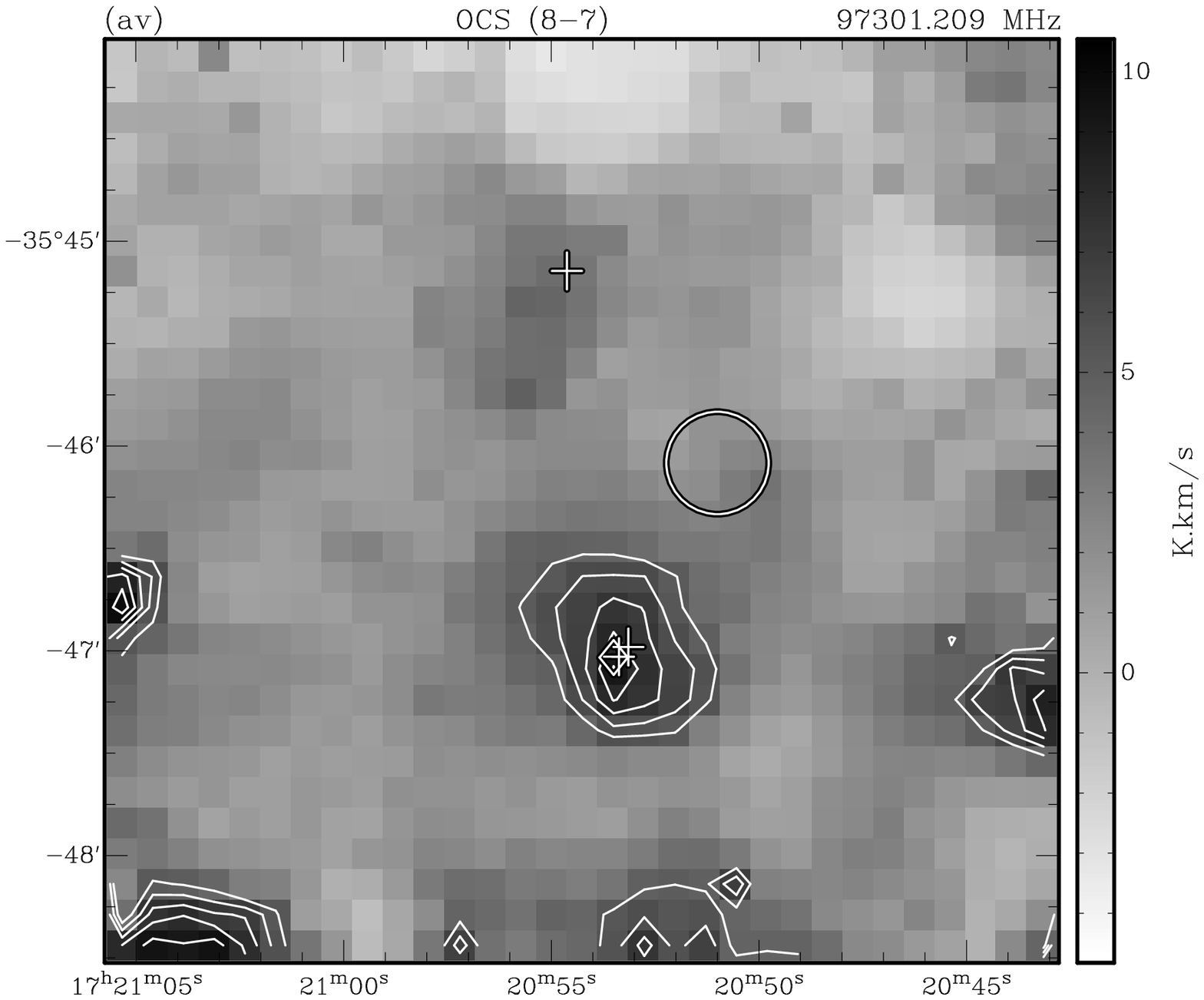}\\
\end{tabular}

\contcaption{{\bf (aq)} SO (3$_2$--2$_1$) --- sulfur monoxide. Contours start
at 5$\sigma$ and increase in 1$\sigma$ steps, where 1$\sigma$
is 0.6\,K\,km\,s$^{-1}$.
{\bf (ar)} HC$_3$N (10--9) --- cyanoacetylene. Contours start
at 5$\sigma$ and increase in 2$\sigma$ steps, where 1$\sigma$
is 1.2\,K\,km\,s$^{-1}$.
{\bf (as)} HC$_3$N (11--10) --- cyanoacetylene. Contours start
at 5$\sigma$ and increase in 2$\sigma$ steps, where 1$\sigma$
is 1.1\,K\,km\,s$^{-1}$.
{\bf (at)} HC$_3$N (12--11) --- cyanoacetylene. Contours start
at 5$\sigma$ and increase in 2$\sigma$ steps, where 1$\sigma$
is 1.1\,K\,km\,s$^{-1}$.
{\bf (au)} OCS (7--6) --- carbonyl sulfide. Contours start
at 5$\sigma$ and increase in 1$\sigma$ steps, where 1$\sigma$
is 0.6\,K\,km\,s$^{-1}$.
{\bf (av)} OCS (8--7) --- carbonyl sulfide. Contours start
at 5$\sigma$ and increase in 1$\sigma$ steps, where 1$\sigma$
is 1.1\,K\,km\,s$^{-1}$. Strong ``emission'' is seen around edges of
the bottom half of the image,
which is most likely due to poor weather, and is not real emission.
}
\end{figure*}

\begin{figure*}
\begin{tabular}{cc}
\includegraphics[width=0.45\textwidth]{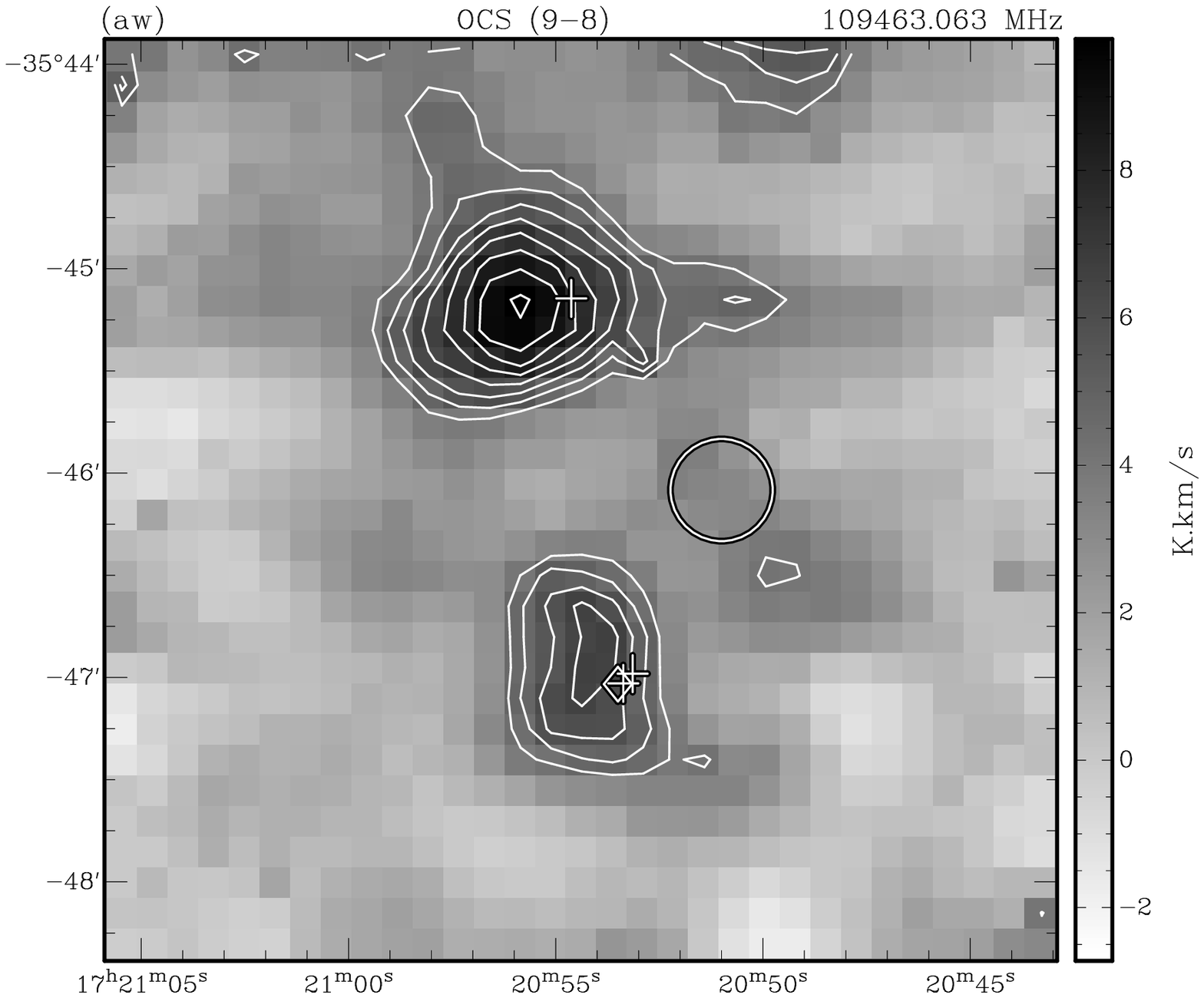}&
\includegraphics[width=0.45\textwidth]{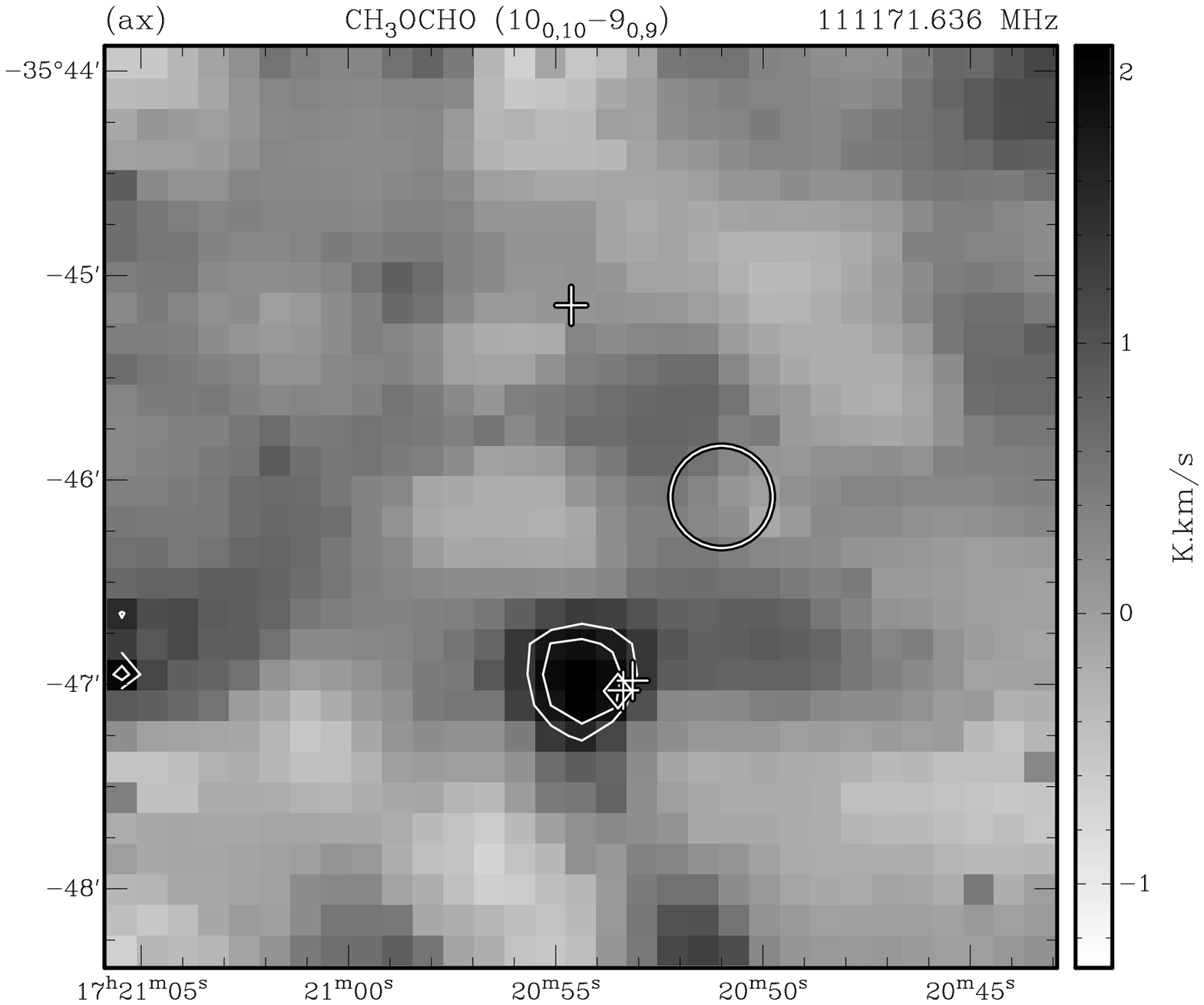}\\
\end{tabular}

\contcaption{{\bf (aw)} OCS (9--8) --- carbonyl sulfide. Contours start
at 5$\sigma$ and increase in 1$\sigma$ steps, where 1$\sigma$
is 0.7\,K\,km\,s$^{-1}$. ``Emission'' is seen around the northern edge of 
the image, which is most likely due to poor weather, and is not real emission.
{\bf (ax)} CH$_3$OCHO-A (10$_{0,10}$--9$_{0,9}$) --- methyl formate. Contours start
at 5$\sigma$ and increase in 1$\sigma$ steps, where 1$\sigma$
is 0.3\,K\,km\,s$^{-1}$.
}
\end{figure*}

\section{Discussion}


\subsection{Morphological analysis}
\label{morph}
Since the spectral resolution was relatively coarse, we
refrain from an analysis of spectrally resolved data-cubes and only show
and interpret integrated images of the different molecular species and
isotopologues. Furthermore, the angular resolution of the Mopra
telescope in the 3\,mm band is $\sim 36''$ (at 90\,GHz) corresponding
to a linear resolution of $\sim$65000\,AU or $\sim$0.3\,pc. This
spatial resolution is not sufficient to analyze spatial differences
within one or the other massive star-forming region. However, it
allows us to investigate the molecular global properties, similarities
and differences between the different massive star-forming regions
present in our field of view. In the following, we will qualitatively
describe the global spatial characteristics of the different molecular
species and its isotopologues.

{\bf H41$\alpha$:} The only covered hydrogen recombination line that
we could image and is
found as expected toward the UCH{\sc ii} region associated with
NGC\,6334\,I.

{\bf C$_2$H:} C$_2$H shows a peculiar spatial morphology since its main
emission peak is associated with NGC\,6334\,I(N) and appears as a weak
hole at the position of NGC\,6334\,E. Furthermore, extended emission
closely follows the dust continuum ridge shown in Figure \ref{continuum}.
This molecule shows no clear peak toward the southern source NGC\,6334I. The dearth
of emission coincident with both NGC\,6334\,E and I suggests that this molecule
is destroyed close to \hii region environments.

Unsaturated hydrocarbons like C$_2$H are known to be strong at early
evolutionary stages (eg. \citealt{millar85,turner99}).
During the warm-up of the cores it reacts quickly with oxygen to
form CO through ion-molecule chemistry (eg. \citealt{turner99,herbst86}).
These models are consistent with our finding of the
C$_2$H peak close to NGC\,6334\,I(N) and much less emission toward the hot core
NGC\,6334\,I or the \hii~region NGC\,6334\,E. Furthermore, \citet{beuther08}
confirm that during ongoing evolution C$_2$H abundances decrease toward star
formation sites. Thus, C$_2$H is a good tracer of gas
that is not directly associated with active star formation.

{\bf CN:} The radical CN appears to be smoothly distributed over a region covering
NGC\,6334\,I, E and I(N). One transition is slightly stronger at NGC\,6334\,I, and the other
is slightly stronger at NGC\,6334\,I(N). However, these are weak peaks, over the extended
emission and so it is not possible to tell if the difference in relative intensities is
real or due to noise variations. CN, like C$_2$H, appears to be a good tracer of gas not
directly associated with active star formation.

{\bf $^{12}$CO, $^{13}$CO, C$^{18}$O and C$^{17}$O:} Carbon monoxide
and its isotopologues are easily detected in our field of view.
However, the spatial distribution is surprising because
it does not trace clearly any of the two main massive star-forming
regions, NGC\,6334\,I or I(N). The main emission peak is located
approximately $50''$ north-west of NGC\,6334\,I, closer to the
UCH{\sc ii} region NGC\,6334\,E (but not coincident). We note that this
peak does not appear pronounced in any other spectral line imaged in this work,
nor does it feature prominently in the dust continuum map. Since only CO (1--0)
transitions are prominent at this position (17 20 48.45, -35 46 30) and
these transitions have a low effective critical density ($\sim10^2$\,cm$^{-3}$),
we believe this line of sight contains a high column density of low density gas
without any significant high density centre that might contain star formation.
While $^{12}$CO, $^{13}$CO and C$^{18}$O show weaker secondary peaks
toward NGC\,6334\,I and I(N), these are barely detectable in the rarest
isotopologue C$^{17}$O. Based on the mm continuum maps (e.g., \citealt{sandell00},
see also Figure \ref{continuum}) it is obvious that the highest gas column
densities are toward NGC\,6334\,I and I(N), therefore, non-detection of
these peaks in rare CO isotopologues cannot be explained by column
density effects.
The temperatures toward NGC\,6334\,I and I(N) exceed 100\,K
\citep{beuther05} whereas the excitation temperatures
of the upper energy levels of the $J=1-0$ carbon monoxide transitions
are only of the order 5\,K. With a typical Boltzmann distribution, one
would expect, for example, the $J=6-5$ transitions to peak closer to the
warm regions NGC\,6334\,I and I(N).

An additional interesting feature in the CO maps is the
more extended north-south ridge. It would be useful to have larger
maps to investigate to what extend this ridge is continuing, but a
comparison with the mm continuum map indicates that this CO ridge has
no direct counterpart in the dense gas and dust component.

{\bf N$_2$H$^+$:} The well-known tracer of star formation at
young evolutionary stages N$_2$H$^+$ exhibits its
strongest peak toward the northern region NGC\,6334\,I(N). However, the
molecule has additional -- although progressively weaker -- emission
peaks associated with the more evolved region NGC\,6334\,I. N$_2$H$^+$
emission is at a local minimum at the position of NGC\,6334\,E, indicating that
this ion tends to avoid the \hii region environment. In addition to this,
N$_2$H$^+$ clearly shows extended emission, which is positionally coincident
with the main filament shown in dust continuum emission (Figure \ref{continuum}).
Of all the spectral line species covered in this work, N$_2$H$^+$ most closely
follows the dust continuum emission, showing it to be a good tracer
of cold and dense material, including in star forming regions.

{\bf HCO$^+$ and H$^{13}$CO$^+$:} The HCO$^+$ map clearly shows
its main emission peaks associated with the dense cores NGC\,6334\,I and
I(N). It shows an additional secondary peak associated with the main
carbon monoxide peaks further to the east, as well as hints of the
larger scale north-south ridge discussed for CO and its isotopologues above.
For H$^{13}$CO$^+$ we cannot identify the north-south ridge anymore
but the other features resemble those of the main isotopologue. It is
interesting to note that the main intensity peaks for both
isotopologues are associated with NGC\,6334\,I(N) and not the more
prominent hot-core type region NGC\,6334\,I. This is reminiscent to the
HC$_3$N map published by \citet{sollins04}.

{\bf CH$_3$OH, thermal and class {\sc i}/{\sc ii} masers:} The spectral
setup covers many CH$_3$OH and class {\sc i} and class {\sc ii} maser
lines as shown in Table \ref{tab1} (see also \citealt{mueller04}). It is impossible to
definitively claim detection of masers given the current observations with
a beam larger than $30''$, and a spectral resolution of a few km/s. However,
we can comment on the likelihood of any transition showing masing activity, based
on the relative intensities of lines at the positions of I and I(N). It is likely,
then, that NGC\,6334\,I(N) shows maser emission in the Class I transitions at 84.521\,GHz
and 95.169GHz, and in the Class II transition at 108.894\,GHz. \citet{valtts00}
confirms the masing nature of the Class I transition at 95.169\,GHz.
NGC\,6334\,I, on the other hand,
does not show strong emission in any of the masing transitions, except perhaps the
Class II transition at 107.013\,GHz. \citet{cragg01} and \citet{valtts99}
have both observed this Class II transition towards NGC\,6334\,I and confirm it is a maser.

The thermal methanol transitions appear to divide into two types, based on the strength
of their emission. Methanol transitions showing strong emission at 95.914, the quadruplet
at 96.7 and at 97.583\,GHz all exhibit the bulk of their emission from NGC\,6334\,I(N).
The methanol transitions at 100.639, 107.160 and 111.290\,GHz all show weak emission
that is only detected towards NGC\,6334\,I.

{\bf CH$_3$CN:} Methyl cyanide is detected in the rotational transitions
$J=5_K-4_K$ and $6_K-5_K$.
Both integrated intensity maps show two comparably strong peaks toward
NGC\,6334\,I and I(N). Methyl cyanide is potentially useful for temperature
estimates, by comparing intensity ratios of components in the $K$-ladders
\citep{thorwirth03}.
However, the low spectral resolution of these data do not allow us
to distinguish individual $K$-ladder elements. Of all the species mapped in
this work, CH$_3$CN appears to be the one that is most concentrated towards
the two regions of active star formation: NGC\,6334\,I and NGC\,6334\,I(N).

{\bf $^{12}$CS, $^{13}$CS, C$^{34}$S  and C$^{33}$S:}
Carbon sulfide shows an interesting spatial variation going from the
main isotopologue CS step by step through the rarer isotopologues
$^{13}$CS and C$^{34}$S to C$^{33}$S. While the former main
isotopologue peaks toward the southern hot core region NGC\,6334\,I, the
northern peak NGC\,6334\,I(N) is of comparable brightness for
$^{13}$CS and C$^{34}$S. This is likely due to the main CS isotopologue
being optically thick, which is then tracing a warmer environment around
NGC\,6334\,I. The optically thin tracers $^{13}$CS and C$^{34}$S show
approximately equal brightnesses in I and I(N), suggesting the
amount of CS is approximately equal in each. However, the rare
isotopologue C$^{33}$S is brighter in I. We would expect C$^{33}$S
to look similar to $^{13}$CS and C$^{34}$S. This suprising result
needs to be investigated further.

{\bf SO, OCS, H$_2$CS and NS:} Several sulfur-bearing species have been
observed, and all of them show two peaks associated with NGC\,6334\,I
and I(N). The general trend appears to be that the emission is stronger towards
NGC\,6334\,I(N), rather than I. The one exception to this rule is OCS (8-7), where
very little emission is detected towards NGC\,6334\,I(N), even though NGC\,6334\,I shows a
bright peak. This is an unexpected result, especially in light of the fact that
the two other OCS transitions clearly show NGC\,6334\,I(N) as a strong source of OCS
emission.

{\bf HC$_3$N, HNCO, HNC, HN$^{13}$C, HCN, H$^{13}$CN:} There are
several hydrogen-carbon-nitrogen chain molecules observed, and all of
them show two strong peaks associated with NGC\,6334\,I and I(N). Except
for HCN where the two peaks are of approximately the same strength, in
all other lines the region I(N) is clearly the dominant one. For
HC$_3$N this has already previously been reported by
\citet{sollins04}. HC$_3$N is unusual in that it shows extended emission
which appears to follow the main filament seen in dust continuum emission
(Figure \ref{continuum}), to the west of NGC\,6334\,E. Furthermore, the HC$_3$N emission
appears to surround NGC\,6334\,E, but very little emission is seen inside the
ring denoting the extent of this \hii region. Since its morphology closely resembles
that of the dust continuum emission, we find that HC$_3$N is a good tracer
of quiescent, dense gas, as is the case for N$_2$H$^+$.

{\bf SiO:} As previously reported by \citet{megeath99}, SiO emission
is found strongly toward NGC\,6334\,I(N) and is weaker
elsewhere in the field of view. In NGC\,6334\,I(N)
the SiO emission is clearly associated with its molecular outflow(s)
(\citealt{megeath99}, Hunter et al. in prep.). Previous observations
of the G333 giant molecular cloud \citet{lo07} have found SiO emission
from a region that shows similar properties to NGC\,6334\,I(N) with
an outflow associated with a very early stage of high mass star formation.

{\bf CH$_3$OCHO:} The dense core tracing molecule methyl formate is
detected only toward the southern hot core region NGC\,6334\,I.

\begin{table*}
 \centering
 \begin{minipage}{140mm}
  \caption{Molecular lines mapped towards the NGC\,6334\,I/I(N) region in the present study. Transition
frequencies (MHz) were taken from the CDMS \citep{mueller01,mueller05} in most cases and the JPL
catalogue \citep{pickett98}. Also given are the integrated intensities for each transition, as measured
at NGC\,6334\,I and I(N) by integrating the emission over one beam, as was done for the spectra presented
in Figure \ref{spectra}. 3$\sigma$-upper limits are given for those transitions not detected.}
\label{tab1}
  \begin{tabular}{lrrccc}
  \hline
   Molecule & Transition & Frequency & Methanol & \multicolumn{2}{c}{Integrated Intensity} \\
            &            &   (MHz)   &   Maser  & \multicolumn{2}{c}{(K\,km/s)}\\
            &            &           &   Class  & NGC\,6334\,I & NGC\,6334\,I(N) \\
 \hline
H                & 41$\alpha$                        &  92034.442 &    &  26 &  $<10$\footnote{$3\sigma$-upper limit} \\
C$_2$H           & $N=1-0$, $J=3/2-1/2$, $F=2-1$     &  87316.898 &    &  66 &  60 \\
C$_2$H           & $N=1-0$, $J=3/2-1/2$, $F=1-0$     &  87328.585 &    &  38 &  32 \\
C$_2$H           & $N=1-0$, $J=1/2-1/2$, $F=1-1$     &  87401.989 &    &  35\footnote{blend with C$_2$H $N=1-0$, $J=1/2-1/2$, $F=0-1$ at 87407.165\,MHz} &  23$^b$ \\
CN               & $N=1-0$, $J=1/2-1/2$, $F=1/2-1/2$ & 113123.370 &    & 8.9 &  19 \\
CN               & $N=1-0$, $J=1/2-1/2$, $F=1/2-3/2$ & 113144.157 &    &  38 &  34 \\
CN               & $N=1-0$, $J=1/2-1/2$, $F=3/2-1/2$ & 113170.492 &    &  61 &  64 \\
CN               & $N=1-0$, $J=1/2-1/2$, $F=3/2-3/2$ & 113191.279 &    &  56 &  46 \\
H$^{13}$CN       & $J=1-0$                           &  86339.922 &    &  61 &  57 \\
HCN              & $J=1-0$                           &  88631.602 &    & 350 & 300 \\
HN$^{13}$C       & $J=1-0$                           &  87090.850 &    &  19 &  20 \\
HNC              & $J=1-0$                           &  90663.568 &    & 100 & 110 \\
C$^{18}$O        & $J=1-0$                           & 109782.173 &    &  69 &  69 \\
$^{13}$CO        & $J=1-0$                           & 110201.354 &    & 360 & 350 \\ 
C$^{17}$O        & $J=1-0$                           & 112359.284 &    &  19 &  21 \\
$^{12}$CO        & $J=1-0$                           & 115271.202 &    & 930 & 680 \\
N$_2$H$^+$       & $J=1-0$                           &  93173.392 &    & 210 & 440 \\
H$^{13}$CO$^+$   & $J=1-0$                           &  86754.288 &    &  16 &  24 \\
HCO$^+$          & $J=1-0$                           &  89188.525 &    & 170 & 150 \\
CH$_3$OH--$E$    & $J_{K_a,K_c}=5_{-1,5}-4_{0,4}$    &  84521.169 &  I &  80 & 180 \\ 
CH$_3$OH--$A^+$  & $J_{K_a,K_c}=8_{0,8}-7_{1,7}$     &  95169.463 &  I &  41 &  88 \\ 
CH$_3$OH--$A^+$  & $J_{K_a,K_c}=2_{1,2}-1_{1,1}$     &  95914.309 &    &  16 &  32 \\ 
CH$_3$OH--$A^+$  & $J_{K_a,K_c}=2_{0,2}-1_{0,1}$     &  96741.375 &    & 150 & 370 \\ 
CH$_3$OH--$A^-$  & $J_{K_a,K_c}=2_{1,1}-1_{1,0}$     &  97582.804 &    & 9.5 &  24 \\ 
CH$_3$OH--$E$    & $J_{K_a,K_c}=13_{2,11}-12_{3,9}$  & 100638.900 &    & 4.9 &  $<0.9^a$ \\ 
CH$_3$OH--$E$    & $J_{K_a,K_c}=11_{-1,11}-10_{-2,9}$& 104300.414 &  I & 5.5 &  $<1.8^a$ \\ 
CH$_3$OH--$A^+$  & $J_{K_a,K_c}=3_{1,3}-4_{0,4}$     & 107013.803 & II &  15 & 2.7 \\ 
CH$_3$OH--$E$    & $J_{K_a,K_c}=15_{-2,14}-15_{1,14}$& 107159.820 &    & 8.2 &  $<1.8^a$ \\ 
CH$_3$OH--$E$    & $J_{K_a,K_c}=0_{0,0}-1_{-1,1}$    & 108893.963 & II &  18 &  28 \\ 
CH$_3$OH--$A^+$  & $J_{K_a,K_c}=7_{2,5}-8_{1,8}$     & 111289.550 &    & 1.9 &  $<1.2^a$ \\ 
CH$_3$CN         & $J_K=5_0-4_0$                     &  91987.091 &    &  41\footnote{Integrated emission for all $J=5-4$ transitions} &  35$^c$ \\
CH$_3$CN         & $J_K=5_1-4_1$                     &  91985.318 &    &     &     \\ 
CH$_3$CN         & $J_K=5_2-4_2$                     &  91979.998 &    &     &     \\
CH$_3$CN         & $J_K=5_3-4_3$                     &  91971.134 &    &     &     \\
CH$_3$CN         & $J_K=6_0-5_0$                     & 110383.504 &    &  30\footnote{Integrated emission for all $J=6-5$ transitions} &  32$^d$ \\
CH$_3$CN         & $J_K=6_1-5_1$                     & 110381.376 &    &     &     \\ 
CH$_3$CN         & $J_K=6_2-5_2$                     & 110374.993 &    &     &     \\
CH$_3$CN         & $J_K=6_3-5_3$                     & 110364.358 &    &     &     \\
HNCO             & $J_{K_a,K_c}=4_{0,4}-3_{0,3}$     &  87925.237 &    &  $<1.8^a$ &  25 \\
HNCO             & $J_{K_a,K_c}=5_{0,5}-4_{0,4}$     & 109905.749 &    &  12 &  23 \\
$^{13}$CS        & $J=2-1$                           &  92494.308 &    &  36 &  32 \\
C$^{34}$S        & $J=2-1$                           &  96412.950 &    &  73 &  61 \\
C$^{33}$S        & $J=2-1$                           &  97172.064 &    &  21 &  14 \\
CS               & $J=2-1$                           &  97980.953 &    & 290 & 210 \\
SiO              & $J=2-1$                           &  86846.960 &    &  13 &  41 \\
NS               & $^2\Pi_{1/2}$ N=2-1, J=5/2-3/2    & 115153.835 &    & 8.5 &  17 \\ 
                 & F=7/2-5/2, $c$-state              &            &    &     &     \\ 
\hline
\end{tabular}
\end{minipage}
\end{table*}

\begin{table*}
 \centering
 \begin{minipage}{140mm}
  \contcaption{}
  \begin{tabular}{lrrccc}
  \hline
   Molecule & Transition & Frequency & Methanol & \multicolumn{2}{c}{Integrated Intensity} \\
            &            &   (MHz)   &   Maser  & \multicolumn{2}{c}{(K\,km/s)}\\
            &            &           &   Class  & NGC\,6334\,I & NGC\,6334\,I(N) \\
 \hline
H$_2$CS          & $J_{K_a,K_c}=3_{1,3}-2_{1,2}$     & 101477.810 &    & $>$24& $>$20\\
H$_2$CS          & $J_{K_a,K_c}=3_{1,2}-2_{1,1}$     & 104617.040 &    &  27 &  34 \\
SO               & $N_J=2_2-1_1$                     &  86093.950 &    &  22 &  19 \\
SO               & $N_J=2_3-1_2$                     &  99299.870 &    &  53 &  69 \\
SO               & $N_J=3_2-2_1$                     & 109252.220 &    &  18 &  21 \\ 
HC$_3$N          & $J=10-9$                          &  90979.023 &    &  38 &  48 \\
HC$_3$N          & $J=11-10$                         & 100076.392 &    &  27 &  38 \\
HC$_3$N          & $J=12-11$                         & 109173.634 &    &  33 &  59 \\
OCS              & $J=7-6$                           &  85139.103 &    &  12 &  17 \\
OCS              & $J=8-7$                           &  97301.209 &    &  19 &  19 \\
OCS              & $J=9-8$                           & 109463.063 &    &  19 &  25 \\
CH$_3$OCHO$^e$\footnotetext[5]{Blend of A and E types}  & $J_{K_a,K_c}=10_{0,10}-9_{0,9}$   & 111171.636 &    & 4.5 &  $<0.9^a$ \\

\hline
\end{tabular}
\end{minipage}
\end{table*}

\subsection{Comparison of NGC\,6334I and I(N)}
\label{comp}
In \S\ref{morph}, we described the morphology of emission for molecules, or groups of
molecules. Here we synthesise this information to present a comparison of the two
star forming sites NGC\,6334\,I and I(N). Based on previous work on this region (eg.
\citealt{walsh98}), it
is clear that NGC\,6334\,I appears to be more evolved than I(N): it contains
a prominent \uchii~region, as well as a strong infrared source, whereas only weak radio
continuum \citep{rodriguez07} and infrared emission \citep{walsh99}
has been found in the vicitiy of I(N). We also find evidence for an evolutionary
difference in our data. Both the CS/N$_2$H$^+$ and CS/HNCO ratios are much larger for
NGC\,6334\,I than for I(N). Previous work on the CS/N$_2$H$^+$ ratio \citep{zinchenko09}
and CS/HNCO ratio \citep{martin08} have indicated that CS is likely
to be enhanced in abundance towards photodissociation regions (PDRs), whilst both
N$_2$H$^+$ and HNCO are likely to be reduced in abundance towards PDRs. Thus, NGC\,6334\,I
shows PDR-like ratios, presumably due to the \uchii~region, whereas abundances of both
N$_2$H$^+$ and HNCO do not appear to be reduced, compared to CS in NGC\,6334\,I(N), where
a PDR is yet to form.

When comparing the non-masing lines of CH$_3$OH, we find that the emission is stronger
towards NGC\,6334\,I in the 100.638, 107.159 and 111.289\,GHz transitions. CH$_3$OH emission
is stronger towards NGC\,6334\,I(N) in the 95.915, 96.741 (quadruplet) and 97.582\,GHz
transitions. The stronger transitions in NGC\,6334\,I all have upper energy levels between 103
and 304\,K, whilst all the stronger transitions in NGC\,6334\,I(N) have upper energy levels
no greater than 22\,K. Thus, we can attribute the stronger emission in NGC\,6334\,I to hot
CH$_3$OH and the stronger emission in NGC\,6334\,I(N) to cold CH$_3$OH emission. Whilst
previous work has shown that both NGC\,6334\,I and I(N) have hot components, with
temperatures in excess of 400\,K \citep{beuther07}, it is clear that the large-scale
CH$_3$OH observed here has only had time to heat up in NGC\,6334\,I.

We detect CH$_3$OCHO only towards NGC\,6334\,I. This molecule is thought to be produced only after
a young star or protostar has heated up the surrounding material to above
100\,K \citep{garrod06}. This provides further evidence of the more evolved state
of NGC\,6334\,I over that of I(N).

\subsection{Line ratio maps}
The ratio of integrated intensities for different spectral lines can potentially
be used to identify physical and chemical characteristics of regions within the map.
For example, a simple comparison of ratios for different lines of the same species
might be used to map physical quantities such as density and temperature. However,
most species detected in multiple spectral lines (ie. CN, CH$_3$CN, HNCO, SO,
HC$_3$N and OCS) show only a narrow range of energy levels over the multiple
spectral lines. Thus only small changes in line ratios are
expected from different spectral lines of the same species.
In addition to this, small differences in the relative position of
maps made at different frequencies can lead to artifacts that can dominate line ratios.
In the present study, these complicating factors make such line ratio maps too
unreliable to interpret. The only exception to this is CH$_3$OH, where clear
differences in the intensities of lines are seen between NGC\,6334\,I and I(N),
as discussed in \S\ref{morph}.

\begin{figure*}
\begin{tabular}{cc}
\includegraphics[width=0.45\textwidth]{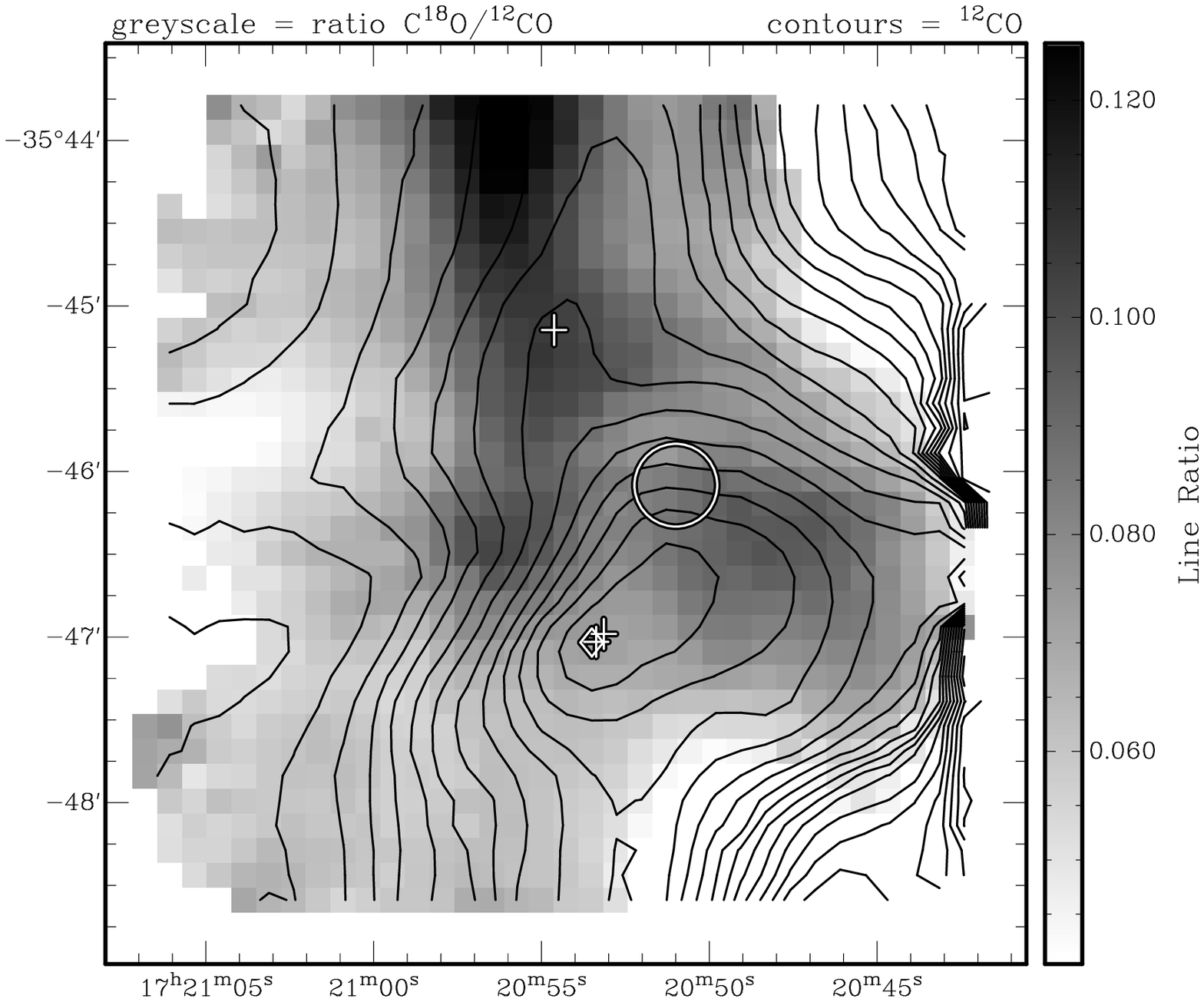}&
\includegraphics[width=0.45\textwidth]{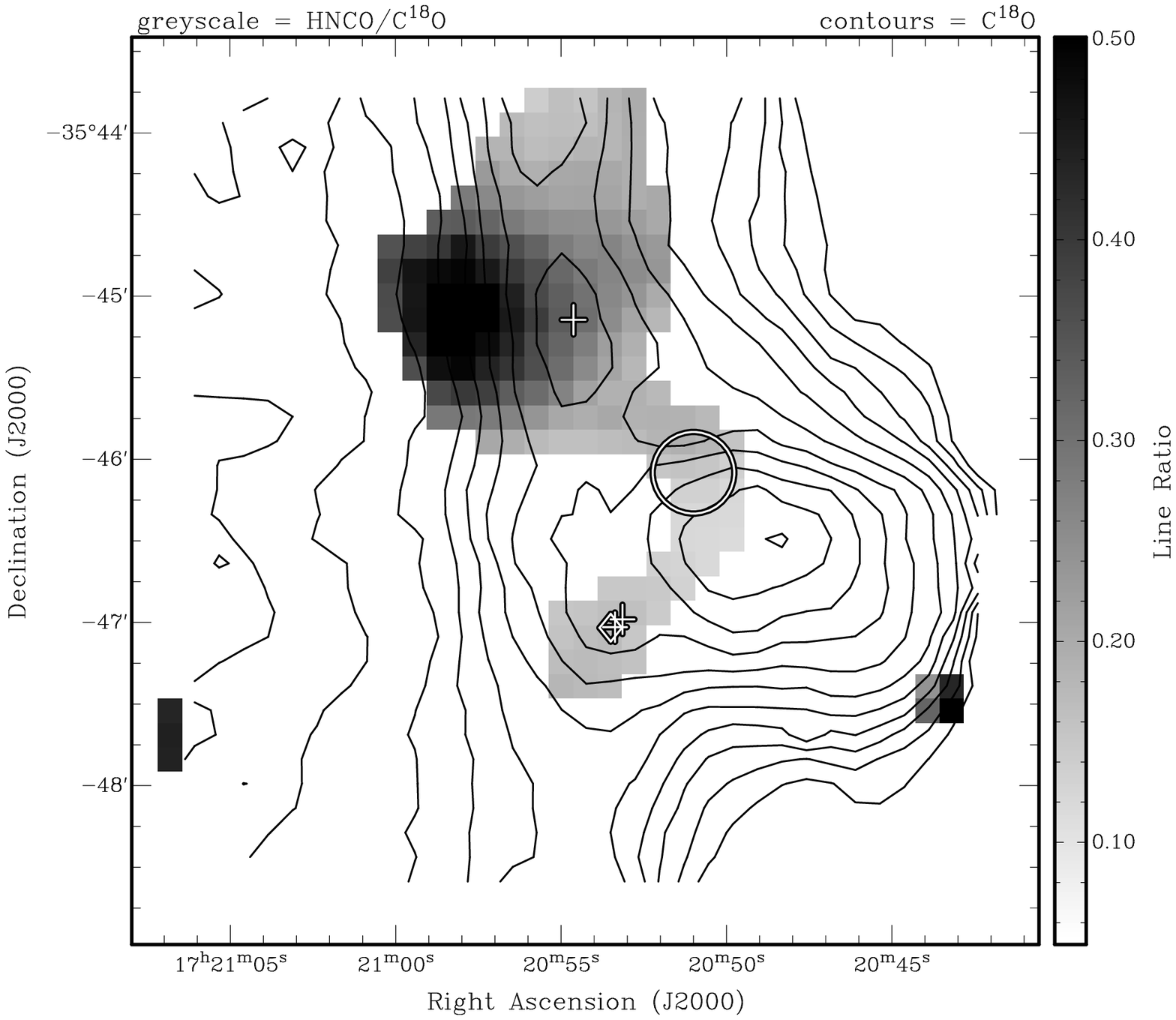}\\
\includegraphics[width=0.45\textwidth]{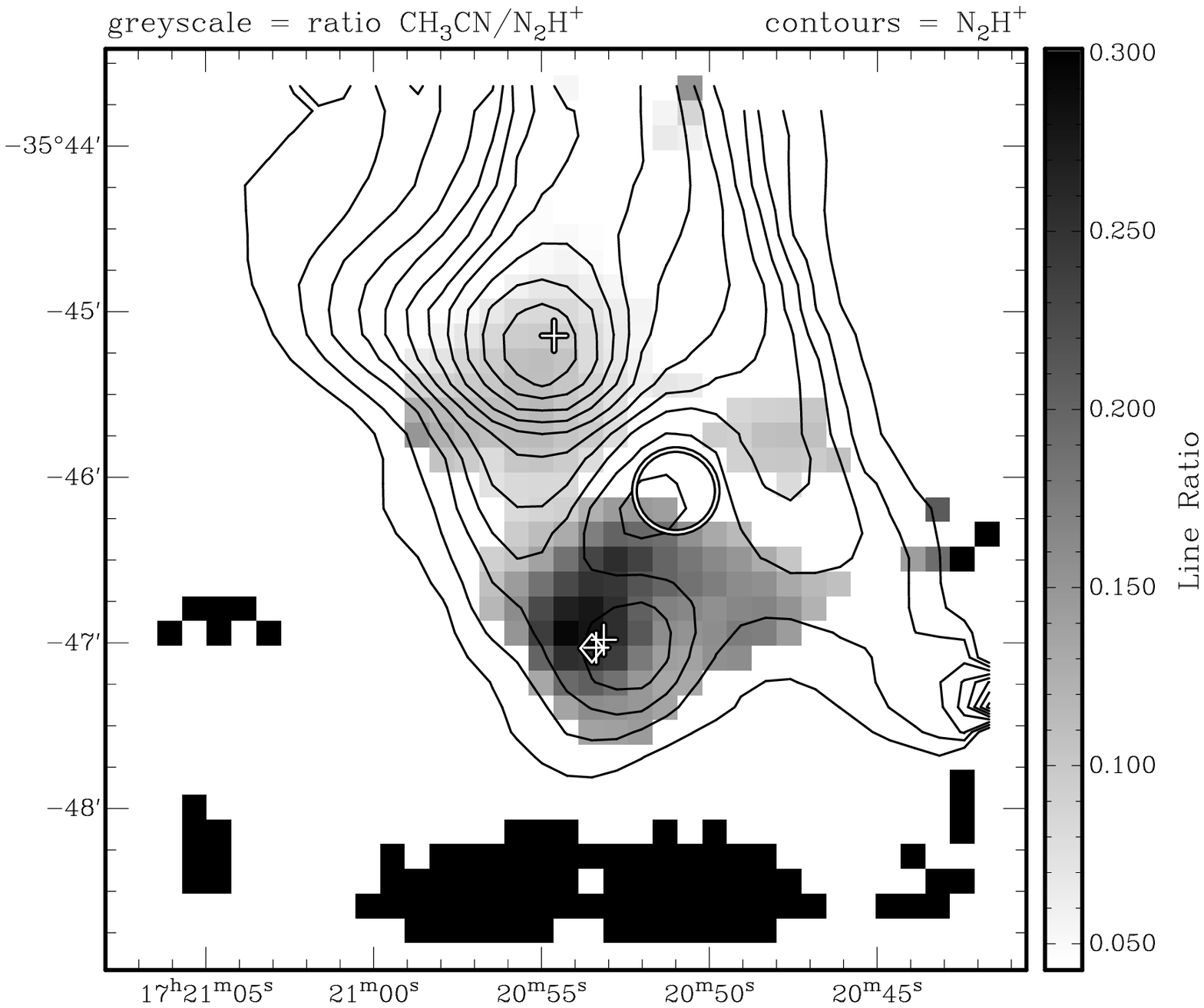}&
\includegraphics[width=0.45\textwidth]{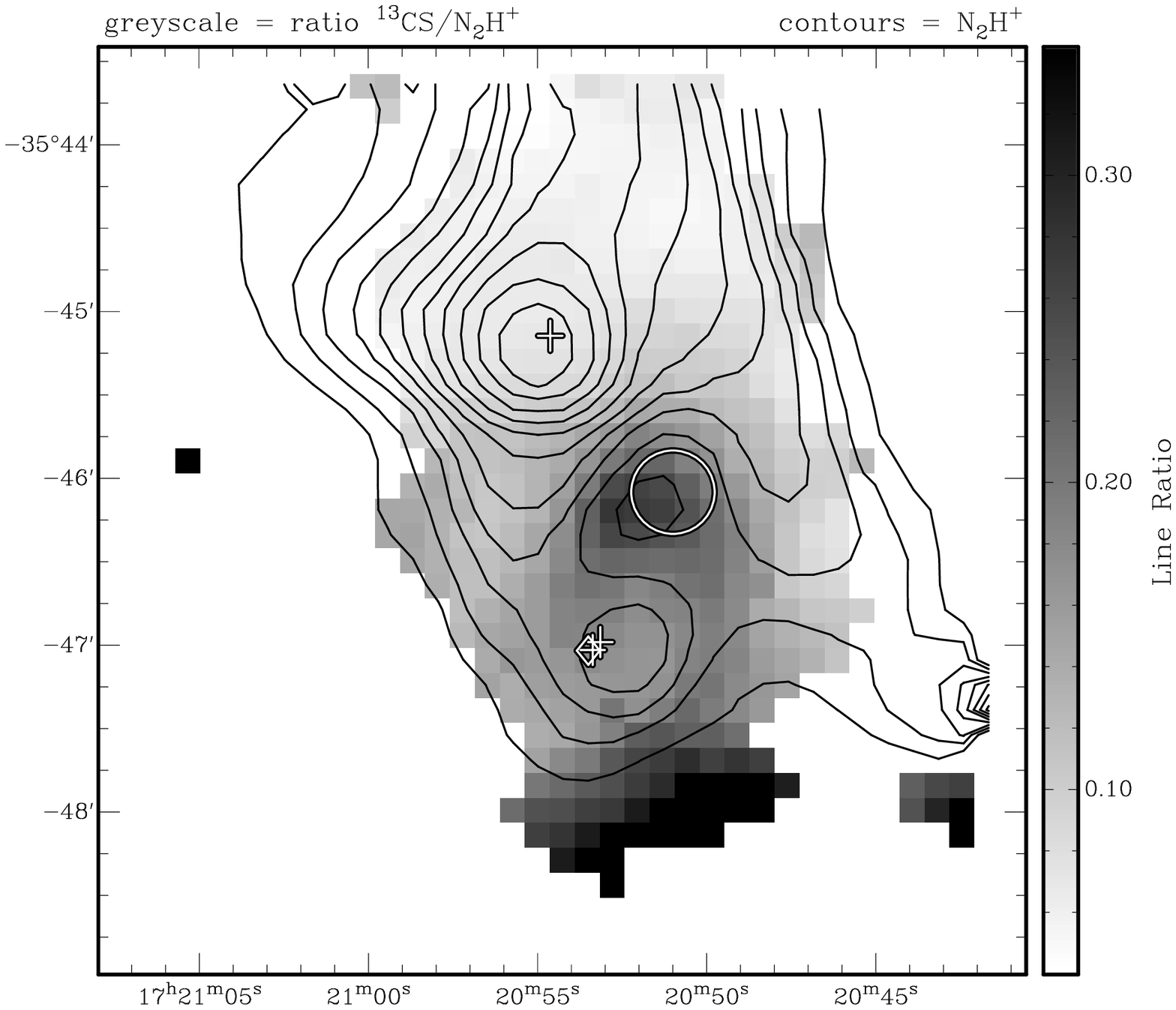}\\
\end{tabular}

\caption{Line ratio maps.
Upper-left shows the C$^{18}$O (1--0)/$^{12}$CO (1--0) ratio in greyscale and
$^{12}$CO (1--0) integrated intensity in contours.
Upper-right shows the HNCO (5$_{0,5}$--4$_{0,4}$)/C$^{18}$O (1--0)
ratio in greyscale and C$^{18}$O (1--0) integrated intensity in contours.
Lower-left shows the CH$_3$CN (5--4)/N$_2$H$^+$ (1--0) ratio in greyscale and
N$_2$H$^+$ integrated intensity in contours.
Lower-right shows the $^{13}$CS (2--1)/N$_2$H$^+$ (1--0) ratio in greyscale and
N$_2$H$^+$ (1--0) integrated intensity in contours. Dark areas show high line
ratios and light areas show low line ratios. The same contour levels as for
Figure \ref{images} have been used. Each ratio map is masked, based on the
weaker image (the numerator), to reduce contributions due to non-real emission.
However, non-real emission is seen at the bottom of the line ratio map for
CH$_3$CN/N$_2$H$^+$ map, below the N$_2$H$^+$ contours.}
\label{linerats}
\end{figure*}

It is possible to compare line ratios between species, that have been
observed simultaneously, as this eliminates any relative positional offsets, as
well as any varying effects of weather.
In Figure \ref{linerats}, we show images based on the ratio of line integrated
intensities for different species.

The upper-left image shows the C$^{18}$O/$^{12}$CO line ratio map.
The greatest ratio is found surrounding NGC\,6334\,I(N). A high ratio
between these two lines means that the $^{12}$CO line is optically thick.
We also see a secondary
peak in the ratio to the south--west of NGC\,6334\,E, which is close to the
peak $^{12}$CO peak integrated intensity (shown in contours). At this position,
it is likely that there is a high column density of gas. At NGC\,6334\,I, the
line ratio is relatively low. The $^{12}$CO gas is likely optically thick
here since it is one of the centres of high column density gas in many other lines.
The relatively low ratio, compared to that found in NGC\,6334\,I(N) indicates that
the $^{12}$CO emission is likely to be optically thick along most lines of sight
cvered in these observations.

The upper-right image shows the HNCO/C$^{18}$O line ratio map. As mentioned in
\S\ref{comp}, HNCO tends to be destroyed in PDRs. Thus, the line ratio map broadley
shows two regions in the map: a region to the east of NGC\,6334\,I(N) which shows
a high ratio (ie. not a PDR) and a region including NGC\,6334\,I and E, which
appears to be a good candidate PDR, although low signal in HNCO limits our
interpretation beyond a narrow strip joining NGC\,6334\,I and E.

The lower-left image shows the CH$_3$CN/N$_2$H$^+$ line ratio map. As mentioned in
\S\ref{morph}, the CH$_3$CN emission is strongly confined to the centres of active
star formation: NGC\,6334\,I and I(N), whereas N$_2$H$^+$ tends to trace ubiquitous
and quiescent gas. The line ratio map clearly shows low line ratio towards
NGC\,6334\,I, in contrast to a high ratio seen at NGC\,6334\,I(N). This demonstrates
that I(N) is a more quiescent region, presumably with star formation at a younger
stage.

The lower-right image shows the $^{13}$CS/N$_2$H$^+$ line ratio map. Here, we see
the highest line ratio at the position of the \hii~region NGC\,6334\,E. This is
probably the result of N$_2$H$^+$ being destroyed by the \hii~region.

\section{Conclusions}
We have mapped a $5^\prime \times 5^\prime$ region, encompassing the high mass star formation
sites NGC\,6334\,I and I(N). We have covered most of the 3\,mm spectral window with these
observations, using the Mopra radiotelescope. Our observations do not have sufficient
spectral resolution to resolve spectral lines, but we compare morphologies of the various
lines detected. We detect emission from 19 different moleules, ions and radicals, including
multiple transitions of the same species and transitions from various isotopologues to yield
a total of 52 spectral line detections.

We find that CH$_3$CN most closely follows the sites of active star formation, although these sites
feature prominently most other observed species. In contrast, emission from both CN and C$_2$H
appear to be widespread and are good tracers of gas that is not associated with active star
formation. N$_2$H$^+$ and HC$_3$N morphologies closely resemble that of dust continuum emission,
indicating that these two species are good tracers of column density, associated with both
star forming sites and quiescent gas.

Interpretation of the emission from common species and their isotopologues (ie. CO, HCO$^+$, HCN
and HNC) are difficult to interpret, which is likely due to some transitions being optically thick.
All CO isotopologues show a peak of emission about 45\arcsec~to the south-west of the \hii~region
NGC\,6334\,E. However, this peak is not prominent in any other tracer, including dust continuum
emission. We are currently unable to provide an explanation for this.

CH$_3$OH emission is seen in many transitions, including known Class I and II masers. Due to
the low spatial resolution of these observations, we cannot conclude that we have detected any
new masers amongst these new transitions. Thermal CH$_3$OH transitions show a marked difference
in their ocurrence: transitions with upper level kinetic temperatures above 100\,K are found
to be stronger towards NGC\,6334\,I and transitions with upper level kinetic temperatures
below 22\,K are found to be stronger towards NGC\,6334\,I(N). This supports the interpretation
that NGC\,6334\,I is a more evolved site for star formation than I(N).

\section*{Acknowledgments}
The authors would like to thank G\"{o}ran Sandell for providing the SCUBA continuum
images shown in Figure \ref{continuum}. The Mopra Telescope is part of the Australia Telescope and is
funded by the Commonwealth of Australia for operation as a
National Facility managed by CSIRO. The University of New South
Wales Mopra Spectrometer Digital Filter Bank used for the observations
with the Mopra Telescope was provided with support
from the Australian Research Council, together with the University
of New South Wales, University of Sydney, Monash University and the CSIRO.
H.B. acknowledges financial support by the Emmy-Noether-Program of the
Deutsche Forschungsgemeinschaft (DFG, grant BE2578). The authors would like to
thank the anonymous referee whose comments and suggestions have greatly
improved the quality of this paper.


\bsp

\label{lastpage}

\end{document}